\documentclass[%
 reprint,
 amsmath,amssymb,
 aps,
]{revtex4-2}
\usepackage{float}
\usepackage{graphicx}
\usepackage{dcolumn}
\usepackage{bm}
\usepackage{hyperref}

\begin{document}

\preprint{APS/123-QED}

\title{Lower tensor-to-scalar ratio as a possible signature of modified gravity}

\author{Abra\~{a}o J. S. Capistrano}\email{capistrano@ufpr.br}
\affiliation{Universidade Federal do Paran\'{a}, Departmento de Engenharia e Exatas, Rua Pioneiro, 2153, Palotina, 85950-000, Paraná/PR, Brasil\\
Federal University of Latin American Integration (UNILA), Applied physics graduation program, Avenida Tarqu\'{i}nio Joslin dos Santos, 1000-Polo Universit\'{a}rio, Foz do Igua\c{c}u, 85867-670, Paran\'{a}/PR,Brasil}

\author{Rafael C. Nunes}\email{rafadcnunes@gmail.com} 
\affiliation{Universidade Federal do Rio Grande do Sul, Instituto de F\'{i}sica, Porto Alegre, 91501-970, Rio Grande do Sul/RS, Brasil\\
Instituto Nacional de Pesquisas Espaciais, Divis\~ao de Astrof\'isica, Avenida dos Astronautas 1758,S\~ao Jos\'e dos Campos, 12227-010, S\~ao Paulo/SP, Brasil}

\author{Lu\'{i}s A. Cabral}\email{cabral@uft.edu.br}
\affiliation{Universidade Federal do Norte do Tocantins, Araguaína, 77824-838, Tocantins/TO, Brasil}
\date{\today}

\begin{abstract}
This paper simplifies the induced four-dimensional gravitational equations originating from a five-dimensional bulk within the framework of Nash's embeddings, incorporating them into a well-known $\mu-\Sigma$ modified gravity (MG) parametrization. By leveraging data from Planck Public Release 4 (PR4), BICEP/Keck Array 2018, Planck cosmic microwave background lensing, and baryon acoustic oscillation observations, we establish a stringent lower limit for the tensor-to-scalar ratio parameter: $r < 0.0303$ at a confidence level (CL) of 95\%. This finding suggests the presence of extrinsic dynamics influencing standard four-dimensional cosmology. Notably, this limit surpasses those typically obtained through Bayesian analysis using Markov Chain Monte Carlo (MCMC) techniques, which yield $r<0.038$, or through the frequentist profile likelihood method, which yields $r<0.037$ at 95\% CL.
\end{abstract}

\maketitle


\section{Introduction}\label{sec1}

As largely accepted, inflation became one of the cornerstone of modern cosmology. It does not only solve the flatness and horizon problems, but also states the quantum seeds of cosmological fluctuations that eventually drove the universe to evolve~\cite{PhysRevD.23.347,Starobinsky:1979ty,Mukhanov:1981xt,Linde:1983gd,HAWKING1982295,HAWKING1983180,STAROBINSKY1982175,PhysRevLett.49.1110,PhysRevD.28.679} throughout a brief and rapid exponential expansion of the universe right after Big Bang. This inflationary period is thought to have smoothed out the early universe's irregularities and laid the foundation for the large-scale structure we observe today. It is reinforced by the anisotropy seen in Cosmic Microwave Background (CMB) observations~\cite{refplanck2018}. While the precise mechanisms fueling inflation remain elusive, numerous competing theoretical models have emerged~\citep{Giovannini1999,Giovannini1999b,Giovannini2003,cognola2008,Myrzakulov2015,Salvio2017,Oikonomou2017,Agarwal2017,Keskin2018,Salvio2019,Castello2021}, each vying to explain this fundamental cosmic process. However, despite the ongoing debate, the overarching framework consistently yields predictions that align remarkably well with cosmological observations (see \cite{ellis2023inflation, achucarro2022inflation} for a review).

A significant portion of scientific inquiry into the origins of the universe revolves around scrutinizing and characterizing the statistical properties of primordial density perturbations, particularly through the analysis of the statistical two-point function. Empirical evidence suggests that these fluctuations adhere closely to Gaussian distribution and exhibit near-scale invariance. Consequently, we can effectively capture their statistical behavior using a power-law power spectrum governed by two crucial parameters. These parameters, integral to the $\Lambda$CDM standard cosmological model, are the scalar amplitude $A_s$, representing the perturbation amplitude, and the spectral tilt $n_s$, governing the scale dependency of the density perturbation power spectrum. Remarkably, even with just these two parameters, we can glean insights into certain facets of inflationary dynamics today, shedding light on the energy scales pivotal during the early epochs of the universe's evolution. Statistical analyses from canonical inflationary models constrain $n_s = 0.9649 \pm 0.0044$ and $\ln(10^{10} A_s) = 3.044 \pm 0.014 $ at the 68\% confidence level (CL) using Planck-CMB data \cite{2020_CMB}. Conversely, analysis from ACT-CMB indicates agreement with a Harrison-Zel'dovich primordial spectrum, where $n_s = 1.009 \pm 0.015$ \cite{Giar_2023}. This introduces tension with Planck-CMB measurements. For further discussions on this topic, see also \cite{Di_Valentino_2022,Di_Valentino_2023,Calderon_2023}.

Another pivotal parameter arising from inflationary theories is the tensor-to-scalar ratio $r$, indicative of primordial gravitational waves. Multiple CMB experiments have already imposed stringent upper bounds on the amplitude of the tensor spectrum. Notably, the BICEP/Keck collaboration has established the most stringent constraint to date, setting $r < 0.036$ at a 95\% CL \cite{Ade_2021}, effectively challenging certain classes of single-field monomial models. A frequentist proﬁle likelihood method~\cite{Campeti_2022} was also used to investigate the discrepancy with Bayesian analysis using MCMC, identified by the SPIDER collaboration~\cite{SPIDER2022}. As a result, they found an upper limit of $r<0.037$ at 95$\%$ C.L.

In this paper, we propose the possibility for the extrinsic curvature, thought as the orthogonal component of the gravitational field besides the metric $g_{\mu\nu}$, works as an inflaton field. Differently from traditional extra-dimensional braneworld models~\cite{add,RS,RS1,dgp}, the dynamics of extrinsic curvature is considered.  Our \textit{leitmotiv} focus on how a dynamic embedding should affect the physical universe and may provide new insights on the current problems in cosmology. As a main character in this framework, the extrinsic curvature should not be restricted to the analysis of the background geometry but it should also include its perturbative dynamics~\cite{Deruelle,battyecarterPRD,bayttecarterPLB}. To do so, we start from the perturbations of the geometry adopting Nash-Greene theorem~\cite{nash56,greene70} that provides a general structure for embedding between non-Riemannian geometries~\cite{MAIA20029,GDE, Maia_2007,gde2,sepangi,sepangi1,sepangi2,capistrano2021,capistrano2022,capistrano_inflation2023}. This is rather different from common approaches of brane-world perturbative models
in which the perturbations are triggered from  confined sources. Then the dynamics of the extrinsic curvature itself is replaced. A common practice in these models is to rely on junction conditions such as Israel-Darmois-Lanczos~(IDL) condition~\cite{israel}. Commonly used in Randall-Sundrum models~\cite{RS,RS1}, IDL condition replaces the extrinsic curvature by an algebraic relation with the energy-momentum tensor. It was showed it belongs to a very special case~\cite{BATTYE2001331,GDE}, and, in general, IDL condition can be completely removed or replaced~\cite{tsujikawa}. 

The paper is organized in sections. The second section verses on the essentials of embeddings as a mathematical background and summarizes our cosmological model with both background and perturbed cosmological equations. Moreover, the effective fluid approach is presented to develop a more realistic model to compare with observations. In the third section, we present a simpler version of the model in the form of a Modified Gravity (MG) framework. The fourth section is devoted to the analysis of extrinsic curvature as an inflaton field in which we investigate the slow-roll conditions for the extrinsic  potential. In sections V and IV, we examine the model in contrast with cosmological data with the latest NPIPE Planck DR4 likelihoods~\cite{Carron_2022,Rosenberg:2022sdy}, the BICEP2/Keck collaboration \cite{BICEPKeck} and a junction of Large-scale structure (LSS) catalogue with 6dF Galaxy Survey~\cite{6dFGalaxy}, the Seventh Data Release of SDSS Main Galaxy Sample (SDSS DR7 MGS)~\citep{Ross:2014qpa} and clustering measurements of the Extended Baryon Oscillation Spectroscopic Survey (eBOSS) associated with the SDSS's Sixteenth Data Release~\citep{Alam:2016hwk}.  The related joint likelihood analysis is computed with MGCAMB-II~\cite{mgcamb2023} patch by means of Cobaya~\cite{cobaya} sampler. In the final section, we conclude with our remarks and prospects.

\section{Embeddings as a theoretical background}
We present the main results from previous works~\cite{MAIA20029,GDE,Maia_2007,gde2,capistrano2022}. Once a general arbitrary D-dimensional case is possible~\cite{MAIA20029,GDE}, for comparison purposes in the recent literature, we define our model in five dimensions that has a five-dimensional bulk $V_5$ with a embedded four-dimensional geometry $V_4$.   

A gravitational action $S$ is defined in such a way
\begin{equation}\label{eq:action}
S= -\frac{1}{2\kappa^2_5} \int \sqrt{|\mathcal{G}|}{\;}^5\mathcal{R}d^{5}x - \int \sqrt{|\mathcal{G}|}\mathcal{L}^{*}_{m}d^{5}x\;,
\end{equation}
where $\kappa^2_5$ is a fundamental energy scale on the embedded space, the curly $^5\mathcal{R}$ curvature means the five-dimensional Ricci scalar, and the matter Lagrangian $\mathcal{L}^{*}_{m}$ denotes the confined matter fields on a four-dimensional embedded space-time. 

The non-perturbed extrinsic  curvature $\bar{k}_{\mu\nu}$ is defined as~\cite{eisen},
\begin{equation}
\bar{k}_{\mu\nu} =  -\mathcal{X}^A_{,\mu}\;\bar{\eta}^B_{,\nu} \mathcal{G}_{AB}\;, \label{eq:extrinsic} 
\end{equation}
where Eq.(\ref{eq:extrinsic}) shows the projection of the  variation of the set of normal unitary vector $\bar{\eta}^B$ onto the tangent plane  orthogonal  to  the embedded space $V_4$. In other words, the variation of  $\bar{\eta}^B$ leads to the bending of $V_4$ and it has its tangent components with coefficients $\bar{k}_{\mu\nu}$. The embedding coordinate $\mathcal{X}$ defines a regular local map $\mathcal{X}: V_4 \rightarrow  V_5$ and must satisfy the  embedding  equations
\begin{equation} \label{eq:X}
\mathcal{X}^A_{,\mu} \mathcal{X}^B_{,\nu}\mathcal{G}_{AB}=g_{\mu\nu},\;  \mathcal{X}^A_{,\mu}\bar{\eta}^B \mathcal{G}_{AB}=0,  \;  \bar{\eta}^A \bar{\eta}^B \mathcal{G}_{AB}=1,
\end{equation}
where  we have  denoted by  $\mathcal{G}_{AB}$ the metric components of the bulk $V_5$ in  arbitrary  coordinates. The embedding frame is defined by the set of coordinates $\{\mathcal{X}^A,\bar{\eta}^A\}$ that composes a Gaussian reference frame. Throughout  the  paper, except when  explicitly  stated in  contrary, the overbar sign indicates a background (non-perturbed) quantity.

The bulk metric $\mathcal{G}_{AB}$ is defined as
\begin{eqnarray}
  \mathcal{G}_{AB} &=& \left(
          \begin{array}{cc}
            g_{\mu\nu} & 0 \\
            0 & 1 \\
          \end{array}
        \right)\;.
   \label{eq:metricbulk}
\end{eqnarray}
Concerning notation, capital Latin indices run from 1 to 5. Small case Latin indices refer to the only one extra dimension considered. All Greek indices refer to the embedded space-time counting from 1 to 4. 

In this paper, we use Nash theorem of 1956~\cite{nash56}. The seminal result of this theorem shows how to produce orthogonal perturbations from the background metric $\bar{g}_{\mu\nu}$ given by
\begin{equation}\label{eq:nashdeformation}
 \bar{k}_{\mu\nu}=-\frac{1}{2}\frac{\partial \bar{g}_{\mu\nu} }{\partial y}\;.
\end{equation}
where $y$ is an arbitrary spatial coordinate along the orthogonal direction to the tangent plane. This mechanism avoids false perturbations due to the possibility to induce coordinate gauges. Therefore, the physical effects of the extrinsic curvature associated with Eq.(\ref{eq:nashdeformation}) represent an acceleration tangent to the four-dimensional space-time that always points to the concave side of the curve.  As a result, it induces a Riemann stretching on the space-time geometry, which may be related to the accelerated expansion of the universe \cite{GDE,gde2}. 

As a consequence of Eq.(\ref{eq:nashdeformation}), we have how new geometries $g_{\mu\nu}$ are generated from small perturbations on the background metric increments $\delta g_{\mu\nu}$ as
\begin{equation}\label{eq:nashdeformation1}
g_{\mu\nu}=\;\bar{g}_{\mu\nu}+\delta g_{\mu\nu}\;.
\end{equation}
Moreover, one obtains the perturbed extrinsic curvature $k_{\mu\nu}$ as
\begin{equation}\label{eq:curvextrperturb}
k_{\mu\nu}=\;\bar{k}_{\mu\nu}+\delta k_{\mu\nu}\;,
\end{equation}
where $\delta k_{\mu\nu}=-2\delta y~g^{\sigma\rho}k_{\mu\sigma}k_{\nu\rho}$. In principle, this is a continuous process to any arbitrary perturbation increments of superior orders of $\delta g_{\mu\nu}$ and $\delta k_{\mu\nu}$. For our purposes, we keep the perturbations at linear order as show in Eq.(\ref{eq:nashdeformation1}) and (\ref{eq:curvextrperturb}).

The increments $\delta g_{\mu\nu}$ and $\delta k_{\mu\nu}$ are the Nash's fluctuations later applied to non-Riemannian metrics by Greene~\cite{greene70}. The deformation formula in~Eq.(\ref{eq:nashdeformation}) is also a pivot element to obtain solutions of the Gauss and Codazzi equations 
\begin{eqnarray}
\nonumber&&^5{\cal R}_{ABCD}\mathcal{Z}^A_{,\alpha}\mathcal{Z}^B_{,\beta}\mathcal{Z}^C_{,\gamma}
\mathcal{Z}^D_{,\delta}= \bar{R}_{\alpha\beta\gamma\delta}\;+\\
&&\hspace{4cm}+\;(\bar{k}_{\alpha\gamma}\;\bar{k}_{\beta\delta}\!-\!\;\bar{k}_{\alpha\delta}\;\bar{k}_{\beta\gamma})\!\;,\label{eq:G1}\\
&&^5{\cal R}_{ABCD}\mathcal{Z}^A_{,\alpha} \mathcal{Z}^B_{,\beta}\mathcal{Z}^C_{,\gamma}\eta^D=\;\bar{k}_{\alpha[\beta;\gamma]} \;, \label{eq:C1}
\end{eqnarray}
where $^5{\cal R}_{ABCD}$ is the five-dimensional Riemann tensor and $\bar{R}_{\alpha\beta\gamma\delta}$ is the background four-dimensional Riemann tensor. The perturbed embedding coordinate $\mathcal{Z}^{A}_{,\mu}$ is defined as $\mathcal{Z}^{A}_{,\mu}=  \mathcal{X}^{A}_{,\mu} + \delta y \;  \eta^{A}_{,\mu}\;$. The normal vector $\eta^{A}$ is invariant under perturbations, i.e., $\eta^{A}= \bar{\eta}^{A}$. The semicolon in~Eq.(\ref{eq:C1}) denotes covariant derivative with respect to the metric and the brackets apply the covariant derivatives to the adjoining indices. In Eq.(\ref{eq:G1}), the Gauss equation entwines the bulk Riemann curvature as  a  reference standpoint for the  Riemann  curvature  of  the   embedded  space-time. This relation is complemented by the  Codazzi equation in Eq.(\ref{eq:C1}), which shows the  projection of  the  Riemann  tensor  of  the bulk space  along the  orthogonal  direction  results in the variation of  the  extrinsic  curvature.

In order to guarantee the possibility to generate new geometries, in five-dimensions, the  set of  coordinates  $\mathcal{Z}^A$  also needs to  satisfy embedding  equations similar  to  Eq.(\ref{eq:X}):
\begin{equation}
\mathcal{Z}^A{}_{,\mu} \mathcal{Z}^B{}_{,\nu} \mathcal{G}_{AB}=g_{\mu\nu},\;  \mathcal{Z}^A{}_{,\mu}\eta^B \mathcal{G}_{AB}=0,  \;  \eta^A \eta^B \mathcal{G}_{AB}=1\;, \label{eq:Z}
\end{equation} 
then, Eq.(\ref{eq:nashdeformation1}) and (\ref{eq:curvextrperturb}) are valid.

In a similar fashion as Kaluza-Klein  and brane-world models,  we consider that the  bulk   $V_5$  has its dynamics governed by  the  higher-dimensional Einstein's  equations
\begin{equation}
^5{\mathcal R}_{AB} -\frac{1}{2} \,{^5{\mathcal R}} {\mathcal G}_{AB}  =G_{*}  T^*_{AB} \label{eq:BE0}
\end{equation}
where  $G_*$  is the  new  gravitational  constant  and $T^*_{AB}$ denotes the components  of  the energy-momentum tensor of the  material sources. Those sources are confined to four dimensions. It  is  a  consequence  of the isomorphism between the three-form, from the derivative of the Yang-Mills curvature, and one-form current. This is  valid only  in  four  dimensions. Thus,  all  known  observable  sources  of gravitation  composing the generic energy-momentum tensor $T^*_{AB}$ are  also  confined. This outcome is independent of  the variation of extra coordinate  $y$. Otherwise speaking, the four-dimensionality of space-time is a consequence of the invariance of Maxwell's equations under the Poincaré group. It is well known that any gauge theory can be mathematically constructed or extended in a higher dimensional space, but in the present framework, the four-dimensionality of the embedded space-time will suffice based on what experimentally high-energy tests suggest~\cite{Lim2014,Mieling2022,Feng2024}. Moreover, in  the  Gaussian  frame $\{\mathcal{Z}^A,\eta^A\}$ of  any perturbed  space-time, we can write  the confinement condition of the energy-momentum  tensor  source $T^*_{AB}$ of the bulk Einstein's equation in Eq.(\ref{eq:BE0})  with the projections
\begin{equation}
8\pi G T_{\mu\nu} =G_* \mathcal{Z}^A_{,\mu}\mathcal{Z}^B_{,\nu}T^*_{AB}, \mathcal{Z}^A_{,\mu}\eta^B T^*_{AB}=0, \eta^A  \eta^B T^*_{AB}=0, \label{eq:confinement}
\end{equation}

In this framework, the matter content is localized in the $V_4$ embedded space due to the fact that Nash's deformation formula in Eq.(\ref{eq:nashdeformation}) imposes a geometric constraint on the confined sources. Any deformation is not arbitrary and can be generated by smooth perturbations along the direction $\delta y$ orthogonal to the embedded space $V_4$. In five dimensions, this process is simplified and just one deformation parameter suffices to locally deform the embedded background. In general, the curvature radii $l^a$ of the embedded background correspond to the direction in which the embedded space-time deviates  more sharply from the tangent plane and are the solutions of the homogeneous equation
\begin{equation}\label{eq:curvradii}
\mbox{det}(g_{\mu\nu}- l^a k_{\mu\nu a})=0\;.
\end{equation}
This is a local invariant property of the embedded space-time and does not depend on the chosen Gaussian system. The smallest solution $l$ provides 
\begin{equation}\label{eq:curvradii2}
\frac{1}{l}=\left(g_{\mu\nu} \mathcal{G}^{AB} l_{A}^{\mu} l_{B}^{\nu}\right)^{-1/2}\;.
\end{equation}
Otherwise speaking, $l$ constrains a local limit for the region in which the bulk is accessed by the gravitons. Then, one can find the typical length $d$~\cite{MAIA20029} of the n-extra dimensional space accessed by gravitons as
\begin{equation}\label{eq:typicallenght}
d=\frac{M_{Pl}^{2/n}}{M_{\ast}^{1+(2/n)}}\frac{1}{(1+\frac{M_{Pl}^2}{M_{e}^2})^{1/n}}   \;,
\end{equation}
where $M_{\ast}$ and $M_{Pl}$ are respectively the fundamental and effective Planck scales. The extrinsic scale $M_e$ is given by
\begin{equation}\label{eq:typicallenght2}
\frac{1}{M_{e}}=\int (K^2+h^2)\sqrt{g}d^4x\;,
\end{equation}
where $K^2$ is the Gaussian curvature and $h^2$ is the mean curvature. Hence, for smooth oscillations of the embedded background, the limit imposed by Eq.(\ref{eq:typicallenght}) with the $M_e$ scale, prevents the leak of energy of the confined sources to higher dimensional space but it allows the graviton oscillations. This eliminates the necessity of  introduction of a radion field, e.g., commonly adopted in most popular braneworld models~\cite{add,RS,RS1,dgp}.

\subsection{Background cosmology}
The background cosmology is defined as usual by means of the line element of Friedmann-Lemaître-Robertson-Walker (FLRW) four-dimensional metric as
\begin{equation}
  ds^2= -dt^2 + a^2\left(dr^2+r^2d\theta^2 + r^2\sin^2\theta d\phi^2 \right)\;,
\end{equation}
where the scale factor is denoted by $a\equiv a(t)$ and $t$ is the physical time. Once the embedding is properly set, we impose that the bulk geometry is a solution of Einstein’s equations given by Eq.(\ref{eq:BE0}). Thus, from~Eq.(\ref{eq:G1}) and ~Eq.(\ref{eq:C1}), and the confinement condition in~Eq.(\ref{eq:confinement}), one obtains the tangent components of non-perturbed field equations as
\begin{eqnarray}
  \bar{G}_{\mu\nu}- \bar{Q}_{\mu\nu} &=&-8\pi G \bar{T}_{\mu\nu}\;,  \label{eq:noperttensoreq}\\
  \bar{k}_{\mu[\nu;\rho]} &=& 0 \;,
  \label{eq:nopervecteq}
\end{eqnarray}
where $\bar{T}_{\mu\nu}$ is the non-perturbed energy-momentum tensor of the confined perfect fluid and $G$ is the gravitational Newtonian constant. The background tensors $\bar{G}_{\mu\nu}$ and $\bar{Q}_{\mu\nu}$ represent the four dimensional Einstein tensor and the {\it extrinsic deformation tensor}, respectively. For an arbitrary D-dimensional case, see the detail derivation of~Eq.(\ref{eq:noperttensoreq}) and~Eq.(\ref{eq:nopervecteq}) in Ref.~\cite{GDE}.

In addition, the non-perturbed $\bar{Q}_{\mu\nu}$ in Eq.(\ref{eq:noperttensoreq}) is defined as
\begin{equation}\label{eq:qmunu}
  \bar{Q}_{\mu\nu}=\bar{k}^{\rho}_{\mu}\bar{k}_{\rho\nu}- \bar{k}_{\mu\nu }h -\frac{1}{2}\left(K^2-h^2\right)\bar{g}_{\mu\nu}\;,
\end{equation}
where $h^2=h\! \cdot \! h$ denotes the mean curvature with $h= \;\bar{g}^{\mu\nu}\;\bar{k}_{\mu\nu}$. The Gaussian curvature is denoted by $K^{2}=\bar{k}^{\mu\nu}\bar{k}_{\mu\nu}$. A direct consequence of the previous definition in~Eq.(\ref{eq:qmunu}) is that the deformation tensor $\bar{Q}_{\mu\nu}$ is a conserved quantity such as
\begin{equation}\label{eq:qmunuconserv}
  \bar{Q}_{\mu\nu;\mu}=0\;.
\end{equation}

Solving the trace of Codazzi equation in~Eq.(\ref{eq:nopervecteq}) and Eq.(\ref{eq:qmunu}), the following components were found~\cite{GDE} 
\begin{eqnarray}
\label{eq:BB}
&&\bar{k}_{ij}=\frac{b}{a^2}\bar{g}_{ij},\;\;i,j=1,2,3,\\
&&\bar{k}_{44}=\frac{-1}{\dot{a}}\frac{d}{dt}\frac{b}{a}\\
 &&
 \bar{k}_{44}=-\frac{b}{a^{2}}\left(\frac{B}{H}-1\right)\;, \\
&&K^{2}=\frac{b^2}{a^4}\left( \frac{B^2}{H^2}-2\frac BH+4\right),
 \;\, h=\frac{b}{a^2}\left(\frac BH+2\right),\label{eq:hk}\\
&&\bar{Q}_{ij}= \frac{b^{2}}{a^{4}}\left( 2\frac{B}{H}-1\right)
\bar{g}_{ij},\;\bar{Q}_{44} = -\frac{3b^{2}}{a^{4}},
  \label{eq:Qab}\\
&&\bar{Q}= -(K^2 -h^2) =\frac{6b^{2}}{a^{4}} \frac{B}{H}\;, \label{Q}
 \end{eqnarray}
where $\bar{Q}$ denotes the deformation scalar defined in a standard way, i.e., by the contraction $\bar{g}^{\mu\nu}\bar{Q}_{\mu\nu}=\bar{Q}$. One important definition is the evolution of the bending function $b(t)\equiv b= k_{11}=k_{22}=k_{33}$ driven by extrinsic geometry. Thus, we defined $B=B(t)\equiv \frac{\dot{b}}{b}=(db/dt)/b$ as a copy of the Hubble parameter $H\equiv H(t)=\frac{\dot{a}}{a}=(da/dt)/a$. As a consequence, the $B(t)$ function borrows the same unit of $H$. 

An important aspect is that in five dimensions, the trace of Codazzi equation in~Eq.(\ref{eq:nopervecteq}) is homogeneous which turns the solution for the bending function $b(t)$ arbitrary.  To remove such arbitrariness, we need to state the dynamics of the extrinsic curvature, since the  metric and the extrinsic  curvature are independent variables that must satisfy the Gauss~(Eq.(\ref{eq:G1})) and Codazzi~(Eq.(\ref{eq:C1})) equations.  Thus, one counts a  total  of 20 unknowns  $g_{\mu\nu}$   and  $k_{\mu\nu}$,  against  only   15  dynamical  equations. This requires to consider $k_{\mu\nu}$ as a source for the missing equations. A   well  known  theorem due to   S. Gupta~\cite{Gupta} states  that any symmetric  rank-2  tensor satisfies an  Einstein-like  system of  equations, having  the  Pauli-Fierz  equation  as its linear  approximation. As proposed  by Isham, Salam, and Strathdee~\cite{salam} in the context of strong  gravity, they propose a f-meson spin-2 field that would act as  an intermediate  field between  gravitation  and hadron particles. In a previous publication \cite{gde2}, an Einstein-like  dynamical  equation  for  the   extrinsic  curvature was adapted  from  the  original  equation  of  S. Gupta. Using Gupta's theorem, as shown in~\cite{gde2,capistrano2022}, one simply obtains
\begin{equation}\label{eq:b(t)funct}
b(t)=b_0 a(t)^{\beta_0}\;,
\end{equation}
where the term $b_0$ is the current value of the bending function and $\beta_0$ is an integration constant. As we are going to show, it will be associated with the fluid parameter $w$. Accordingly, the Friedmann equation in terms of redshift can be written in a form
\begin{eqnarray}\label{eq:nonpertfriedtotal3}
\nonumber&&\left(\frac{H}{H_0}\right)^2=\Omega_{m(0)}(1+z)^3 +\Omega_{rad0} (1+z)^{4}\\
&&\hspace{4cm}+\Omega_{ext(0)}(1+z)^{4-2\beta_0},
\end{eqnarray}
where $\Omega_{m(0)}$ denotes the current cosmological parameter for matter density content. The radiation content is denoted by  $\Omega_{rad(0)}=\Omega_{m0}z_{eq}$, wherein the equivalence number for the expansion factor $a_{eq}$ is 
\begin{equation}\label{eq:equiv re}
  a_{eq}= \frac{1}{1+z_{eq}}=\frac{1}{(1 + 2.5\times 10^4 \Omega_{m(0)} h^2 (T_{cmb}/2.7)^{-4})}\;,
\end{equation}
where $z_{eq}$ is the equivalence redshift. The CMB temperature is adopted for the value $T_{cmb} = 2.7255 K$ and the Hubble factor $h=0.67$~\cite{refplanck2018}. 
The term $\Omega_{ext(0)}$ stands for the density parameter associated with the extrinsic curvature. For a flat universe, $\Omega_{ext(0)}= 1- \Omega_{m(0)}-\Omega_{rad(0)}$. Moreover, the current extrinsic cosmological parameter $\Omega_{ext(0)}$ is defined as
\begin{equation}\label{eq:extomega}
\Omega_{ext(0)}=\frac{8\pi G}{3H_0^2}\rho_{ext(0)}\equiv \frac{b_0^2}{H_0^2a_0^{\beta_0}}\;,
\end{equation}
where $a_0$ sets the current value of the scale factor and $\rho_{ext(0)}\equiv \frac{3}{8\pi G}b^2_0 $ denotes the current extrinsic density.

\subsection{Perturbations in Conformal Newtonian gauge}
The perturbed equations are necessary for the right estimation of cosmological parameters. In this framework, the relevant modifications of field equations under cosmological perturbations are applied to LHS of~Eq.(\ref{eq:noperttensoreq}) with the presence of $\bar{Q}_{\mu\nu}$, and $\bar{G}_{\mu\nu}$ and $\bar{T}_{\mu\nu}$ tensors that are treated in a very standard fashion as shown in Ref.~\cite{gde2}. Thus, a fluid with pressure $P$ and density $\bar{\rho}$, the perturbed components of the stress-tensor energy $\delta T_{\mu\nu}$ are given as
\begin{eqnarray}
  \delta T^i_j &=& \delta p\;\delta^i_j+ \Sigma_{j}^{i}\;,\\
  \delta T^4_4 &=& -\delta \rho\;, \\
  \delta T^4_i &=& \frac{1}{a}(\bar{\rho}+P)\delta u_{\parallel i}\;,
\end{eqnarray}
where $\delta u_{\parallel i}$ is the tangent velocity potential of the fluid. The anisotropic stress tensor is defined as $\Sigma_{j}^{i}= T^{i}_{j}-\delta^{i}_{j}T^{k}_{k}/3$. Moreover, the only relevant field equations that propagate cosmological perturbations are given by
\begin{equation} 
\delta G^{\mu}_{\nu}-\delta Q^{\mu}_{\nu}= -8\pi G\delta T^{\mu}_{\nu}\;. \label{eq:perturbgravitensor} 
\end{equation}

In addition, we have to determine the perturbed extrinsic terms given by $\delta Q_{\mu\nu}$. Using the Nash relation of~Eq.(\ref{eq:nashdeformation}) and~Eq.(\ref{eq:curvextrperturb}), one obtains
\begin{equation}\label{perturbkmunu}
\delta{k}_{\mu\nu}=\;\bar{g}^{\sigma\rho}\;\bar{k}_{\mu\sigma}\;\delta g_{\nu\rho}\;.
\end{equation}
This is a pivot result since it shows how the effects of the extrinsic quantities are projected onto the perturbed four embedded space-time and on how Nash flow of~Eq.(\ref{eq:nashdeformation}) is connected to cosmological perturbations. Hence, the perturbations of the deformation tensor $\bar{Q}_{\mu\nu}$ are given in a form
\begin{equation}\label{Qmunu}
\delta Q_{\mu\nu}= -\frac{3}{2}(K^2-h^2)\delta g_{\mu\nu} \;.
\end{equation}

Alternatively, we construct a relation of the equations in a fluid approach writing the gravi-tensor equation in a general form as
\begin{equation}
  G_{\mu\nu}=-8\pi G T^{total}_{\mu\nu}\;,\label{eq:BE3}
\end{equation}
where the tensors are written by the composition of their background and perturbed components. The related Einstein tensor is written as $G_{\mu\nu} = \bar{G}_{\mu\nu}+\delta G_{\mu\nu}$, $T^{total}_{\mu\nu}=T_{\mu\nu}+ T^{ex}_{\mu\nu}$, where $T_{\mu\nu} = \bar{T}_{\mu\nu}+\delta T_{\mu\nu}$ and $T^{ex}_{\mu\nu} = \bar{T}^{ex}_{\mu\nu}+\delta T^{ex}_{\mu\nu}$. 

In the conformal Newtonian gauge, the FLRW metric is given by
\begin{equation}\label{eq:scalarpertmetric2}
ds^2 = a^2 [-(1+ 2\Psi) d\tau^2 + (1-2\Phi) dx^i dx_i] \;,
\end{equation}
where $\Psi=\Psi(\vec{x},\tau)$ and $\Phi=\Phi(\vec{x},\tau)$ denote the Newtonian potential and the Newtonian curvature in conformal time $\tau$ that is defined as $d\tau=dt/a(t)$. Hence, we determine the confined matter fields represented by the non-perturbed stress energy tensor $\bar{T}_{\mu\nu}$ in a co-moving fluid such as
\begin{equation}\label{eq:stresstensor}
\bar{T}_{\mu\nu}=\left(\bar{\rho}+P\right) U_{\mu}U_{\nu} + P \bar{g}_{\mu\nu}\;;\;U_{\mu}=\delta^4_{\mu}\;,
\end{equation}
where $U_{\mu}$ denotes the co-moving velocity. The related conservation equation is given by
\begin{equation}\label{eq:nonpertdensity}
  \bar{\rho} + 3H\left(\bar{\rho} + P\right)=0\;.
\end{equation}
From the perturbed conservation equation $\delta{T}_{\mu\nu;\nu}=0$, the evolution equations can be obtained for the ``contrast'' matter density $\delta_m$ and fluid velocity $\theta$ such as
\begin{eqnarray}\label{eq:perturbmatter}
&&\dot{\delta_m} = -(1+w)(\theta-3\dot{\Phi})-3H(c_s^2-w)\delta_m\;, \\
\nonumber&&\dot{\theta} = -H(1-3w)\theta- \frac{\dot{w}}{1+w}\theta+\frac{c_s^2}{1+w}k^2\delta_m+\\
&&\hspace{5cm}-k^2\sigma+k^2\Psi\;, 
\end{eqnarray}
where $\theta=ik^ju_j$, $w$ is the fluid parameter $w=\frac{P}{\bar{\rho}}$, $c^2_s$ is the sound velocity defined as $c^2_s=\frac{\delta P}{\delta \rho}$ and $\sigma$ is the anisotropic stress. The dot symbol denotes the ordinary derivative with respect to conformal time $\tau$. 

To avoid divergences when Equation of State~(EoS) crosses $w=-1$, eventually, the scalar velocity $V=(1+w)\theta$ should be defined~\cite{sapone2009,nesseris2019}. Hence, we have the equations
\begin{eqnarray}\label{eq:pertmatter}
&&  \delta'_m = 3(1+w)\Phi'- \frac{V}{a^2H}-\frac{3}{a}\left(\frac{\delta P}{\bar{\rho}}-w\delta_m\right)\;, \\
\nonumber&&  V' = -(1-3w)\frac{V}{a}+\frac{k^2}{a^2H}\frac{\delta P}{\bar{\rho}}+(1+w)\frac{k^2}{a^2H}\Psi+\\
&&\hspace{4.5cm}-\frac{k^2}{a^2H}(1+w)\sigma,
\label{eq:velocitypertmatter}
\end{eqnarray}
where the prime symbol $'$ denotes the ordinary derivative with respect to scale factor as  $'=\frac{d }{da}$.

For the induced extrinsic part, we have that the term $8\pi G\bar{T}^{ext}_{\mu\nu}\doteq \bar{Q}_{\mu\nu}$ which is written as copy of a perfect fluid as
\begin{equation}\label{eq:energy tensor2}
-8\pi G\bar{T}^{ext}_{\mu\nu}=(\bar{p}_{ext}+\bar{\rho}_{ext})U_{\mu}U_{\nu}+\bar{p}_{ext}\;\bar{g}_{\mu\nu},\;\;\;U_{\mu}=\delta_{\mu}^{4}\;,
\end{equation}
where $U_{\mu}$ is the co-moving four-velocity. Since $T_{\mu\nu;\nu}=0$ and $T^{ext}_{\mu\nu;\nu}=0$ (as a consequence of~Eq.(\ref{eq:qmunuconserv})), $T^{total}_{\mu\nu}$ is conserved, accordingly. Hence, the conservation equation for extrinsic quantities is given by
\begin{equation}\label{eq:pertdensity2}
  \frac{d\bar{\rho}_{ext}}{dt} + 3H\left(\bar{\rho}_{ext} + \bar{p}_{ext}\right)=0\;,
\end{equation}
where $\bar{\rho}_{ext}$ and $\bar{p}_{ext}$ denote the non-perturbed extrinsic density and extrinsic pressure, respectively. The time-time and space-time components of $T^{ext}_{\mu\nu}$ can be set as
\begin{eqnarray}
&&-8\pi G T^{ext}_{44} = \bar{T}^{ext}_{44} + \delta T^{ext}_{44}=\bar{Q}_{44}+\delta Q_{44}\;, \label{eq:extrqmunu1}\\
&&-8\pi G T^{ext}_{i4} = \bar{T}^{ext}_{i4} + \delta T^{ext}_{i4}= \bar{Q}_{i4}+\delta Q_{i4}=0\;.\label{eq:extrqmunu2}
\end{eqnarray}

The modified Friedman equation from is written as
\begin{equation}\label{eq:Friedman2}
H^2=\frac{8}{3}\pi G \left(\bar{\rho}_{m}+\bar{\rho}_{rad}+\bar{\rho}_{ext}\right)\;\;,
\end{equation}
where $\bar{\rho}_{ext}(a)$ is given by
\begin{equation}\label{eq:extdensitya1}
\bar{\rho}_{ext}(a)=\bar{\rho}_{ext}(0)a^{2\beta_0-4}\;,
\end{equation}
with $\bar{\rho}_{ext}(0)=\frac{3}{8\pi G} b_0^2$. Once $\bar{\rho}_{ext}(a)$ is already determined, the ``extrinsic'' pressure can be calculated using Eqs.(\ref{eq:pertdensity2}) and (\ref{eq:extdensitya1}) to obtain
\begin{equation}\label{eq:extpressure}
\bar{p}_{ext}(a)=\frac{1}{3}\left(1-2\beta_0\right)\bar{\rho}_{ex}(0)a^{2\beta_0-4}\;.
\end{equation}
Concerning the Null Energy Conditions~(NEC), for positive values of $\beta_0$, NEC is satisfied. If $\beta_0=2$, the extra term in the modified Friedmann equation in~Eq.(\ref{eq:nonpertfriedtotal3}) mimics a cosmological constant. If $\beta_0>2$, the effective behavior becomes phantom-like pattern. However, the condition $\beta_0>1$ is a must if the universe must accelerate (apart from being consistent with NEC). Thus, we define an effective EoS with an ``extrinsic fluid'' parameter $w_{ext}$ by the definition $w_{ext}=\frac{\bar{p}_{ext}}{\bar{\rho}_{ext}}$ to obtain
\begin{equation}\label{eq:wext}
w_{ext}=-1+\frac{1}{3}\left(4-2\beta_0\right)\;.
\end{equation}
which when 
\begin{equation}\label{eq:wext2}
\beta_0=2-\frac{3}{2}\left(1+w\right), 
\end{equation}
one has the fluid correspondence $w_{ext}=w$. This allows us to express all the relevant quantities in a fluid approach. For instance, the dimensionless Hubble parameter $E(z)=\frac{H(z)}{H_0}$ is written as
\begin{eqnarray}\label{eq:dimenHub1}
\nonumber&&E^2(z)=\Omega_{m(0)}(1+z)^3+\Omega_{rad(0)} (1+z)^{4}+ \\
&&\hspace{4cm}+\Omega_{ext(0)}(1+z)^{3(1+w)}.
\end{eqnarray}
Just for referencing purposes, we name this model as $\beta$-model. 

From Eq.(\ref{Qmunu}), we calculate the set of non-zero components of $\delta Q_{\mu\nu}$ using the background relations in Eqs.(\ref{eq:BB}), (\ref{eq:hk}), (\ref{eq:Qab}) and (\ref{Q}). Then, one obtains
\begin{eqnarray}
\delta Q^{4}_{4}&=& \gamma_0 a^{-(1+3w)}\Psi \delta^4_4\;,\\
\delta Q^{i}_{j}&=& \gamma_0 a^{-(1+3w)}\Phi \delta^i_j\;,\\
\delta Q^{i}_{i}&=& 3\gamma_0 a^{-3(1+w)} \Phi\;,\\
\delta Q&=& \gamma_0 (3\Psi-\Phi)a^{-3(1+w)}\;.
\end{eqnarray}
For the sake of notation and using Eq.(\ref{eq:extomega})  in the foregoing set of equations, $\gamma_0$ denotes
\begin{equation}\label{eq:gamma0}
\gamma_0=18b_0^2\beta_0=9 H_0^2\Omega_{ext(0)}\gamma_s\;,
\end{equation}
where $\beta_0$ was simplified by the introduction of the $\gamma_s$ parameter that is regarded as a relic from extrinsic geometry to maintain the characteristic of the parameter $\gamma_0$. Moreover, it turns $\gamma_0$ in Eq.(\ref{eq:gamma0}) independent of the nature of the fluid characterized by $w$ parameter, as expected. This is also necessary to maintain the RG correspondence that is reached when extrinsic curvature vanishes, i.e., $\gamma_s\rightarrow 0$. 

From Eq.(\ref{eq:BE3}) one writes the gauge invariant perturbed field equations in the Fourier \emph{k}-space wave modes as
\begin{eqnarray}
&&k^2\Phi_k + 3 \mathcal{H} \left(\Phi^{'}_k + \Psi_k \mathcal{H} \right)= -4\pi G a^2 \delta \rho_k + \chi(a)\Psi_k , \label{tensorcompo00kspace}\\
&&\Phi^{'}_{k}+ \mathcal{H}\Psi_k= -4\pi G a^2(\bar{\rho}+P) \frac{\theta}{k^2}\;,\label{tensorcomp0ikspace}\\
&&\mathcal{D}_k + \frac{k^2}{3}(\Phi_k-\Psi_k)= -\frac{4}{3}\pi G a^2 \delta \bar{P}-\frac{1}{2}a^2 \delta Q^{i}_{i}\;, \label{tensorcompijkspace}\\
&&k^2(\Phi_{k}-\Psi_k)= 12\pi G a^2(\bar{\rho}+P) \sigma,\label{offdiagonal}
\end{eqnarray}
where $\theta= ik^j\delta u_{\parallel j}$ denotes the divergence of fluid velocity in \emph{k}-space, and $\mathcal{D}_k$ denotes $\mathcal{D}_k=\Phi^{''}_{k} + \mathcal{H}(2\Phi_k+\Psi_k)' + (\mathcal{H}^2+2\mathcal{H}')\Psi_k$. The function $\chi(a)$ is expressed in terms of the cosmological parameters and reads \begin{equation}\label{eq:dimenHub}
\chi(a)= \frac{9}{2}\gamma_s \frac{H_0^2}{\Omega_{rad}(a)}\Omega_{rad}(0) \Omega_{ext}(a) \;.
\end{equation}

\section{Embedding as a modified gravity model}
The use of the set of perturbed equations in Eqs.(\ref{tensorcompo00kspace}), (\ref{tensorcomp0ikspace}, (\ref{tensorcompijkspace}) and (\ref{offdiagonal}) is simplified after some algebra into the following set of equations 
\begin{eqnarray}
&&k^2\Psi_k = -4\pi G a^2 \mu(a,k)\rho \Delta\;,\label{muequation1}\\
&&k^2(\Phi_{k}+\Psi_k)= -8\pi G a^2 \Sigma(a,k)\rho\Delta\;,\label{sigmaequation1}
\end{eqnarray}
where $\rho \Delta= \bar{\rho}\delta+ 3\frac{\mathcal{H}}{k}(\bar{\rho}+P)\theta$. The set of Eqs.(\ref{muequation1}) and (\ref{sigmaequation1}) are valid for all times. When anisotropic stress is neglected, $\mu(a,k)$ and $\Sigma(a,k)$ functions can be written as
\begin{eqnarray}
&&\mu(a,k) = \frac{1}{1-\frac{\chi(a)}{k^2}} \;,\label{muequation}\\
&&\Sigma(a,k)= \frac{1}{2}\left[1+\mu(a,k)\left(1+\frac{\chi(a)}{k^2}\right)\right]\;.\label{sigmaequation}
\end{eqnarray}

Using the definition of the slip function $\gamma(a,k)=\frac{\Phi}{\Psi}$, from Eqs. (\ref{muequation}) and (\ref{sigmaequation}), one easily obtains
\begin{equation}\label{eq:mg3functions}
\Sigma(a,k)= \frac{1}{2}\mu(a,k)(1+\gamma(a,k))\;.
\end{equation}
When the extrinsic term $\gamma_s\rightarrow 0$ in order to recover GR correspondence, one obtains the standard GR limit as  $\Sigma(a,k)=\mu(a,k)$ and $\gamma(a,k)=1$. Thus, the set of basic equations is complete with the matter perturbation equations in Eqs.(\ref{eq:pertmatter}) and (\ref{eq:velocitypertmatter}), and the evolution equation of $\Phi$.  To obey solar constraints, $\gamma_s$ in $\mu(a,k)$ function must comply with the condition 
\begin{equation}\label{eq:gammas2}
\gamma_s< \frac{0.222 k^2_p}{H_0^2\Omega_{ext(0)}}\;,
\end{equation}
at pivot scale wave-number $k_p$. For instance, adopting baseline mean values of 68$\%$ intervals of base-$\Lambda$CDM model from Planck TT,TE,EE+lowE+lensing~\cite{refplanck2018}, we obtain $\gamma_s<3.219\times 10^{-8}$, which means that deviation of modified gravity should be pronounced below that cutoff.

\section{Extrinsic curvature as an effective inflaton field}
In previous publications~\citep{gde2,capistrano2021,capistrano2022}, we explored some of the consequences of the Nash's embedding in the context of dark energy problem. The appearance of the extrinsic energy density $\bar{\rho}_{ext}$ drives the universe to the late accelerated expansion and Friedmann equation is written in a shorter form as
\begin{equation}\label{eq:Friedman3}
H^2=\frac{\kappa^2}{3}\left(\bar{\rho}+\bar{\rho}_{ext}\right)\;,
\end{equation}
with $\kappa=8\pi G$ and $\bar{\rho}=\bar{\rho}_{m}+\bar{\rho}_{rad}$. In terms of inflationary cosmology, such energy density $\bar{\rho}_{ext}$ should provide a response in a form of a scalar field potential $V(\phi)$ generated by a spatially homogeneous ``extrinsic'' scalar field $\phi$~\citep{capistrano_inflation2023}.  
Thus, one defines a Lagrangian $\mathcal{L}_{\phi}$ such as
\begin{equation}
\mathcal{L}_{\phi}= \frac{1}{2} \dot{\phi}^2 - V(\phi)\;, \label{eq:lagrangianphi} 
\end{equation}
where the time derivative is represented by the dot symbol. Immediately, one writes the related energy momentum tensor as
\begin{equation}
T^{\phi}_{\mu\nu}= \partial_{\mu}\phi~\partial_{\nu}\phi + g_{\mu\nu}\left(\frac{1}{2} \dot{\phi}^2 - V(\phi)\right)\;. \label{eq:tmunuphi} 
\end{equation}
From conservation of Eq.(\ref{eq:tmunuphi}),  the inflation dynamics is coupled to the background evolution, and we obtain the relations
\begin{eqnarray}
&& \bar{\rho}_{ext}= \frac{\dot{\phi}^2}{2} + V(\phi)\;, \label{eq:densphi} \\
&& \bar{p}_{ext}= \frac{\dot{\phi}^2}{2} - V(\phi)\;,\label{eq:pressphi}
\end{eqnarray}
where $\bar{\rho}_{ext}$ and $\bar{p}_{ext}$ denote the  energy density and pressure generated by extrinsic curvature as a function of the extrinsic scalar field $\phi$, respectively.

As shown in ref~\citep{capistrano_inflation2023}, using Eqs.(\ref{eq:extdensitya1}), (\ref{eq:extpressure}), (\ref{eq:wext}), (\ref{eq:densphi}) and (\ref{eq:pressphi}), by direct integration, one obtains the potential $\phi(a)$ as
\begin{equation}\label{eq:phi}
\phi(a)= \sqrt{|3(1+w)|}~M_{pl}\ln{a}\;.
\end{equation}
Heron we denote the reduced Planck mass as $M_{pl}=\frac{1}{\sqrt{8\pi G}}=c=1$. The potential $V(\phi)$ is also obtained straightforwardly  
\begin{equation}\label{eq:scalarpot01}
V(\phi)= V_{0} e^{-\alpha_0 \phi}\;,
\end{equation}
where $\alpha_0=\sqrt{|3 (1+w)|}$. This kind of exponential model is well-known in literature and is commonly referred as \textit{power law inflation}~(PLI)\cite{Abbott1984,Lucchin1984,sahni1988,MARTIN201475}. Such potential was also studied in the context of M-theory~\cite{Becker2005} and RSII scenarios~\cite{Bennai2006}. Due to the strong constraints imposed by Planck data~\cite{planckinflation} on PLI, we follow the generalization proposed in Refs.(\citep{Alcaniz_2007,Santos_2018}) named as $\beta$-exponential inflation. In Ref.(\citep{Alcaniz_2007}), the authors introduced a class of potentials in a form
\begin{eqnarray}
&& V(\phi)= V_{0} \exp_{1-\beta}(-\lambda \phi/M_{pl})\;,\label{eq:betainflation} \\
&&\;\;\;\;\;\; = V_{0}\left[1+\beta\left(-\lambda \phi/M_{pl}\right)\right]^{1/\beta} \nonumber \;.
\end{eqnarray}
In general, the function $\exp_{1-\beta}(f)=\left[1+\beta f\right]^{1/\beta}$ may be $1+\beta f >0$ or, otherwise, $\exp_{1-\beta}(f)=0$. Then, Eq.(\ref{eq:betainflation}) should satisfy the identities $\exp_{1-\beta}(\ln_{1-\beta}(f))=f $ and $\ln_{1-\beta}(f)+\ln_{1-\beta}(g)=\ln_{1-\beta}(fg)-\beta\left[\ln_{1-\beta}(f)\ln_{1-\beta}(g)\right]$ for any $g<0$. The term $\ln_{1-\beta}(f)=(f^{\beta}-1)/\beta$ is referred as the generalized logarithm function. To our purposes, Eq.(\ref{eq:scalarpot01}) can be generalized to Eq.(\ref{eq:betainflation}) in a form
\begin{equation}\label{eq:scalarpot02}
V(\phi)= V_{0}\left[1-\beta\alpha_0\phi\right]^{1/\beta}\;.
\end{equation}
Assuming that the field starts rolling in a local minimum in such $\frac{\partial^2 V(\phi)}{\partial^2 \phi}=0$ for any $\beta \neq 1$, starting at $\phi=\phi_{\min}=1$, one obtains the condition $\beta=\frac{1}{\alpha_0}>0$. Thus, we can write Eq.(\ref{eq:scalarpot02}) as
\begin{equation}\label{eq:scalarpot03}
V(\phi)= V_{0}\left[1-\phi\right]^{\alpha_0}\;.
\end{equation}
Then, one defines the pair of slow-roll parameters
\begin{eqnarray}
&&\varepsilon=\frac{1}{2\kappa}\left(\frac{V_{,\phi}}{V}\right)^2\;,\label{eq:slowpair1}\\
&&\eta=\frac{1}{\kappa}\frac{V_{,\phi\phi}}{V}\;.\label{eq:slowrpair2}
\end{eqnarray}
In order to submit the model to the scrutiny of observational data, they are expressed as 
\begin{eqnarray}
\label{eq:tilt}
&&\eta_s=1-6\varepsilon+2\eta\;,\\
&&r=16\varepsilon\;,\label{eq:tensor}
\end{eqnarray}
where $\eta_s$ is the spectral tilt and $r$ is the tensor-to-scalar ratio. The relation between the parameters is given by 
\begin{eqnarray}
\label{eq:tilt2}
&&\eta_s=1-\frac{\alpha_0(\alpha_0+2)}{(1-\phi_{\star})^2}\;,\\
&&r=\frac{8\alpha_0^2}{(1-\phi_{\star})^2}\;,\label{eq:tensor2}
\end{eqnarray}
where $\phi_{\star}$ is the field before the end of inflation given by
\begin{equation}\label{eq:phistar}
\phi_{\star}= 1 -\alpha_0\sqrt{0.5+ \frac{2}{\alpha_0} N}\;.
\end{equation}
The quantity $N$ denotes the number of e-folds before the end of inflation defined as
\begin{equation}\label{eq:efolds}
N= \int^{\phi_{\star}}_{\phi_{end}} \frac{d\phi}{2\sqrt{\varepsilon}}= \frac{\phi_{\star}^2}{2\alpha_0}-\frac{\phi_{\star}}{\alpha_0}+\frac{1}{2\alpha_0} - \frac{\alpha_0}{4}\;.
\end{equation}
In the end of inflation, the field $\phi_{end}$ is calculated from the condition $\varepsilon(\phi_{end})\sim 1$, and one finds
\begin{equation}\label{eq:phiend}
\phi_{end}= 1-\frac{\alpha_0\sqrt{2}}{2}\;.
\end{equation}
\begin{table*}
\centering
\begin{tabular}{|l|l|l|}
\hline
\textbf{Dataset ID} & \textbf{Description} & \textbf{Reference} \\
\hline
\textit{sixdf\_2011\_bao} & 6dF Galaxy Survey & \cite{6dFGalaxy} \\
\hline
\textit{sdss\_dr7\_mgs} & Seventh Data Release of SDSS Main Galaxy Sample (SDSS DR7 MGS) & \cite{Ross:2014qpa} \\
\hline
\textit{sdss\_dr16\_baoplus\_lrg} & BOSS DR16 - Luminous Red Galaxies (LRG) & \cite{Alam:2016hwk} \\
\hline
\textit{sdss\_dr16\_baoplus\_elg} & BOSS DR16 - Emission Line Galaxies (ELG) & \cite{Alam:2020sor} \\
\hline
\textit{sdss\_dr16\_baoplus\_qso} & BOSS DR16 - Quasars (QSO) & \cite{Alam:2020sor} \\
\hline
\textit{sdss\_dr16\_baoplus\_lyauto} & BOSS DR16 - Lyman-alpha forest auto-correlation & \cite{Alam:2020sor} \\
\hline
\textit{sdss\_dr16\_baoplus\_lyxqso} & BOSS DR16 - Lyman-alpha forest x Quasar cross-correlation & \cite{Alam:2020sor} \\
\hline
\end{tabular}
\caption{Datasets from SDSS employed in our analysis.}
\label{tab:datasets}
\end{table*}

Once the slow-roll parameters as a function of $\phi$ are given by
\begin{eqnarray}
&&\varepsilon(\phi)=\frac{\alpha_0^2}{2(1-\phi)^2}\;,\label{eq:slowpair2}\\
&&\eta(\phi)=\frac{\alpha_0(\alpha_0-2)}{2(1-\phi_{\star})^2}\;.\label{eq:slowrpair3}
\end{eqnarray}
\begin{figure}[H]
    \includegraphics[width=3.3in, height=3in]{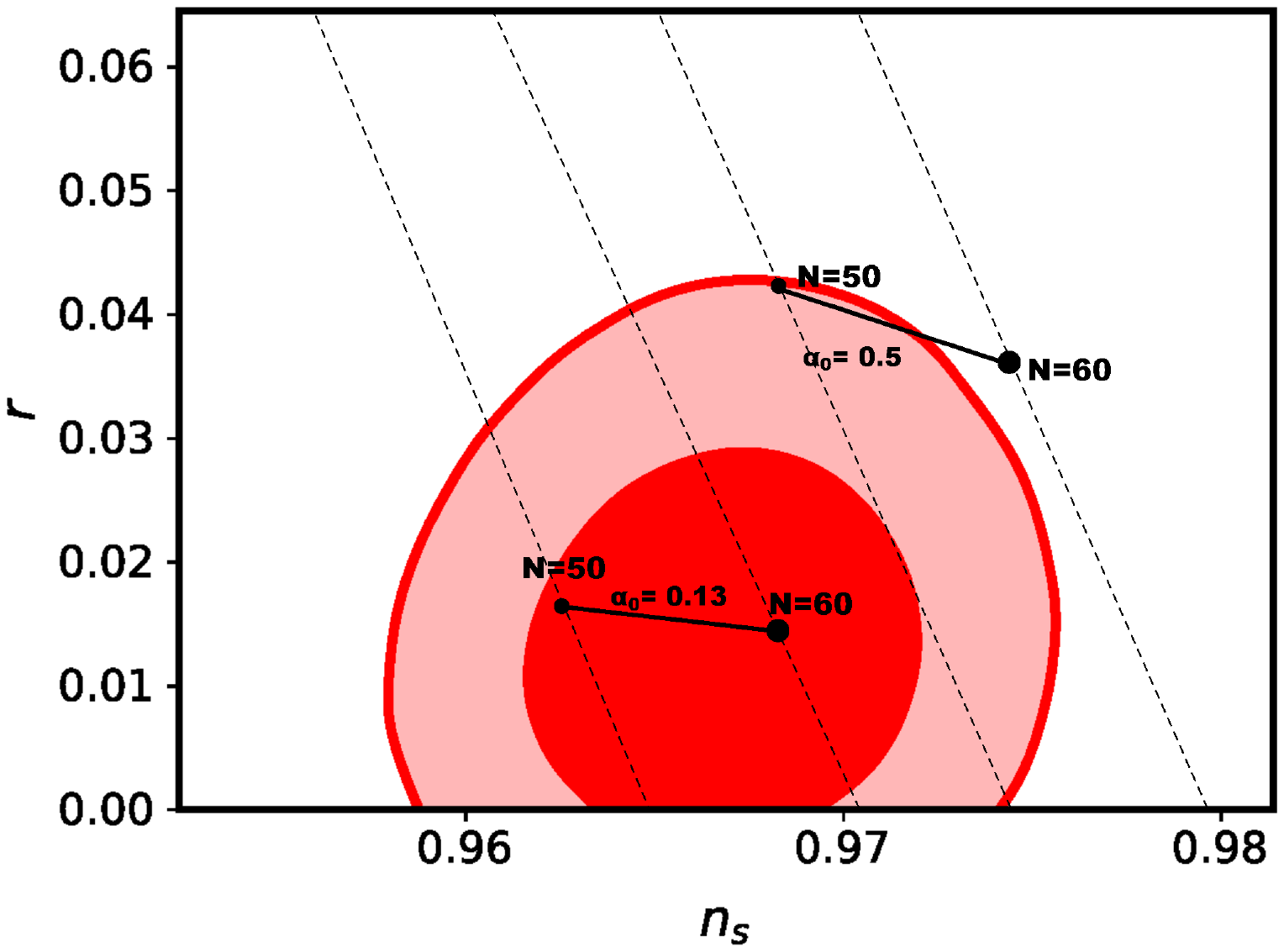}
    \caption{The $\eta_s-r$ plane for the range of values of the parameter $\alpha_0$ in Eq.(\ref{eq:scalarpot01}),for the number of e-folds $N=50$ and $N=60$. The contours correspond to joint fitting Planck+BK18+LSS(68$\%$ and 95$\%$~C.L.)~\cite{BICEPKeck} at the pivot scale $k_p= 0.05$Mpc$^{-1}$.} \label{fig:01}
\end{figure}

Figure (\ref{fig:01}) shows the $\eta_s-r$ plane of selected values of the parameter $\alpha_0$ in Eq.(\ref{eq:scalarpot01}) for the number of e-folds $N=50$ and $N=60$. The 68$\%$ and 95$\%$~C.L. contours correspond to joint fitting Planck(2018)+BK18+LSS extracted from publicly available CosmoMC chains~\cite{BICEPKeck}. Examining the behavior of the $\alpha_0$ parameter, we end up in the conclusions that higher values of $\alpha_0$ parameter for prediction of $r$ are in agreement with observations but compromise the prediction of the spectral index $\eta_s$. In contrast, lower values of $\alpha_0$, worth noting that the $\alpha_0=0$ mimics $\Lambda$CDM, are consistent with Planck data at the 1-$\sigma$ C.L. for the tensor-to-scalar ratio $r$. In this case, $\alpha_0=0.13$ and $\alpha_0=0.5$ suggest that $w=-0.9944$ and $w=-0.9167$, respectively. On the other hand, when considering the combinations with \texttt{HiLLiPoP} and {LoLLiPoP} likelihoods~\cite{planckinflation} with BICEP2/Keck 2015 data~\cite{BK15} as hlpTT+lowT+lowlEB and hlpTT+lowT+lowlEB+BK15, $\alpha_0=0.5$ is compatible at the 1-$\sigma$ and 2-$\sigma$~C.L.

\section{Data and methodology}
\begingroup
\begin{table*}
\caption {Summary of $68\%$ and $95\%$ confidence level limits for the parameters of interest obtained from Planck+BK18 and Planck+BK18+LSS joint data at pivot scale $k_p=0.05$Mpc$^{-1}$.} 
\begin{ruledtabular}
\label{tab:main}
\begin{tabular}{cllll}
 Parameter & \multicolumn{2}{c}{Planck+BK18} & \multicolumn{2}{c}{Planck+BK18+LSS}    \\ 
\cline{2-3} \cline{4-5} 
    &  68$\%$ C.L.   & 95$\%$ C.L.                          
	&  68$\%$ C.L.   & 95$\%$ C.L.                                       \\    \hline
{\boldmath$n_\mathrm{s}$}  &  $0.9646\pm 0.0031          $   & $0.9646^{+0.0057}_{-0.0060}$ & $0.9648\pm 0.0030$          & $0.9648^{+0.0056}_{-0.0060}$ \\     
{\boldmath$r         $}    &  $0.0142^{+0.0049}_{-0.012} $   & $< 0.0307                  $ & $0.0139^{+0.0044}_{-0.013}$ & $< 0.0303                  $\\
{\boldmath$H_0       $}    &  $66.981\pm 0.078           $   & $66.98^{+0.15}_{-0.15}     $ & $67.001\pm 0.078          $ & $67.00^{+0.15}_{-0.15}     $\\
$\Omega_\mathrm{m}   $     &  $0.31794\pm 0.00074        $   & $0.3179^{+0.0014}_{-0.0015}$ & $0.31776\pm 0.00074       $ & $0.3178^{+0.0014}_{-0.0014}$\\
$\mu-1               $     &  $-0.01^{+0.17}_{-0.21}     $   & $-0.01^{+0.40}_{-0.34}     $ & $0.02\pm 0.15             $ & $0.02^{+0.30}_{-0.28}      $\\
$\Sigma-1            $     &  $-0.053^{+0.051}_{-0.042}  $   & $-0.053^{+0.093}_{-0.098}  $ & $-0.049^{+0.048}_{-0.042} $ & $-0.049^{+0.083}_{-0.091}  $\\
$10^{8}\gamma_s      $     &  $-0.3^{+3.1}_{-3.8}        $   & $-0.3^{+7.2}_{-6.2}        $ & $0.3\pm 2.7               $ & $0.3^{+5.4}_{-5.0}         $\\
\end{tabular}
\end{ruledtabular}
\end{table*}
\endgroup

To delineate constraints on the free parameter of our model, we examine several datasets individually as well as in various combinations. Specifically, we will consider:

\begin{itemize}
\item Utilizing the latest NPIPE Planck DR4 likelihoods~\cite{Carron_2022,Rosenberg:2022sdy} of \textit{Planck} 2018 legacy data release, our analysis incorporates Cosmic Microwave Background (CMB) measurements. These include the high-$\ell$ \texttt{Plik} TT likelihood spanning the multipole range $30 \leq \ell \leq 2508$, as well as TE and EE measurements within the multipole range $30 \leq \ell \leq 1996$. Additionally, we incorporate low-$\ell$ TT-only ($2 \leq \ell \leq 29$) and EE-only ($2 \leq \ell \leq 29$) likelihoods. Furthermore, our dataset encompasses CMB NPIPE Planck lensing power spectrum measurements, collectively referred as the \texttt{Planck} dataset.

\item The B-mode polarization data from the BICEP2/Keck collaboration \cite{BICEPKeck}. 
We refer to this data set as \texttt{BK18}.

\item  We refer to this data set as \texttt{LSS}. We consider 6dF Galaxy Survey~\cite{6dFGalaxy}, the Seventh Data Release of SDSS Main Galaxy Sample (SDSS DR7 MGS)~\citep{Ross:2014qpa} and clustering measurements of the Extended Baryon Oscillation Spectroscopic Survey (eBOSS) associated with the SDSS's Sixteenth Data Release~\citep{Alam:2016hwk}. This collection encompasses data from Luminous Red Galaxies (LRG), Emission Line Galaxies (ELG), Quasars (QSO), the Lyman-alpha forest auto-correlation (\textit{lyauto}), and the Lyman-alpha forest x Quasar cross-correlation (\textit{lyxqso}). Table~\ref{tab:datasets} details the diverse BAO components considered in this work.

\begin{itemize}
 \item The Hubble distance at redshift $z$:

\begin{equation}
D_H(z) = \frac{c}{H(z)}, 
\end{equation}
where $H(z)$ is the Hubble parameter. 

\item  The comoving angular diameter distance, $D_{M}(z)$, which also only depends on the expansion history:

\begin{equation}
  D_M(z) = \frac{c}{H_0} \int_0^z dz' \frac{H_0}{H(z')}.
\end{equation}

\item The spherically-averaged BAO distance:

\begin{equation}
D_V(z) = r_d [z D_{M}^2(z) D_H(z) ]^{1/3},
\end{equation}
where $r_d$ is the BAO scale, of which in our analyzes we treat it as a derived parameter.

\end{itemize}

For the growth measurements, the growth function $f$ can be expressed as a differential in the amplitude of linear matter fluctuations on a comoving scale of 8 $h^{-1}$ Mpc, $\sigma_8(z)$, in the form 

\begin{equation}
  f(z) = \frac{\partial \ln \sigma_8}{\partial \ln a}.
\end{equation}
The RSD measurements provide constraints on the quantity $f(z)\sigma_8(z)$. The $\sigma_8(z)$ depends on the matter power spectrum, $P(k,z)$, which is calculated by default in the Boltzmann code. Both $f(z)$ and $\sigma_8(z)$  are sensitive to variations in the effective gravitational coupling and the light deflection parameter, which play a crucial role in Poisson and lensing equations in MG models.
 \end{itemize}

\begin{table}
\centering
\begin{tabular}{l|l}
\hline \hline
Parameter                                   & Prior        \\ \hline
$ \Omega_b h^2$                             & $\mathcal{U}[0.017,0.027]$ \\
$ \Omega_c h^2$                             & $\mathcal{U}[0.09,0.15]$ \\
$\theta_\mathrm{MC}$                        & $\mathcal{U}[0.0103, 0.0105]$       \\
$\tau_{\rm reio}$                           & $\mathcal{N}[0.065,0.0015]$   \\
$\log(10^{10} A_\mathrm{s})$                & $\mathcal{U}[2.6,3.5]$       \\
$n_{s}$                                     & $\mathcal{U}[0.9, 1.1]$     \\
$10^{8}$ $\gamma_{s}$       & $\mathcal{U}[-1, 1]$     \\
\hline
\end{tabular}
\caption{The cosmological parameters along with their respective priors employed in the parameter estimation analysis. }
\label{tab:priors}
\end{table}

\begin{figure}[H]
    \includegraphics[width=3.35in, height=3.5in]{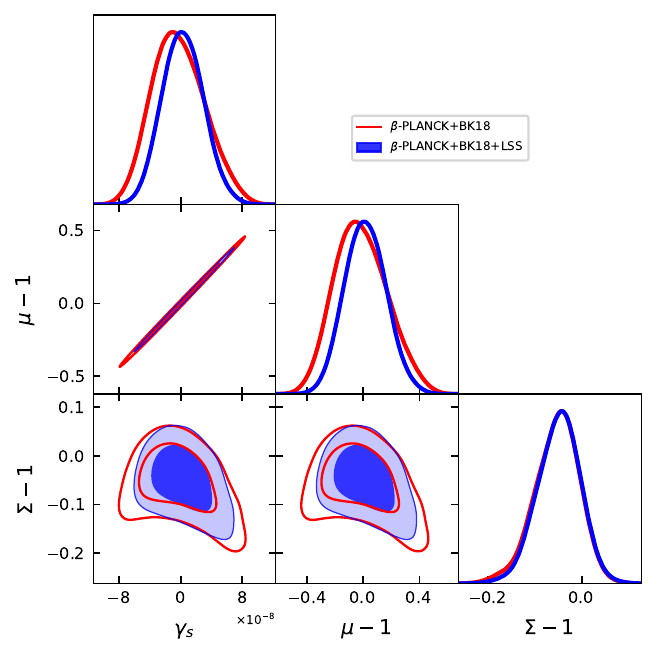}
    \caption{Triangle plot contours with MG parameters $\mu-\Sigma$ and the $\gamma_s$ parameter of the $\beta$-model for the combined analyses of Planck+BK18~(red line) and Planck+BK18+LSS~(blue line).} \label{fig:03}
\end{figure}

The joint likelihood analysis is conducted using MGCAMB-II~\cite{mgcamb2023} through the Cobaya~\cite{cobaya} sampler, employing the $\mu-\Sigma$ parametrization defined as follows:
\begin{eqnarray}
&&\mu(a,k)= 1 + \mu_0 \frac{\Omega_{DE}}{\Omega_{DE(0)}}\;,\label{eq:mucamb1}\\
&& \Sigma(a,k)= 1 + \Sigma_0 \frac{\Omega_{DE}}{\Omega_{DE(0)}} \;.\label{eq:sigmacamb1}
\end{eqnarray}
Here, $\Omega_{DE}(a)$ is denoted as the extrinsic contribution $\Omega_{ext}(a)$ which means that $\Omega_{DE(0)}=\Omega_{ext}(0)=1-\Omega_{m(0)}$. The form of Eqs. (\ref{eq:mucamb1}) and (\ref{eq:sigmacamb1}) in this parametrization is obtained from expanding the function $\chi(a)<<1$ in the denominator of Eq. (\ref{muequation}). We assume a pivot scale fixed at $a_p=10^{-4}$ and $k_p=0.05$Mpc$^{-1}$. It is worth noting that the anisotropic stress is not considered in our analysis.

The priors on the baseline parameters utilized in our analysis are detailed in Table \ref{tab:priors}. In all runs, we ensure a Gelman-Rubin convergence criterion of $R - 1 < 0.03$. In the subsequent section, we will unveil the outcomes of our Bayesian analysis and delve into their implications.

\section{Results}
We commence our analysis by scrutinizing the constraints derived exclusively from the joint analysis of Planck and BK18 datasets. The primary statistical findings concerning the cosmological parameters of interest are summarized in Table \ref{tab:main}.

We analyze the results by examining
the parameters that indicate deviations from the stan standard $\Lambda$CDM cosmology. Figure (\ref{fig:03}) illustrates the parameter space for $\gamma_s$, $\mu$, and $\Sigma$.

\begin{figure}[H]
    \includegraphics[width=3.4in, height=2.7in]{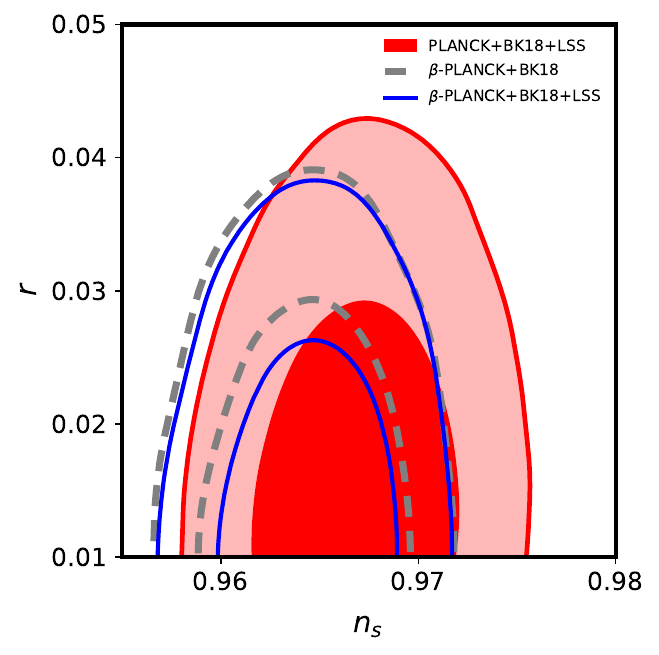}
    \caption{The contours in the $n_s-r$ plane, delineating 68\% and 95\% confidence levels (C.L.), represent the combined analyses of Planck+BK18~(grey dashed line contour) and Planck+BK18+LSS~(blue line contour) using the $\beta$-model. The red contour illustrates the joint fitting of Planck+BK18+LSS extracted from publicly available CosmoMC chains~\cite{BICEPKeck}.} \label{fig:02}
\end{figure}

Figure (\ref{fig:02}) shows the $n_s-r$ plane. It is noteworthy to observe that the tensor-to-scalar ratio effects arise from the modified gravity scenario outlined in this study. In canonical parametric $P(k)$ inflation and $\Lambda$CDM dynamics, $r < 0.036$ at 95\% CL~ \cite{Ade_2021} is found, whereas for the present modified scenario, we find a $r < 0.0307$ at 95\% CL with Planck+BK18 data. This represents a variation of $\Delta r = 0.03$ from standard model tightening the constraint on $r$ suggesting a signature of a MG model rather than $\Lambda$CDM.  At pivot scale $k_p=0.05$Mpc$^{-1}$, our result is compatible with the upper bounds predicted by current Planck data~\cite{planckinflation} with $r < 0.032$ at $95\%$CL from $E$ and $B-$mode spectra BK18~\cite{BICEPKeck} and LSS data~\cite{Alam:2020sor}. In addition, our upper constraint on $r$ with $r < 0.0307$ at $95\%$CL is also tighter than the standard joint data Planck+BK18 with $r<0.036$ and NPIPE~(PR4) with $r < 0.056$ at 95\% CL. Our results indicate a lower value for $r$ as compared with the observations from the PR4 BB spectrum for multipoles between $l= 2$ and $l= 150$ with $r = 0.033\pm 0.069$~\cite{planckinflation}. In contrast with the frequentist proﬁle likelihood method~\cite{Campeti_2022} with an upper limit of $r<0.037$ at $95\%$ CL, we also obtain a tighter constraint on $r$ for the same combination of Planck, BK18 and LSS with $r<0.0303$ at $95\%$ CL.  In this context, our inflationary model partially constraints data better than the $\Lambda$CDM model does.

Considering the scalar spectral index $n_s$, we obtained a close value of $n_s= 0.9641\pm 0.0031$ with Planck+BK18+LSS data at $68\%$ CL to based-$\Lambda$CDM Planck TT,TE,EE+lowE+lensing with $n_s = 0.9649\pm 0.0042$ at $68\%$ CL and $n_s = 0.9665\pm 0.0038$ at $68\%$ CL when considering Planck TT,TE,EE+lowE+lensing+LSS. On the other hand, the small differences between the $n_s$ value seem to affect the value of $H_0$ which is roughly~$0.5\%$ lower than the Planck-$\Lambda$CDM case. From Planck+BK18+LSS dateset, in the analysis of slow-roll parameters, the model prefers the case where $\alpha_0=0.13$ gives $w=-0.9944$ as shown in Figure (\ref{fig:01}) for the number of e-folds $N=50$ and $N=60$. Most importantly, the model does not require a larger number of e-folds to provide a tighter constraint on $r$. We also verified that the inclusion of LSS data does not significantly change the value of $n_s$ but influences the $H_0$ values, which may be improved with upcoming new constraints on the reionization optical depth whose uncertainties may provoke a large impact on fundamental cosmological parameters such as $n_s$~\cite{Sailer_2022}.
\\
\section{Final Remarks}

In this paper, we have derived the gravitational equations within four dimensions by inducing them from a five-dimensional bulk, employing Nash's embeddings framework and incorporating them into a well-established $\mu$-$\Sigma$ representation. From the analysis of the slow-roll conditions, we have obtained a PLI model that was generalized to the $\beta$-exponential inflation~\cite{Alcaniz_2007,Santos_2018}. We have obtained $w=-0.9944$ for the number of e-folds in the range $N=50$ and $N=60$. Interestingly, this model also sits within the previous reasonable expected number e-folds compatible with the tighter restriction on $r$, which is important to maintain the window of solving the horizon problem. This enables us to assess and investigate the impact of the model on linear perturbations in the CMB data. Apart from the values of the tensor-to-ratio parameter, our primary analyses revealed similar predictions to the ones of the $\Lambda$CDM model. Additionally, we have quantified the model's predictions concerning inflationary dynamics. By utilizing data from CMB-PR4, BICEP/Keck Array 2018, and certain LSS measurements, we have established a tighter upper limit of $r< 0.0303$ at 95\% C.L. As compared with $\Lambda$CDM model, such apparent improvement of tensor-to-ratio parameter and a similar but lower value of the scalar spectral index $n_s$ values may suggest a glimpse of a modified gravity signature which may be improved in future experiments.

\begin{acknowledgments}
AJSC thanks the financial support from the Conselho Nacional de Desenvolvimento Cient\'{i}fico e Tecnol\'{o}gico (CNPq, National Council for Scientific and Technological Development) for partial financial support under the project No. 305881/2022-1, and The Funda\c{c}\~{a}o da Universidade Federal do Paran\'{a} (FUNPAR, Paran\'{a} Federal University Foundation) by public notice 04/2023-Pesquisa/PRPPG/UFPR- for partial financial support under process nº 23075.019406/2023-92. R.C.N thanks the financial support from CNPq under the project No. 304306/2022-3, and the Fundação de Amparo à pesquisa do Estado do RS (FAPERGS, Research Support Foundation of the State of RS) for partial financial support under the project No. 23/2551-0000848-3. 
\end{acknowledgments}

\bibliography{nash_infl}

\begin{thebibliography}{81}%
\makeatletter
\providecommand \@ifxundefined [1]{%
 \@ifx{#1\undefined}
}%
\providecommand \@ifnum [1]{%
 \ifnum #1\expandafter \@firstoftwo
 \else \expandafter \@secondoftwo
 \fi
}%
\providecommand \@ifx [1]{%
 \ifx #1\expandafter \@firstoftwo
 \else \expandafter \@secondoftwo
 \fi
}%
\providecommand \natexlab [1]{#1}%
\providecommand \enquote  [1]{``#1''}%
\providecommand \bibnamefont  [1]{#1}%
\providecommand \bibfnamefont [1]{#1}%
\providecommand \citenamefont [1]{#1}%
\providecommand \href@noop [0]{\@secondoftwo}%
\providecommand \href [0]{\begingroup \@sanitize@url \@href}%
\providecommand \@href[1]{\@@startlink{#1}\@@href}%
\providecommand \@@href[1]{\endgroup#1\@@endlink}%
\providecommand \@sanitize@url [0]{\catcode `\\12\catcode `\$12\catcode `\&12\catcode `\#12\catcode `\^12\catcode `\_12\catcode `\%12\relax}%
\providecommand \@@startlink[1]{}%
\providecommand \@@endlink[0]{}%
\providecommand \url  [0]{\begingroup\@sanitize@url \@url }%
\providecommand \@url [1]{\endgroup\@href {#1}{\urlprefix }}%
\providecommand \urlprefix  [0]{URL }%
\providecommand \Eprint [0]{\href }%
\providecommand \doibase [0]{https://doi.org/}%
\providecommand \selectlanguage [0]{\@gobble}%
\providecommand \bibinfo  [0]{\@secondoftwo}%
\providecommand \bibfield  [0]{\@secondoftwo}%
\providecommand \translation [1]{[#1]}%
\providecommand \BibitemOpen [0]{}%
\providecommand \bibitemStop [0]{}%
\providecommand \bibitemNoStop [0]{.\EOS\space}%
\providecommand \EOS [0]{\spacefactor3000\relax}%
\providecommand \BibitemShut  [1]{\csname bibitem#1\endcsname}%
\let\auto@bib@innerbib\@empty
\bibitem [{\citenamefont {Guth}(1981)}]{PhysRevD.23.347}%
  \BibitemOpen
  \bibfield  {author} {\bibinfo {author} {\bibfnamefont {A.~H.}\ \bibnamefont {Guth}},\ }\bibfield  {title} {\bibinfo {title} {Inflationary universe: A possible solution to the horizon and flatness problems},\ }\href {https://doi.org/10.1103/PhysRevD.23.347} {\bibfield  {journal} {\bibinfo  {journal} {Phys. Rev. D}\ }\textbf {\bibinfo {volume} {23}},\ \bibinfo {pages} {347} (\bibinfo {year} {1981})}\BibitemShut {NoStop}%
\bibitem [{\citenamefont {Starobinsky}(1979)}]{Starobinsky:1979ty}%
  \BibitemOpen
  \bibfield  {author} {\bibinfo {author} {\bibfnamefont {A.~A.}\ \bibnamefont {Starobinsky}},\ }\bibfield  {title} {\bibinfo {title} {{Spectrum of relict gravitational radiation and the early state of the universe}},\ }\href@noop {} {\bibfield  {journal} {\bibinfo  {journal} {JETP Lett.}\ }\textbf {\bibinfo {volume} {30}},\ \bibinfo {pages} {682} (\bibinfo {year} {1979})}\BibitemShut {NoStop}%
\bibitem [{\citenamefont {Mukhanov}\ and\ \citenamefont {Chibisov}(1981)}]{Mukhanov:1981xt}%
  \BibitemOpen
  \bibfield  {author} {\bibinfo {author} {\bibfnamefont {V.~F.}\ \bibnamefont {Mukhanov}}\ and\ \bibinfo {author} {\bibfnamefont {G.~V.}\ \bibnamefont {Chibisov}},\ }\bibfield  {title} {\bibinfo {title} {{Quantum Fluctuations and a Nonsingular Universe}},\ }\href@noop {} {\bibfield  {journal} {\bibinfo  {journal} {JETP Lett.}\ }\textbf {\bibinfo {volume} {33}},\ \bibinfo {pages} {532} (\bibinfo {year} {1981})}\BibitemShut {NoStop}%
\bibitem [{\citenamefont {Linde}(1983)}]{Linde:1983gd}%
  \BibitemOpen
  \bibfield  {author} {\bibinfo {author} {\bibfnamefont {A.~D.}\ \bibnamefont {Linde}},\ }\bibfield  {title} {\bibinfo {title} {{Chaotic Inflation}},\ }\href {https://doi.org/10.1016/0370-2693(83)90837-7} {\bibfield  {journal} {\bibinfo  {journal} {Phys. Lett. B}\ }\textbf {\bibinfo {volume} {129}},\ \bibinfo {pages} {177} (\bibinfo {year} {1983})}\BibitemShut {NoStop}%
\bibitem [{\citenamefont {Hawking}(1982)}]{HAWKING1982295}%
  \BibitemOpen
  \bibfield  {author} {\bibinfo {author} {\bibfnamefont {S.}~\bibnamefont {Hawking}},\ }\bibfield  {title} {\bibinfo {title} {The development of irregularities in a single bubble inflationary universe},\ }\href {https://doi.org/https://doi.org/10.1016/0370-2693(82)90373-2} {\bibfield  {journal} {\bibinfo  {journal} {Physics Letters B}\ }\textbf {\bibinfo {volume} {115}},\ \bibinfo {pages} {295} (\bibinfo {year} {1982})}\BibitemShut {NoStop}%
\bibitem [{\citenamefont {Hawking}\ and\ \citenamefont {Moss}(1983)}]{HAWKING1983180}%
  \BibitemOpen
  \bibfield  {author} {\bibinfo {author} {\bibfnamefont {S.}~\bibnamefont {Hawking}}\ and\ \bibinfo {author} {\bibfnamefont {I.}~\bibnamefont {Moss}},\ }\bibfield  {title} {\bibinfo {title} {Fluctuations in the inflationary universe},\ }\href {https://doi.org/https://doi.org/10.1016/0550-3213(83)90319-X} {\bibfield  {journal} {\bibinfo  {journal} {Nuclear Physics B}\ }\textbf {\bibinfo {volume} {224}},\ \bibinfo {pages} {180} (\bibinfo {year} {1983})}\BibitemShut {NoStop}%
\bibitem [{\citenamefont {Starobinsky}(1982)}]{STAROBINSKY1982175}%
  \BibitemOpen
  \bibfield  {author} {\bibinfo {author} {\bibfnamefont {A.}~\bibnamefont {Starobinsky}},\ }\bibfield  {title} {\bibinfo {title} {Dynamics of phase transition in the new inflationary universe scenario and generation of perturbations},\ }\href {https://doi.org/https://doi.org/10.1016/0370-2693(82)90541-X} {\bibfield  {journal} {\bibinfo  {journal} {Physics Letters B}\ }\textbf {\bibinfo {volume} {117}},\ \bibinfo {pages} {175} (\bibinfo {year} {1982})}\BibitemShut {NoStop}%
\bibitem [{\citenamefont {Guth}\ and\ \citenamefont {Pi}(1982)}]{PhysRevLett.49.1110}%
  \BibitemOpen
  \bibfield  {author} {\bibinfo {author} {\bibfnamefont {A.~H.}\ \bibnamefont {Guth}}\ and\ \bibinfo {author} {\bibfnamefont {S.-Y.}\ \bibnamefont {Pi}},\ }\bibfield  {title} {\bibinfo {title} {Fluctuations in the new inflationary universe},\ }\href {https://doi.org/10.1103/PhysRevLett.49.1110} {\bibfield  {journal} {\bibinfo  {journal} {Phys. Rev. Lett.}\ }\textbf {\bibinfo {volume} {49}},\ \bibinfo {pages} {1110} (\bibinfo {year} {1982})}\BibitemShut {NoStop}%
\bibitem [{\citenamefont {Bardeen}\ \emph {et~al.}(1983)\citenamefont {Bardeen}, \citenamefont {Steinhardt},\ and\ \citenamefont {Turner}}]{PhysRevD.28.679}%
  \BibitemOpen
  \bibfield  {author} {\bibinfo {author} {\bibfnamefont {J.~M.}\ \bibnamefont {Bardeen}}, \bibinfo {author} {\bibfnamefont {P.~J.}\ \bibnamefont {Steinhardt}},\ and\ \bibinfo {author} {\bibfnamefont {M.~S.}\ \bibnamefont {Turner}},\ }\bibfield  {title} {\bibinfo {title} {Spontaneous creation of almost scale-free density perturbations in an inflationary universe},\ }\href {https://doi.org/10.1103/PhysRevD.28.679} {\bibfield  {journal} {\bibinfo  {journal} {Phys. Rev. D}\ }\textbf {\bibinfo {volume} {28}},\ \bibinfo {pages} {679} (\bibinfo {year} {1983})}\BibitemShut {NoStop}%
\bibitem [{\citenamefont {Aghanim}\ \emph {et~al.}(2020{\natexlab{a}})\citenamefont {Aghanim} \emph {et~al.}}]{refplanck2018}%
  \BibitemOpen
  \bibfield  {author} {\bibinfo {author} {\bibfnamefont {N.}~\bibnamefont {Aghanim}} \emph {et~al.} (\bibinfo {collaboration} {PLANCK}),\ }\bibfield  {title} {\bibinfo {title} {{Planck 2018 results - VI. Cosmological parameters}},\ }\href {https://doi.org/10.1051/0004-6361/201833910} {\bibfield  {journal} {\bibinfo  {journal} {A\&A}\ }\textbf {\bibinfo {volume} {641}},\ \bibinfo {pages} {A6} (\bibinfo {year} {2020}{\natexlab{a}})},\ \Eprint {https://arxiv.org/abs/1807.06209} {arXiv:1807.06209 [astro-ph.CO]} \BibitemShut {NoStop}%
\bibitem [{\citenamefont {Giovannini}(1999{\natexlab{a}})}]{Giovannini1999}%
  \BibitemOpen
  \bibfield  {author} {\bibinfo {author} {\bibfnamefont {M.}~\bibnamefont {Giovannini}},\ }\bibfield  {title} {\bibinfo {title} {{Spikes in the relic graviton background from quintessential inflation}},\ }\href {https://doi.org/10.1088/0264-9381/16/9/308} {\bibfield  {journal} {\bibinfo  {journal} {Class. Quant. Grav.}\ }\textbf {\bibinfo {volume} {16}},\ \bibinfo {pages} {2905} (\bibinfo {year} {1999}{\natexlab{a}})},\ \Eprint {https://arxiv.org/abs/hep-ph/9903263} {arXiv:hep-ph/9903263} \BibitemShut {NoStop}%
\bibitem [{\citenamefont {Giovannini}(1999{\natexlab{b}})}]{Giovannini1999b}%
  \BibitemOpen
  \bibfield  {author} {\bibinfo {author} {\bibfnamefont {M.}~\bibnamefont {Giovannini}},\ }\bibfield  {title} {\bibinfo {title} {{Production and detection of relic gravitons in quintessential inflationary models}},\ }\href {https://doi.org/10.1103/PhysRevD.60.123511} {\bibfield  {journal} {\bibinfo  {journal} {Phys. Rev. D}\ }\textbf {\bibinfo {volume} {60}},\ \bibinfo {pages} {123511} (\bibinfo {year} {1999}{\natexlab{b}})},\ \Eprint {https://arxiv.org/abs/astro-ph/9903004} {arXiv:astro-ph/9903004} \BibitemShut {NoStop}%
\bibitem [{\citenamefont {Giovannini}(2003)}]{Giovannini2003}%
  \BibitemOpen
  \bibfield  {author} {\bibinfo {author} {\bibfnamefont {M.}~\bibnamefont {Giovannini}},\ }\bibfield  {title} {\bibinfo {title} {{Low scale quintessential inflation}},\ }\href {https://doi.org/10.1103/PhysRevD.67.123512} {\bibfield  {journal} {\bibinfo  {journal} {Phys. Rev. D}\ }\textbf {\bibinfo {volume} {67}},\ \bibinfo {pages} {123512} (\bibinfo {year} {2003})},\ \Eprint {https://arxiv.org/abs/hep-ph/0301264} {arXiv:hep-ph/0301264} \BibitemShut {NoStop}%
\bibitem [{\citenamefont {Cognola}\ \emph {et~al.}(2008)\citenamefont {Cognola}, \citenamefont {Elizalde}, \citenamefont {Nojiri}, \citenamefont {Odintsov}, \citenamefont {Sebastiani},\ and\ \citenamefont {Zerbini}}]{cognola2008}%
  \BibitemOpen
  \bibfield  {author} {\bibinfo {author} {\bibfnamefont {G.}~\bibnamefont {Cognola}}, \bibinfo {author} {\bibfnamefont {E.}~\bibnamefont {Elizalde}}, \bibinfo {author} {\bibfnamefont {S.}~\bibnamefont {Nojiri}}, \bibinfo {author} {\bibfnamefont {S.~D.}\ \bibnamefont {Odintsov}}, \bibinfo {author} {\bibfnamefont {L.}~\bibnamefont {Sebastiani}},\ and\ \bibinfo {author} {\bibfnamefont {S.}~\bibnamefont {Zerbini}},\ }\bibfield  {title} {\bibinfo {title} {Class of viable modified $f(r)$ gravities describing inflation and the onset of accelerated expansion},\ }\href {https://doi.org/10.1103/PhysRevD.77.046009} {\bibfield  {journal} {\bibinfo  {journal} {Phys. Rev. D}\ }\textbf {\bibinfo {volume} {77}},\ \bibinfo {pages} {046009} (\bibinfo {year} {2008})}\BibitemShut {NoStop}%
\bibitem [{\citenamefont {Myrzakulov}\ \emph {et~al.}(2015)\citenamefont {Myrzakulov}, \citenamefont {Sebastiani},\ and\ \citenamefont {Vagnozzi}}]{Myrzakulov2015}%
  \BibitemOpen
  \bibfield  {author} {\bibinfo {author} {\bibfnamefont {R.}~\bibnamefont {Myrzakulov}}, \bibinfo {author} {\bibfnamefont {L.}~\bibnamefont {Sebastiani}},\ and\ \bibinfo {author} {\bibfnamefont {S.}~\bibnamefont {Vagnozzi}},\ }\bibfield  {title} {\bibinfo {title} {{Inflation in $f(R,\phi )$ -theories and mimetic gravity scenario}},\ }\href {https://doi.org/10.1140/epjc/s10052-015-3672-6} {\bibfield  {journal} {\bibinfo  {journal} {Eur. Phys. J. C}\ }\textbf {\bibinfo {volume} {75}},\ \bibinfo {pages} {444} (\bibinfo {year} {2015})},\ \Eprint {https://arxiv.org/abs/1504.07984} {arXiv:1504.07984 [gr-qc]} \BibitemShut {NoStop}%
\bibitem [{\citenamefont {Salvio}(2017)}]{Salvio2017}%
  \BibitemOpen
  \bibfield  {author} {\bibinfo {author} {\bibfnamefont {A.}~\bibnamefont {Salvio}},\ }\bibfield  {title} {\bibinfo {title} {{Inflationary Perturbations in No-Scale Theories}},\ }\href {https://doi.org/10.1140/epjc/s10052-017-4825-6} {\bibfield  {journal} {\bibinfo  {journal} {Eur. Phys. J. C}\ }\textbf {\bibinfo {volume} {77}},\ \bibinfo {pages} {267} (\bibinfo {year} {2017})},\ \Eprint {https://arxiv.org/abs/1703.08012} {arXiv:1703.08012 [astro-ph.CO]} \BibitemShut {NoStop}%
\bibitem [{\citenamefont {Oikonomou}(2017)}]{Oikonomou2017}%
  \BibitemOpen
  \bibfield  {author} {\bibinfo {author} {\bibfnamefont {V.~K.}\ \bibnamefont {Oikonomou}},\ }\bibfield  {title} {\bibinfo {title} {{Viability of the intermediate inflation scenario with F(T) gravity}},\ }\href {https://doi.org/10.1103/PhysRevD.95.084023} {\bibfield  {journal} {\bibinfo  {journal} {Phys. Rev. D}\ }\textbf {\bibinfo {volume} {95}},\ \bibinfo {pages} {084023} (\bibinfo {year} {2017})},\ \Eprint {https://arxiv.org/abs/1703.10515} {arXiv:1703.10515 [gr-qc]} \BibitemShut {NoStop}%
\bibitem [{\citenamefont {Agarwal}\ \emph {et~al.}(2017)\citenamefont {Agarwal}, \citenamefont {Myrzakulov}, \citenamefont {Sami},\ and\ \citenamefont {Singh}}]{Agarwal2017}%
  \BibitemOpen
  \bibfield  {author} {\bibinfo {author} {\bibfnamefont {A.}~\bibnamefont {Agarwal}}, \bibinfo {author} {\bibfnamefont {R.}~\bibnamefont {Myrzakulov}}, \bibinfo {author} {\bibfnamefont {M.}~\bibnamefont {Sami}},\ and\ \bibinfo {author} {\bibfnamefont {N.~K.}\ \bibnamefont {Singh}},\ }\bibfield  {title} {\bibinfo {title} {{Quintessential inflation in a thawing realization}},\ }\href {https://doi.org/10.1016/j.physletb.2017.04.066} {\bibfield  {journal} {\bibinfo  {journal} {Phys. Lett. B}\ }\textbf {\bibinfo {volume} {770}},\ \bibinfo {pages} {200} (\bibinfo {year} {2017})},\ \Eprint {https://arxiv.org/abs/1708.00156} {arXiv:1708.00156 [gr-qc]} \BibitemShut {NoStop}%
\bibitem [{\citenamefont {Keskin}(2018)}]{Keskin2018}%
  \BibitemOpen
  \bibfield  {author} {\bibinfo {author} {\bibfnamefont {A.~I.}\ \bibnamefont {Keskin}},\ }\bibfield  {title} {\bibinfo {title} {{Viable super inflation scenario from F(T) modified teleparallel gravity}},\ }\href {https://doi.org/10.1140/epjc/s10052-018-6199-9} {\bibfield  {journal} {\bibinfo  {journal} {Eur. Phys. J. C}\ }\textbf {\bibinfo {volume} {78}},\ \bibinfo {pages} {705} (\bibinfo {year} {2018})}\BibitemShut {NoStop}%
\bibitem [{\citenamefont {Salvio}(2019)}]{Salvio2019}%
  \BibitemOpen
  \bibfield  {author} {\bibinfo {author} {\bibfnamefont {A.}~\bibnamefont {Salvio}},\ }\bibfield  {title} {\bibinfo {title} {{Quasi-Conformal Models and the Early Universe}},\ }\href {https://doi.org/10.1140/epjc/s10052-019-7267-5} {\bibfield  {journal} {\bibinfo  {journal} {Eur. Phys. J. C}\ }\textbf {\bibinfo {volume} {79}},\ \bibinfo {pages} {750} (\bibinfo {year} {2019})},\ \Eprint {https://arxiv.org/abs/1907.00983} {arXiv:1907.00983 [hep-ph]} \BibitemShut {NoStop}%
\bibitem [{\citenamefont {Castello}\ \emph {et~al.}(2021)\citenamefont {Castello}, \citenamefont {Ili\'c},\ and\ \citenamefont {Kunz}}]{Castello2021}%
  \BibitemOpen
  \bibfield  {author} {\bibinfo {author} {\bibfnamefont {S.}~\bibnamefont {Castello}}, \bibinfo {author} {\bibfnamefont {S.}~\bibnamefont {Ili\'c}},\ and\ \bibinfo {author} {\bibfnamefont {M.}~\bibnamefont {Kunz}},\ }\bibfield  {title} {\bibinfo {title} {{Updated dark energy view of inflation}},\ }\href {https://doi.org/10.1103/PhysRevD.104.023522} {\bibfield  {journal} {\bibinfo  {journal} {Phys. Rev. D}\ }\textbf {\bibinfo {volume} {104}},\ \bibinfo {pages} {023522} (\bibinfo {year} {2021})},\ \Eprint {https://arxiv.org/abs/2104.15091} {arXiv:2104.15091 [astro-ph.CO]} \BibitemShut {NoStop}%
\bibitem [{\citenamefont {Ellis}\ and\ \citenamefont {Wands}(2023)}]{ellis2023inflation}%
  \BibitemOpen
  \bibfield  {author} {\bibinfo {author} {\bibfnamefont {J.}~\bibnamefont {Ellis}}\ and\ \bibinfo {author} {\bibfnamefont {D.}~\bibnamefont {Wands}},\ }\href@noop {} {\bibinfo {title} {Inflation (2023)}} (\bibinfo {year} {2023}),\ \Eprint {https://arxiv.org/abs/2312.13238} {arXiv:2312.13238 [astro-ph.CO]} \BibitemShut {NoStop}%
\bibitem [{\citenamefont {Achúcarro}\ \emph {et~al.}(2022)\citenamefont {Achúcarro}, \citenamefont {Biagetti}, \citenamefont {Braglia}, \citenamefont {Cabass}, \citenamefont {Caldwell}, \citenamefont {Castorina}, \citenamefont {Chen}, \citenamefont {Coulton}, \citenamefont {Flauger}, \citenamefont {Fumagalli}, \citenamefont {Ivanov}, \citenamefont {Lee}, \citenamefont {Maleknejad}, \citenamefont {Meerburg}, \citenamefont {Dizgah}, \citenamefont {Palma}, \citenamefont {Pimentel}, \citenamefont {Renaux-Petel}, \citenamefont {Wallisch}, \citenamefont {Wandelt}, \citenamefont {Witkowski},\ and\ \citenamefont {Wu}}]{achucarro2022inflation}%
  \BibitemOpen
  \bibfield  {author} {\bibinfo {author} {\bibfnamefont {A.}~\bibnamefont {Achúcarro}}, \bibinfo {author} {\bibfnamefont {M.}~\bibnamefont {Biagetti}}, \bibinfo {author} {\bibfnamefont {M.}~\bibnamefont {Braglia}}, \bibinfo {author} {\bibfnamefont {G.}~\bibnamefont {Cabass}}, \bibinfo {author} {\bibfnamefont {R.}~\bibnamefont {Caldwell}}, \bibinfo {author} {\bibfnamefont {E.}~\bibnamefont {Castorina}}, \bibinfo {author} {\bibfnamefont {X.}~\bibnamefont {Chen}}, \bibinfo {author} {\bibfnamefont {W.}~\bibnamefont {Coulton}}, \bibinfo {author} {\bibfnamefont {R.}~\bibnamefont {Flauger}}, \bibinfo {author} {\bibfnamefont {J.}~\bibnamefont {Fumagalli}}, \bibinfo {author} {\bibfnamefont {M.~M.}\ \bibnamefont {Ivanov}}, \bibinfo {author} {\bibfnamefont {H.}~\bibnamefont {Lee}}, \bibinfo {author} {\bibfnamefont {A.}~\bibnamefont {Maleknejad}}, \bibinfo {author} {\bibfnamefont {P.~D.}\ \bibnamefont {Meerburg}}, \bibinfo {author} {\bibfnamefont {A.~M.}\ \bibnamefont {Dizgah}}, \bibinfo {author} {\bibfnamefont {G.~A.}\
  \bibnamefont {Palma}}, \bibinfo {author} {\bibfnamefont {G.~L.}\ \bibnamefont {Pimentel}}, \bibinfo {author} {\bibfnamefont {S.}~\bibnamefont {Renaux-Petel}}, \bibinfo {author} {\bibfnamefont {B.}~\bibnamefont {Wallisch}}, \bibinfo {author} {\bibfnamefont {B.~D.}\ \bibnamefont {Wandelt}}, \bibinfo {author} {\bibfnamefont {L.~T.}\ \bibnamefont {Witkowski}},\ and\ \bibinfo {author} {\bibfnamefont {W.~L.~K.}\ \bibnamefont {Wu}},\ }\href@noop {} {\bibinfo {title} {Inflation: Theory and observations}} (\bibinfo {year} {2022}),\ \Eprint {https://arxiv.org/abs/2203.08128} {arXiv:2203.08128 [astro-ph.CO]} \BibitemShut {NoStop}%
\bibitem [{\citenamefont {Aghanim}\ \emph {et~al.}(2020{\natexlab{b}})\citenamefont {Aghanim}, \citenamefont {Akrami}, \citenamefont {Ashdown}, \citenamefont {Aumont}, \citenamefont {Baccigalupi}, \citenamefont {Ballardini}, \citenamefont {Banday}, \citenamefont {Barreiro}, \citenamefont {Bartolo}, \citenamefont {Basak}, \citenamefont {Battye}, \citenamefont {Benabed}, \citenamefont {Bernard}, \citenamefont {Bersanelli}, \citenamefont {Bielewicz}, \citenamefont {Bock}, \citenamefont {Bond}, \citenamefont {Borrill}, \citenamefont {Bouchet}, \citenamefont {Boulanger}, \citenamefont {Bucher}, \citenamefont {Burigana}, \citenamefont {Butler}, \citenamefont {Calabrese}, \citenamefont {Cardoso}, \citenamefont {Carron}, \citenamefont {Challinor}, \citenamefont {Chiang}, \citenamefont {Chluba}, \citenamefont {Colombo}, \citenamefont {Combet}, \citenamefont {Contreras}, \citenamefont {Crill}, \citenamefont {Cuttaia}, \citenamefont {de~Bernardis}, \citenamefont {de~Zotti}, \citenamefont {Delabrouille}, \citenamefont
  {Delouis}, \citenamefont {Di~Valentino}, \citenamefont {Diego}, \citenamefont {Doré}, \citenamefont {Douspis}, \citenamefont {Ducout}, \citenamefont {Dupac}, \citenamefont {Dusini}, \citenamefont {Efstathiou}, \citenamefont {Elsner}, \citenamefont {Enßlin}, \citenamefont {Eriksen}, \citenamefont {Fantaye}, \citenamefont {Farhang}, \citenamefont {Fergusson}, \citenamefont {Fernandez-Cobos}, \citenamefont {Finelli}, \citenamefont {Forastieri}, \citenamefont {Frailis}, \citenamefont {Fraisse}, \citenamefont {Franceschi}, \citenamefont {Frolov}, \citenamefont {Galeotta}, \citenamefont {Galli}, \citenamefont {Ganga}, \citenamefont {Génova-Santos}, \citenamefont {Gerbino}, \citenamefont {Ghosh}, \citenamefont {González-Nuevo}, \citenamefont {Górski}, \citenamefont {Gratton}, \citenamefont {Gruppuso}, \citenamefont {Gudmundsson}, \citenamefont {Hamann}, \citenamefont {Handley}, \citenamefont {Hansen}, \citenamefont {Herranz}, \citenamefont {Hildebrandt}, \citenamefont {Hivon}, \citenamefont {Huang},
  \citenamefont {Jaffe}, \citenamefont {Jones}, \citenamefont {Karakci}, \citenamefont {Keihänen}, \citenamefont {Keskitalo}, \citenamefont {Kiiveri}, \citenamefont {Kim}, \citenamefont {Kisner}, \citenamefont {Knox}, \citenamefont {Krachmalnicoff}, \citenamefont {Kunz}, \citenamefont {Kurki-Suonio}, \citenamefont {Lagache}, \citenamefont {Lamarre}, \citenamefont {Lasenby}, \citenamefont {Lattanzi}, \citenamefont {Lawrence}, \citenamefont {Le~Jeune}, \citenamefont {Lemos}, \citenamefont {Lesgourgues}, \citenamefont {Levrier}, \citenamefont {Lewis}, \citenamefont {Liguori}, \citenamefont {Lilje}, \citenamefont {Lilley}, \citenamefont {Lindholm}, \citenamefont {López-Caniego}, \citenamefont {Lubin}, \citenamefont {Ma}, \citenamefont {Macías-Pérez}, \citenamefont {Maggio}, \citenamefont {Maino}, \citenamefont {Mandolesi}, \citenamefont {Mangilli}, \citenamefont {Marcos-Caballero}, \citenamefont {Maris}, \citenamefont {Martin}, \citenamefont {Martinelli}, \citenamefont {Martínez-González}, \citenamefont
  {Matarrese}, \citenamefont {Mauri}, \citenamefont {McEwen}, \citenamefont {Meinhold}, \citenamefont {Melchiorri}, \citenamefont {Mennella}, \citenamefont {Migliaccio}, \citenamefont {Millea}, \citenamefont {Mitra}, \citenamefont {Miville-Deschênes}, \citenamefont {Molinari}, \citenamefont {Montier}, \citenamefont {Morgante}, \citenamefont {Moss}, \citenamefont {Natoli}, \citenamefont {Nørgaard-Nielsen}, \citenamefont {Pagano}, \citenamefont {Paoletti}, \citenamefont {Partridge}, \citenamefont {Patanchon}, \citenamefont {Peiris}, \citenamefont {Perrotta}, \citenamefont {Pettorino}, \citenamefont {Piacentini}, \citenamefont {Polastri}, \citenamefont {Polenta}, \citenamefont {Puget}, \citenamefont {Rachen}, \citenamefont {Reinecke}, \citenamefont {Remazeilles}, \citenamefont {Renzi}, \citenamefont {Rocha}, \citenamefont {Rosset}, \citenamefont {Roudier}, \citenamefont {Rubiño-Martín}, \citenamefont {Ruiz-Granados}, \citenamefont {Salvati}, \citenamefont {Sandri}, \citenamefont {Savelainen}, \citenamefont
  {Scott}, \citenamefont {Shellard}, \citenamefont {Sirignano}, \citenamefont {Sirri}, \citenamefont {Spencer}, \citenamefont {Sunyaev}, \citenamefont {Suur-Uski}, \citenamefont {Tauber}, \citenamefont {Tavagnacco}, \citenamefont {Tenti}, \citenamefont {Toffolatti}, \citenamefont {Tomasi}, \citenamefont {Trombetti}, \citenamefont {Valenziano}, \citenamefont {Valiviita}, \citenamefont {Van~Tent}, \citenamefont {Vibert}, \citenamefont {Vielva}, \citenamefont {Villa}, \citenamefont {Vittorio}, \citenamefont {Wandelt}, \citenamefont {Wehus}, \citenamefont {White}, \citenamefont {White}, \citenamefont {Zacchei},\ and\ \citenamefont {Zonca}}]{2020_CMB}%
  \BibitemOpen
  \bibfield  {author} {\bibinfo {author} {\bibfnamefont {N.}~\bibnamefont {Aghanim}}, \bibinfo {author} {\bibfnamefont {Y.}~\bibnamefont {Akrami}}, \bibinfo {author} {\bibfnamefont {M.}~\bibnamefont {Ashdown}}, \bibinfo {author} {\bibfnamefont {J.}~\bibnamefont {Aumont}}, \bibinfo {author} {\bibfnamefont {C.}~\bibnamefont {Baccigalupi}}, \bibinfo {author} {\bibfnamefont {M.}~\bibnamefont {Ballardini}}, \bibinfo {author} {\bibfnamefont {A.~J.}\ \bibnamefont {Banday}}, \bibinfo {author} {\bibfnamefont {R.~B.}\ \bibnamefont {Barreiro}}, \bibinfo {author} {\bibfnamefont {N.}~\bibnamefont {Bartolo}}, \bibinfo {author} {\bibfnamefont {S.}~\bibnamefont {Basak}}, \bibinfo {author} {\bibfnamefont {R.}~\bibnamefont {Battye}}, \bibinfo {author} {\bibfnamefont {K.}~\bibnamefont {Benabed}}, \bibinfo {author} {\bibfnamefont {J.-P.}\ \bibnamefont {Bernard}}, \bibinfo {author} {\bibfnamefont {M.}~\bibnamefont {Bersanelli}}, \bibinfo {author} {\bibfnamefont {P.}~\bibnamefont {Bielewicz}}, \bibinfo {author} {\bibfnamefont
  {J.~J.}\ \bibnamefont {Bock}}, \bibinfo {author} {\bibfnamefont {J.~R.}\ \bibnamefont {Bond}}, \bibinfo {author} {\bibfnamefont {J.}~\bibnamefont {Borrill}}, \bibinfo {author} {\bibfnamefont {F.~R.}\ \bibnamefont {Bouchet}}, \bibinfo {author} {\bibfnamefont {F.}~\bibnamefont {Boulanger}}, \bibinfo {author} {\bibfnamefont {M.}~\bibnamefont {Bucher}}, \bibinfo {author} {\bibfnamefont {C.}~\bibnamefont {Burigana}}, \bibinfo {author} {\bibfnamefont {R.~C.}\ \bibnamefont {Butler}}, \bibinfo {author} {\bibfnamefont {E.}~\bibnamefont {Calabrese}}, \bibinfo {author} {\bibfnamefont {J.-F.}\ \bibnamefont {Cardoso}}, \bibinfo {author} {\bibfnamefont {J.}~\bibnamefont {Carron}}, \bibinfo {author} {\bibfnamefont {A.}~\bibnamefont {Challinor}}, \bibinfo {author} {\bibfnamefont {H.~C.}\ \bibnamefont {Chiang}}, \bibinfo {author} {\bibfnamefont {J.}~\bibnamefont {Chluba}}, \bibinfo {author} {\bibfnamefont {L.~P.~L.}\ \bibnamefont {Colombo}}, \bibinfo {author} {\bibfnamefont {C.}~\bibnamefont {Combet}}, \bibinfo {author}
  {\bibfnamefont {D.}~\bibnamefont {Contreras}}, \bibinfo {author} {\bibfnamefont {B.~P.}\ \bibnamefont {Crill}}, \bibinfo {author} {\bibfnamefont {F.}~\bibnamefont {Cuttaia}}, \bibinfo {author} {\bibfnamefont {P.}~\bibnamefont {de~Bernardis}}, \bibinfo {author} {\bibfnamefont {G.}~\bibnamefont {de~Zotti}}, \bibinfo {author} {\bibfnamefont {J.}~\bibnamefont {Delabrouille}}, \bibinfo {author} {\bibfnamefont {J.-M.}\ \bibnamefont {Delouis}}, \bibinfo {author} {\bibfnamefont {E.}~\bibnamefont {Di~Valentino}}, \bibinfo {author} {\bibfnamefont {J.~M.}\ \bibnamefont {Diego}}, \bibinfo {author} {\bibfnamefont {O.}~\bibnamefont {Doré}}, \bibinfo {author} {\bibfnamefont {M.}~\bibnamefont {Douspis}}, \bibinfo {author} {\bibfnamefont {A.}~\bibnamefont {Ducout}}, \bibinfo {author} {\bibfnamefont {X.}~\bibnamefont {Dupac}}, \bibinfo {author} {\bibfnamefont {S.}~\bibnamefont {Dusini}}, \bibinfo {author} {\bibfnamefont {G.}~\bibnamefont {Efstathiou}}, \bibinfo {author} {\bibfnamefont {F.}~\bibnamefont {Elsner}}, \bibinfo
  {author} {\bibfnamefont {T.~A.}\ \bibnamefont {Enßlin}}, \bibinfo {author} {\bibfnamefont {H.~K.}\ \bibnamefont {Eriksen}}, \bibinfo {author} {\bibfnamefont {Y.}~\bibnamefont {Fantaye}}, \bibinfo {author} {\bibfnamefont {M.}~\bibnamefont {Farhang}}, \bibinfo {author} {\bibfnamefont {J.}~\bibnamefont {Fergusson}}, \bibinfo {author} {\bibfnamefont {R.}~\bibnamefont {Fernandez-Cobos}}, \bibinfo {author} {\bibfnamefont {F.}~\bibnamefont {Finelli}}, \bibinfo {author} {\bibfnamefont {F.}~\bibnamefont {Forastieri}}, \bibinfo {author} {\bibfnamefont {M.}~\bibnamefont {Frailis}}, \bibinfo {author} {\bibfnamefont {A.~A.}\ \bibnamefont {Fraisse}}, \bibinfo {author} {\bibfnamefont {E.}~\bibnamefont {Franceschi}}, \bibinfo {author} {\bibfnamefont {A.}~\bibnamefont {Frolov}}, \bibinfo {author} {\bibfnamefont {S.}~\bibnamefont {Galeotta}}, \bibinfo {author} {\bibfnamefont {S.}~\bibnamefont {Galli}}, \bibinfo {author} {\bibfnamefont {K.}~\bibnamefont {Ganga}}, \bibinfo {author} {\bibfnamefont {R.~T.}\ \bibnamefont
  {Génova-Santos}}, \bibinfo {author} {\bibfnamefont {M.}~\bibnamefont {Gerbino}}, \bibinfo {author} {\bibfnamefont {T.}~\bibnamefont {Ghosh}}, \bibinfo {author} {\bibfnamefont {J.}~\bibnamefont {González-Nuevo}}, \bibinfo {author} {\bibfnamefont {K.~M.}\ \bibnamefont {Górski}}, \bibinfo {author} {\bibfnamefont {S.}~\bibnamefont {Gratton}}, \bibinfo {author} {\bibfnamefont {A.}~\bibnamefont {Gruppuso}}, \bibinfo {author} {\bibfnamefont {J.~E.}\ \bibnamefont {Gudmundsson}}, \bibinfo {author} {\bibfnamefont {J.}~\bibnamefont {Hamann}}, \bibinfo {author} {\bibfnamefont {W.}~\bibnamefont {Handley}}, \bibinfo {author} {\bibfnamefont {F.~K.}\ \bibnamefont {Hansen}}, \bibinfo {author} {\bibfnamefont {D.}~\bibnamefont {Herranz}}, \bibinfo {author} {\bibfnamefont {S.~R.}\ \bibnamefont {Hildebrandt}}, \bibinfo {author} {\bibfnamefont {E.}~\bibnamefont {Hivon}}, \bibinfo {author} {\bibfnamefont {Z.}~\bibnamefont {Huang}}, \bibinfo {author} {\bibfnamefont {A.~H.}\ \bibnamefont {Jaffe}}, \bibinfo {author}
  {\bibfnamefont {W.~C.}\ \bibnamefont {Jones}}, \bibinfo {author} {\bibfnamefont {A.}~\bibnamefont {Karakci}}, \bibinfo {author} {\bibfnamefont {E.}~\bibnamefont {Keihänen}}, \bibinfo {author} {\bibfnamefont {R.}~\bibnamefont {Keskitalo}}, \bibinfo {author} {\bibfnamefont {K.}~\bibnamefont {Kiiveri}}, \bibinfo {author} {\bibfnamefont {J.}~\bibnamefont {Kim}}, \bibinfo {author} {\bibfnamefont {T.~S.}\ \bibnamefont {Kisner}}, \bibinfo {author} {\bibfnamefont {L.}~\bibnamefont {Knox}}, \bibinfo {author} {\bibfnamefont {N.}~\bibnamefont {Krachmalnicoff}}, \bibinfo {author} {\bibfnamefont {M.}~\bibnamefont {Kunz}}, \bibinfo {author} {\bibfnamefont {H.}~\bibnamefont {Kurki-Suonio}}, \bibinfo {author} {\bibfnamefont {G.}~\bibnamefont {Lagache}}, \bibinfo {author} {\bibfnamefont {J.-M.}\ \bibnamefont {Lamarre}}, \bibinfo {author} {\bibfnamefont {A.}~\bibnamefont {Lasenby}}, \bibinfo {author} {\bibfnamefont {M.}~\bibnamefont {Lattanzi}}, \bibinfo {author} {\bibfnamefont {C.~R.}\ \bibnamefont {Lawrence}}, \bibinfo
  {author} {\bibfnamefont {M.}~\bibnamefont {Le~Jeune}}, \bibinfo {author} {\bibfnamefont {P.}~\bibnamefont {Lemos}}, \bibinfo {author} {\bibfnamefont {J.}~\bibnamefont {Lesgourgues}}, \bibinfo {author} {\bibfnamefont {F.}~\bibnamefont {Levrier}}, \bibinfo {author} {\bibfnamefont {A.}~\bibnamefont {Lewis}}, \bibinfo {author} {\bibfnamefont {M.}~\bibnamefont {Liguori}}, \bibinfo {author} {\bibfnamefont {P.~B.}\ \bibnamefont {Lilje}}, \bibinfo {author} {\bibfnamefont {M.}~\bibnamefont {Lilley}}, \bibinfo {author} {\bibfnamefont {V.}~\bibnamefont {Lindholm}}, \bibinfo {author} {\bibfnamefont {M.}~\bibnamefont {López-Caniego}}, \bibinfo {author} {\bibfnamefont {P.~M.}\ \bibnamefont {Lubin}}, \bibinfo {author} {\bibfnamefont {Y.-Z.}\ \bibnamefont {Ma}}, \bibinfo {author} {\bibfnamefont {J.~F.}\ \bibnamefont {Macías-Pérez}}, \bibinfo {author} {\bibfnamefont {G.}~\bibnamefont {Maggio}}, \bibinfo {author} {\bibfnamefont {D.}~\bibnamefont {Maino}}, \bibinfo {author} {\bibfnamefont {N.}~\bibnamefont {Mandolesi}},
  \bibinfo {author} {\bibfnamefont {A.}~\bibnamefont {Mangilli}}, \bibinfo {author} {\bibfnamefont {A.}~\bibnamefont {Marcos-Caballero}}, \bibinfo {author} {\bibfnamefont {M.}~\bibnamefont {Maris}}, \bibinfo {author} {\bibfnamefont {P.~G.}\ \bibnamefont {Martin}}, \bibinfo {author} {\bibfnamefont {M.}~\bibnamefont {Martinelli}}, \bibinfo {author} {\bibfnamefont {E.}~\bibnamefont {Martínez-González}}, \bibinfo {author} {\bibfnamefont {S.}~\bibnamefont {Matarrese}}, \bibinfo {author} {\bibfnamefont {N.}~\bibnamefont {Mauri}}, \bibinfo {author} {\bibfnamefont {J.~D.}\ \bibnamefont {McEwen}}, \bibinfo {author} {\bibfnamefont {P.~R.}\ \bibnamefont {Meinhold}}, \bibinfo {author} {\bibfnamefont {A.}~\bibnamefont {Melchiorri}}, \bibinfo {author} {\bibfnamefont {A.}~\bibnamefont {Mennella}}, \bibinfo {author} {\bibfnamefont {M.}~\bibnamefont {Migliaccio}}, \bibinfo {author} {\bibfnamefont {M.}~\bibnamefont {Millea}}, \bibinfo {author} {\bibfnamefont {S.}~\bibnamefont {Mitra}}, \bibinfo {author} {\bibfnamefont
  {M.-A.}\ \bibnamefont {Miville-Deschênes}}, \bibinfo {author} {\bibfnamefont {D.}~\bibnamefont {Molinari}}, \bibinfo {author} {\bibfnamefont {L.}~\bibnamefont {Montier}}, \bibinfo {author} {\bibfnamefont {G.}~\bibnamefont {Morgante}}, \bibinfo {author} {\bibfnamefont {A.}~\bibnamefont {Moss}}, \bibinfo {author} {\bibfnamefont {P.}~\bibnamefont {Natoli}}, \bibinfo {author} {\bibfnamefont {H.~U.}\ \bibnamefont {Nørgaard-Nielsen}}, \bibinfo {author} {\bibfnamefont {L.}~\bibnamefont {Pagano}}, \bibinfo {author} {\bibfnamefont {D.}~\bibnamefont {Paoletti}}, \bibinfo {author} {\bibfnamefont {B.}~\bibnamefont {Partridge}}, \bibinfo {author} {\bibfnamefont {G.}~\bibnamefont {Patanchon}}, \bibinfo {author} {\bibfnamefont {H.~V.}\ \bibnamefont {Peiris}}, \bibinfo {author} {\bibfnamefont {F.}~\bibnamefont {Perrotta}}, \bibinfo {author} {\bibfnamefont {V.}~\bibnamefont {Pettorino}}, \bibinfo {author} {\bibfnamefont {F.}~\bibnamefont {Piacentini}}, \bibinfo {author} {\bibfnamefont {L.}~\bibnamefont {Polastri}},
  \bibinfo {author} {\bibfnamefont {G.}~\bibnamefont {Polenta}}, \bibinfo {author} {\bibfnamefont {J.-L.}\ \bibnamefont {Puget}}, \bibinfo {author} {\bibfnamefont {J.~P.}\ \bibnamefont {Rachen}}, \bibinfo {author} {\bibfnamefont {M.}~\bibnamefont {Reinecke}}, \bibinfo {author} {\bibfnamefont {M.}~\bibnamefont {Remazeilles}}, \bibinfo {author} {\bibfnamefont {A.}~\bibnamefont {Renzi}}, \bibinfo {author} {\bibfnamefont {G.}~\bibnamefont {Rocha}}, \bibinfo {author} {\bibfnamefont {C.}~\bibnamefont {Rosset}}, \bibinfo {author} {\bibfnamefont {G.}~\bibnamefont {Roudier}}, \bibinfo {author} {\bibfnamefont {J.~A.}\ \bibnamefont {Rubiño-Martín}}, \bibinfo {author} {\bibfnamefont {B.}~\bibnamefont {Ruiz-Granados}}, \bibinfo {author} {\bibfnamefont {L.}~\bibnamefont {Salvati}}, \bibinfo {author} {\bibfnamefont {M.}~\bibnamefont {Sandri}}, \bibinfo {author} {\bibfnamefont {M.}~\bibnamefont {Savelainen}}, \bibinfo {author} {\bibfnamefont {D.}~\bibnamefont {Scott}}, \bibinfo {author} {\bibfnamefont {E.~P.~S.}\
  \bibnamefont {Shellard}}, \bibinfo {author} {\bibfnamefont {C.}~\bibnamefont {Sirignano}}, \bibinfo {author} {\bibfnamefont {G.}~\bibnamefont {Sirri}}, \bibinfo {author} {\bibfnamefont {L.~D.}\ \bibnamefont {Spencer}}, \bibinfo {author} {\bibfnamefont {R.}~\bibnamefont {Sunyaev}}, \bibinfo {author} {\bibfnamefont {A.-S.}\ \bibnamefont {Suur-Uski}}, \bibinfo {author} {\bibfnamefont {J.~A.}\ \bibnamefont {Tauber}}, \bibinfo {author} {\bibfnamefont {D.}~\bibnamefont {Tavagnacco}}, \bibinfo {author} {\bibfnamefont {M.}~\bibnamefont {Tenti}}, \bibinfo {author} {\bibfnamefont {L.}~\bibnamefont {Toffolatti}}, \bibinfo {author} {\bibfnamefont {M.}~\bibnamefont {Tomasi}}, \bibinfo {author} {\bibfnamefont {T.}~\bibnamefont {Trombetti}}, \bibinfo {author} {\bibfnamefont {L.}~\bibnamefont {Valenziano}}, \bibinfo {author} {\bibfnamefont {J.}~\bibnamefont {Valiviita}}, \bibinfo {author} {\bibfnamefont {B.}~\bibnamefont {Van~Tent}}, \bibinfo {author} {\bibfnamefont {L.}~\bibnamefont {Vibert}}, \bibinfo {author}
  {\bibfnamefont {P.}~\bibnamefont {Vielva}}, \bibinfo {author} {\bibfnamefont {F.}~\bibnamefont {Villa}}, \bibinfo {author} {\bibfnamefont {N.}~\bibnamefont {Vittorio}}, \bibinfo {author} {\bibfnamefont {B.~D.}\ \bibnamefont {Wandelt}}, \bibinfo {author} {\bibfnamefont {I.~K.}\ \bibnamefont {Wehus}}, \bibinfo {author} {\bibfnamefont {M.}~\bibnamefont {White}}, \bibinfo {author} {\bibfnamefont {S.~D.~M.}\ \bibnamefont {White}}, \bibinfo {author} {\bibfnamefont {A.}~\bibnamefont {Zacchei}},\ and\ \bibinfo {author} {\bibfnamefont {A.}~\bibnamefont {Zonca}},\ }\bibfield  {title} {\bibinfo {title} {Planck2018 results: Vi. cosmological parameters},\ }\href {https://doi.org/10.1051/0004-6361/201833910} {\bibfield  {journal} {\bibinfo  {journal} {Astronomy \& Astrophysics}\ }\textbf {\bibinfo {volume} {641}},\ \bibinfo {pages} {A6} (\bibinfo {year} {2020}{\natexlab{b}})}\BibitemShut {NoStop}%
\bibitem [{\citenamefont {Giarè}\ \emph {et~al.}(2023)\citenamefont {Giarè}, \citenamefont {Renzi}, \citenamefont {Mena}, \citenamefont {Di~Valentino},\ and\ \citenamefont {Melchiorri}}]{Giar_2023}%
  \BibitemOpen
  \bibfield  {author} {\bibinfo {author} {\bibfnamefont {W.}~\bibnamefont {Giarè}}, \bibinfo {author} {\bibfnamefont {F.}~\bibnamefont {Renzi}}, \bibinfo {author} {\bibfnamefont {O.}~\bibnamefont {Mena}}, \bibinfo {author} {\bibfnamefont {E.}~\bibnamefont {Di~Valentino}},\ and\ \bibinfo {author} {\bibfnamefont {A.}~\bibnamefont {Melchiorri}},\ }\bibfield  {title} {\bibinfo {title} {Is the harrison-zel'dovich spectrum coming back? act preference for $n_s\sim 1$ and its discordance with planck},\ }\href {https://doi.org/10.1093/mnras/stad724} {\bibfield  {journal} {\bibinfo  {journal} {Monthly Notices of the Royal Astronomical Society}\ }\textbf {\bibinfo {volume} {521}},\ \bibinfo {pages} {2911–2918} (\bibinfo {year} {2023})}\BibitemShut {NoStop}%
\bibitem [{\citenamefont {Di~Valentino}\ \emph {et~al.}(2022)\citenamefont {Di~Valentino}, \citenamefont {Giarè}, \citenamefont {Melchiorri},\ and\ \citenamefont {Silk}}]{Di_Valentino_2022}%
  \BibitemOpen
  \bibfield  {author} {\bibinfo {author} {\bibfnamefont {E.}~\bibnamefont {Di~Valentino}}, \bibinfo {author} {\bibfnamefont {W.}~\bibnamefont {Giarè}}, \bibinfo {author} {\bibfnamefont {A.}~\bibnamefont {Melchiorri}},\ and\ \bibinfo {author} {\bibfnamefont {J.}~\bibnamefont {Silk}},\ }\bibfield  {title} {\bibinfo {title} {Health checkup test of the standard cosmological model in view of recent cosmic microwave background anisotropies experiments},\ }\bibfield  {journal} {\bibinfo  {journal} {Physical Review D}\ }\textbf {\bibinfo {volume} {106}},\ \href {https://doi.org/10.1103/physrevd.106.103506} {10.1103/physrevd.106.103506} (\bibinfo {year} {2022})\BibitemShut {NoStop}%
\bibitem [{\citenamefont {Di~Valentino}\ \emph {et~al.}(2023)\citenamefont {Di~Valentino}, \citenamefont {Giarè}, \citenamefont {Melchiorri},\ and\ \citenamefont {Silk}}]{Di_Valentino_2023}%
  \BibitemOpen
  \bibfield  {author} {\bibinfo {author} {\bibfnamefont {E.}~\bibnamefont {Di~Valentino}}, \bibinfo {author} {\bibfnamefont {W.}~\bibnamefont {Giarè}}, \bibinfo {author} {\bibfnamefont {A.}~\bibnamefont {Melchiorri}},\ and\ \bibinfo {author} {\bibfnamefont {J.}~\bibnamefont {Silk}},\ }\bibfield  {title} {\bibinfo {title} {Quantifying the global ``cmb tension'' between the atacama cosmology telescope and the planck satellite in extended models of cosmology},\ }\href {https://doi.org/10.1093/mnras/stad152} {\bibfield  {journal} {\bibinfo  {journal} {Monthly Notices of the Royal Astronomical Society}\ }\textbf {\bibinfo {volume} {520}},\ \bibinfo {pages} {210–215} (\bibinfo {year} {2023})}\BibitemShut {NoStop}%
\bibitem [{\citenamefont {Calderon}\ \emph {et~al.}(2023)\citenamefont {Calderon}, \citenamefont {Shafieloo}, \citenamefont {Kumar~Hazra},\ and\ \citenamefont {Sohn}}]{Calderon_2023}%
  \BibitemOpen
  \bibfield  {author} {\bibinfo {author} {\bibfnamefont {R.}~\bibnamefont {Calderon}}, \bibinfo {author} {\bibfnamefont {A.}~\bibnamefont {Shafieloo}}, \bibinfo {author} {\bibfnamefont {D.}~\bibnamefont {Kumar~Hazra}},\ and\ \bibinfo {author} {\bibfnamefont {W.}~\bibnamefont {Sohn}},\ }\bibfield  {title} {\bibinfo {title} {On the consistency of $\lambda$cdm with cmb measurements in light of the latest planck, act and spt data},\ }\href {https://doi.org/10.1088/1475-7516/2023/08/059} {\bibfield  {journal} {\bibinfo  {journal} {Journal of Cosmology and Astroparticle Physics}\ }\textbf {\bibinfo {volume} {2023}}\bibinfo  {number} { (08)},\ \bibinfo {pages} {059}}\BibitemShut {NoStop}%
\bibitem [{\citenamefont {Ade}\ \emph {et~al.}(2021{\natexlab{a}})\citenamefont {Ade}, \citenamefont {Ahmed}, \citenamefont {Amiri}, \citenamefont {Barkats}, \citenamefont {Thakur}, \citenamefont {Bischoff}, \citenamefont {Beck}, \citenamefont {Bock}, \citenamefont {Boenish}, \citenamefont {Bullock}, \citenamefont {Buza}, \citenamefont {Cheshire}, \citenamefont {Connors}, \citenamefont {Cornelison}, \citenamefont {Crumrine}, \citenamefont {Cukierman}, \citenamefont {Denison}, \citenamefont {Dierickx}, \citenamefont {Duband}, \citenamefont {Eiben}, \citenamefont {Fatigoni}, \citenamefont {Filippini}, \citenamefont {Fliescher}, \citenamefont {Goeckner-Wald}, \citenamefont {Goldfinger}, \citenamefont {Grayson}, \citenamefont {Grimes}, \citenamefont {Hall}, \citenamefont {Halal}, \citenamefont {Halpern}, \citenamefont {Hand}, \citenamefont {Harrison}, \citenamefont {Henderson}, \citenamefont {Hildebrandt}, \citenamefont {Hilton}, \citenamefont {Hubmayr}, \citenamefont {Hui}, \citenamefont {Irwin}, \citenamefont
  {Kang}, \citenamefont {Karkare}, \citenamefont {Karpel}, \citenamefont {Kefeli}, \citenamefont {Kernasovskiy}, \citenamefont {Kovac}, \citenamefont {Kuo}, \citenamefont {Lau}, \citenamefont {Leitch}, \citenamefont {Lennox}, \citenamefont {Megerian}, \citenamefont {Minutolo}, \citenamefont {Moncelsi}, \citenamefont {Nakato}, \citenamefont {Namikawa}, \citenamefont {Nguyen}, \citenamefont {O’Brient}, \citenamefont {Ogburn}, \citenamefont {Palladino}, \citenamefont {Prouve}, \citenamefont {Pryke}, \citenamefont {Racine}, \citenamefont {Reintsema}, \citenamefont {Richter}, \citenamefont {Schillaci}, \citenamefont {Schwarz}, \citenamefont {Schmitt}, \citenamefont {Sheehy}, \citenamefont {Soliman}, \citenamefont {Germaine}, \citenamefont {Steinbach}, \citenamefont {Sudiwala}, \citenamefont {Teply}, \citenamefont {Thompson}, \citenamefont {Tolan}, \citenamefont {Tucker}, \citenamefont {Turner}, \citenamefont {Umiltà}, \citenamefont {Vergès}, \citenamefont {Vieregg}, \citenamefont {Wandui}, \citenamefont
  {Weber}, \citenamefont {Wiebe}, \citenamefont {Willmert}, \citenamefont {Wong}, \citenamefont {Wu}, \citenamefont {Yang}, \citenamefont {Yoon}, \citenamefont {Young}, \citenamefont {Yu}, \citenamefont {Zeng}, \citenamefont {Zhang},\ and\ \citenamefont {Zhang}}]{Ade_2021}%
  \BibitemOpen
\bibfield  {number} {  }\bibfield  {author} {\bibinfo {author} {\bibfnamefont {P.}~\bibnamefont {Ade}}, \bibinfo {author} {\bibfnamefont {Z.}~\bibnamefont {Ahmed}}, \bibinfo {author} {\bibfnamefont {M.}~\bibnamefont {Amiri}}, \bibinfo {author} {\bibfnamefont {D.}~\bibnamefont {Barkats}}, \bibinfo {author} {\bibfnamefont {R.~B.}\ \bibnamefont {Thakur}}, \bibinfo {author} {\bibfnamefont {C.}~\bibnamefont {Bischoff}}, \bibinfo {author} {\bibfnamefont {D.}~\bibnamefont {Beck}}, \bibinfo {author} {\bibfnamefont {J.}~\bibnamefont {Bock}}, \bibinfo {author} {\bibfnamefont {H.}~\bibnamefont {Boenish}}, \bibinfo {author} {\bibfnamefont {E.}~\bibnamefont {Bullock}}, \bibinfo {author} {\bibfnamefont {V.}~\bibnamefont {Buza}}, \bibinfo {author} {\bibfnamefont {J.}~\bibnamefont {Cheshire}}, \bibinfo {author} {\bibfnamefont {J.}~\bibnamefont {Connors}}, \bibinfo {author} {\bibfnamefont {J.}~\bibnamefont {Cornelison}}, \bibinfo {author} {\bibfnamefont {M.}~\bibnamefont {Crumrine}}, \bibinfo {author} {\bibfnamefont
  {A.}~\bibnamefont {Cukierman}}, \bibinfo {author} {\bibfnamefont {E.}~\bibnamefont {Denison}}, \bibinfo {author} {\bibfnamefont {M.}~\bibnamefont {Dierickx}}, \bibinfo {author} {\bibfnamefont {L.}~\bibnamefont {Duband}}, \bibinfo {author} {\bibfnamefont {M.}~\bibnamefont {Eiben}}, \bibinfo {author} {\bibfnamefont {S.}~\bibnamefont {Fatigoni}}, \bibinfo {author} {\bibfnamefont {J.}~\bibnamefont {Filippini}}, \bibinfo {author} {\bibfnamefont {S.}~\bibnamefont {Fliescher}}, \bibinfo {author} {\bibfnamefont {N.}~\bibnamefont {Goeckner-Wald}}, \bibinfo {author} {\bibfnamefont {D.}~\bibnamefont {Goldfinger}}, \bibinfo {author} {\bibfnamefont {J.}~\bibnamefont {Grayson}}, \bibinfo {author} {\bibfnamefont {P.}~\bibnamefont {Grimes}}, \bibinfo {author} {\bibfnamefont {G.}~\bibnamefont {Hall}}, \bibinfo {author} {\bibfnamefont {G.}~\bibnamefont {Halal}}, \bibinfo {author} {\bibfnamefont {M.}~\bibnamefont {Halpern}}, \bibinfo {author} {\bibfnamefont {E.}~\bibnamefont {Hand}}, \bibinfo {author} {\bibfnamefont
  {S.}~\bibnamefont {Harrison}}, \bibinfo {author} {\bibfnamefont {S.}~\bibnamefont {Henderson}}, \bibinfo {author} {\bibfnamefont {S.}~\bibnamefont {Hildebrandt}}, \bibinfo {author} {\bibfnamefont {G.}~\bibnamefont {Hilton}}, \bibinfo {author} {\bibfnamefont {J.}~\bibnamefont {Hubmayr}}, \bibinfo {author} {\bibfnamefont {H.}~\bibnamefont {Hui}}, \bibinfo {author} {\bibfnamefont {K.}~\bibnamefont {Irwin}}, \bibinfo {author} {\bibfnamefont {J.}~\bibnamefont {Kang}}, \bibinfo {author} {\bibfnamefont {K.}~\bibnamefont {Karkare}}, \bibinfo {author} {\bibfnamefont {E.}~\bibnamefont {Karpel}}, \bibinfo {author} {\bibfnamefont {S.}~\bibnamefont {Kefeli}}, \bibinfo {author} {\bibfnamefont {S.}~\bibnamefont {Kernasovskiy}}, \bibinfo {author} {\bibfnamefont {J.}~\bibnamefont {Kovac}}, \bibinfo {author} {\bibfnamefont {C.}~\bibnamefont {Kuo}}, \bibinfo {author} {\bibfnamefont {K.}~\bibnamefont {Lau}}, \bibinfo {author} {\bibfnamefont {E.}~\bibnamefont {Leitch}}, \bibinfo {author} {\bibfnamefont {A.}~\bibnamefont
  {Lennox}}, \bibinfo {author} {\bibfnamefont {K.}~\bibnamefont {Megerian}}, \bibinfo {author} {\bibfnamefont {L.}~\bibnamefont {Minutolo}}, \bibinfo {author} {\bibfnamefont {L.}~\bibnamefont {Moncelsi}}, \bibinfo {author} {\bibfnamefont {Y.}~\bibnamefont {Nakato}}, \bibinfo {author} {\bibfnamefont {T.}~\bibnamefont {Namikawa}}, \bibinfo {author} {\bibfnamefont {H.}~\bibnamefont {Nguyen}}, \bibinfo {author} {\bibfnamefont {R.}~\bibnamefont {O’Brient}}, \bibinfo {author} {\bibfnamefont {R.}~\bibnamefont {Ogburn}}, \bibinfo {author} {\bibfnamefont {S.}~\bibnamefont {Palladino}}, \bibinfo {author} {\bibfnamefont {T.}~\bibnamefont {Prouve}}, \bibinfo {author} {\bibfnamefont {C.}~\bibnamefont {Pryke}}, \bibinfo {author} {\bibfnamefont {B.}~\bibnamefont {Racine}}, \bibinfo {author} {\bibfnamefont {C.}~\bibnamefont {Reintsema}}, \bibinfo {author} {\bibfnamefont {S.}~\bibnamefont {Richter}}, \bibinfo {author} {\bibfnamefont {A.}~\bibnamefont {Schillaci}}, \bibinfo {author} {\bibfnamefont {R.}~\bibnamefont
  {Schwarz}}, \bibinfo {author} {\bibfnamefont {B.}~\bibnamefont {Schmitt}}, \bibinfo {author} {\bibfnamefont {C.}~\bibnamefont {Sheehy}}, \bibinfo {author} {\bibfnamefont {A.}~\bibnamefont {Soliman}}, \bibinfo {author} {\bibfnamefont {T.}~\bibnamefont {Germaine}}, \bibinfo {author} {\bibfnamefont {B.}~\bibnamefont {Steinbach}}, \bibinfo {author} {\bibfnamefont {R.}~\bibnamefont {Sudiwala}}, \bibinfo {author} {\bibfnamefont {G.}~\bibnamefont {Teply}}, \bibinfo {author} {\bibfnamefont {K.}~\bibnamefont {Thompson}}, \bibinfo {author} {\bibfnamefont {J.}~\bibnamefont {Tolan}}, \bibinfo {author} {\bibfnamefont {C.}~\bibnamefont {Tucker}}, \bibinfo {author} {\bibfnamefont {A.}~\bibnamefont {Turner}}, \bibinfo {author} {\bibfnamefont {C.}~\bibnamefont {Umiltà}}, \bibinfo {author} {\bibfnamefont {C.}~\bibnamefont {Vergès}}, \bibinfo {author} {\bibfnamefont {A.}~\bibnamefont {Vieregg}}, \bibinfo {author} {\bibfnamefont {A.}~\bibnamefont {Wandui}}, \bibinfo {author} {\bibfnamefont {A.}~\bibnamefont {Weber}},
  \bibinfo {author} {\bibfnamefont {D.}~\bibnamefont {Wiebe}}, \bibinfo {author} {\bibfnamefont {J.}~\bibnamefont {Willmert}}, \bibinfo {author} {\bibfnamefont {C.}~\bibnamefont {Wong}}, \bibinfo {author} {\bibfnamefont {W.}~\bibnamefont {Wu}}, \bibinfo {author} {\bibfnamefont {H.}~\bibnamefont {Yang}}, \bibinfo {author} {\bibfnamefont {K.}~\bibnamefont {Yoon}}, \bibinfo {author} {\bibfnamefont {E.}~\bibnamefont {Young}}, \bibinfo {author} {\bibfnamefont {C.}~\bibnamefont {Yu}}, \bibinfo {author} {\bibfnamefont {L.}~\bibnamefont {Zeng}}, \bibinfo {author} {\bibfnamefont {C.}~\bibnamefont {Zhang}},\ and\ \bibinfo {author} {\bibfnamefont {S.}~\bibnamefont {Zhang}},\ }\bibfield  {title} {\bibinfo {title} {Improved constraints on primordial gravitational waves using planck, wmap, and bicep/keck observations through the 2018 observing season},\ }\bibfield  {journal} {\bibinfo  {journal} {Physical Review Letters}\ }\textbf {\bibinfo {volume} {127}},\ \href {https://doi.org/10.1103/physrevlett.127.151301}
  {10.1103/physrevlett.127.151301} (\bibinfo {year} {2021}{\natexlab{a}})\BibitemShut {NoStop}%
\bibitem [{\citenamefont {Campeti}\ and\ \citenamefont {Komatsu}(2022)}]{Campeti_2022}%
  \BibitemOpen
  \bibfield  {author} {\bibinfo {author} {\bibfnamefont {P.}~\bibnamefont {Campeti}}\ and\ \bibinfo {author} {\bibfnamefont {E.}~\bibnamefont {Komatsu}},\ }\bibfield  {title} {\bibinfo {title} {New constraint on the tensor-to-scalar ratio from the planck and bicep/keck array data using the profile likelihood},\ }\href {https://doi.org/10.3847/1538-4357/ac9ea3} {\bibfield  {journal} {\bibinfo  {journal} {The Astrophysical Journal}\ }\textbf {\bibinfo {volume} {941}},\ \bibinfo {pages} {110} (\bibinfo {year} {2022})}\BibitemShut {NoStop}%
\bibitem [{\citenamefont {Ade}\ \emph {et~al.}(2022)\citenamefont {Ade}, \citenamefont {Amiri}, \citenamefont {Benton}, \citenamefont {Bergman}, \citenamefont {Bihary}, \citenamefont {Bock}, \citenamefont {Bond}, \citenamefont {Bonetti}, \citenamefont {Bryan}, \citenamefont {Chiang}, \citenamefont {Contaldi}, \citenamefont {Doré}, \citenamefont {Duivenvoorden}, \citenamefont {Eriksen}, \citenamefont {Farhang}, \citenamefont {Filippini}, \citenamefont {Fraisse}, \citenamefont {Freese}, \citenamefont {Galloway}, \citenamefont {Gambrel}, \citenamefont {Gandilo}, \citenamefont {Ganga}, \citenamefont {Gualtieri}, \citenamefont {Gudmundsson}, \citenamefont {Halpern}, \citenamefont {Hartley}, \citenamefont {Hasselfield}, \citenamefont {Hilton}, \citenamefont {Holmes}, \citenamefont {Hristov}, \citenamefont {Huang}, \citenamefont {Irwin}, \citenamefont {Jones}, \citenamefont {Karakci}, \citenamefont {Kuo}, \citenamefont {Kermish}, \citenamefont {Leung}, \citenamefont {Li}, \citenamefont {Mak}, \citenamefont {Mason},
  \citenamefont {Megerian}, \citenamefont {Moncelsi}, \citenamefont {Morford}, \citenamefont {Nagy}, \citenamefont {Netterfield}, \citenamefont {Nolta}, \citenamefont {O’Brient}, \citenamefont {Osherson}, \citenamefont {Padilla}, \citenamefont {Racine}, \citenamefont {Rahlin}, \citenamefont {Reintsema}, \citenamefont {Ruhl}, \citenamefont {Runyan}, \citenamefont {Ruud}, \citenamefont {Shariff}, \citenamefont {Shaw}, \citenamefont {Shiu}, \citenamefont {Soler}, \citenamefont {Song}, \citenamefont {Trangsrud}, \citenamefont {Tucker}, \citenamefont {Tucker}, \citenamefont {Turner}, \citenamefont {van~der List}, \citenamefont {Weber}, \citenamefont {Wehus}, \citenamefont {Wen}, \citenamefont {Wiebe}, \citenamefont {Young},\ and\ \citenamefont {Collaboration}}]{SPIDER2022}%
  \BibitemOpen
  \bibfield  {author} {\bibinfo {author} {\bibfnamefont {P.~A.~R.}\ \bibnamefont {Ade}}, \bibinfo {author} {\bibfnamefont {M.}~\bibnamefont {Amiri}}, \bibinfo {author} {\bibfnamefont {S.~J.}\ \bibnamefont {Benton}}, \bibinfo {author} {\bibfnamefont {A.~S.}\ \bibnamefont {Bergman}}, \bibinfo {author} {\bibfnamefont {R.}~\bibnamefont {Bihary}}, \bibinfo {author} {\bibfnamefont {J.~J.}\ \bibnamefont {Bock}}, \bibinfo {author} {\bibfnamefont {J.~R.}\ \bibnamefont {Bond}}, \bibinfo {author} {\bibfnamefont {J.~A.}\ \bibnamefont {Bonetti}}, \bibinfo {author} {\bibfnamefont {S.~A.}\ \bibnamefont {Bryan}}, \bibinfo {author} {\bibfnamefont {H.~C.}\ \bibnamefont {Chiang}}, \bibinfo {author} {\bibfnamefont {C.~R.}\ \bibnamefont {Contaldi}}, \bibinfo {author} {\bibfnamefont {O.}~\bibnamefont {Doré}}, \bibinfo {author} {\bibfnamefont {A.~J.}\ \bibnamefont {Duivenvoorden}}, \bibinfo {author} {\bibfnamefont {H.~K.}\ \bibnamefont {Eriksen}}, \bibinfo {author} {\bibfnamefont {M.}~\bibnamefont {Farhang}}, \bibinfo {author}
  {\bibfnamefont {J.~P.}\ \bibnamefont {Filippini}}, \bibinfo {author} {\bibfnamefont {A.~A.}\ \bibnamefont {Fraisse}}, \bibinfo {author} {\bibfnamefont {K.}~\bibnamefont {Freese}}, \bibinfo {author} {\bibfnamefont {M.}~\bibnamefont {Galloway}}, \bibinfo {author} {\bibfnamefont {A.~E.}\ \bibnamefont {Gambrel}}, \bibinfo {author} {\bibfnamefont {N.~N.}\ \bibnamefont {Gandilo}}, \bibinfo {author} {\bibfnamefont {K.}~\bibnamefont {Ganga}}, \bibinfo {author} {\bibfnamefont {R.}~\bibnamefont {Gualtieri}}, \bibinfo {author} {\bibfnamefont {J.~E.}\ \bibnamefont {Gudmundsson}}, \bibinfo {author} {\bibfnamefont {M.}~\bibnamefont {Halpern}}, \bibinfo {author} {\bibfnamefont {J.}~\bibnamefont {Hartley}}, \bibinfo {author} {\bibfnamefont {M.}~\bibnamefont {Hasselfield}}, \bibinfo {author} {\bibfnamefont {G.}~\bibnamefont {Hilton}}, \bibinfo {author} {\bibfnamefont {W.}~\bibnamefont {Holmes}}, \bibinfo {author} {\bibfnamefont {V.~V.}\ \bibnamefont {Hristov}}, \bibinfo {author} {\bibfnamefont {Z.}~\bibnamefont {Huang}},
  \bibinfo {author} {\bibfnamefont {K.~D.}\ \bibnamefont {Irwin}}, \bibinfo {author} {\bibfnamefont {W.~C.}\ \bibnamefont {Jones}}, \bibinfo {author} {\bibfnamefont {A.}~\bibnamefont {Karakci}}, \bibinfo {author} {\bibfnamefont {C.~L.}\ \bibnamefont {Kuo}}, \bibinfo {author} {\bibfnamefont {Z.~D.}\ \bibnamefont {Kermish}}, \bibinfo {author} {\bibfnamefont {J.~S.-Y.}\ \bibnamefont {Leung}}, \bibinfo {author} {\bibfnamefont {S.}~\bibnamefont {Li}}, \bibinfo {author} {\bibfnamefont {D.~S.~Y.}\ \bibnamefont {Mak}}, \bibinfo {author} {\bibfnamefont {P.~V.}\ \bibnamefont {Mason}}, \bibinfo {author} {\bibfnamefont {K.}~\bibnamefont {Megerian}}, \bibinfo {author} {\bibfnamefont {L.}~\bibnamefont {Moncelsi}}, \bibinfo {author} {\bibfnamefont {T.~A.}\ \bibnamefont {Morford}}, \bibinfo {author} {\bibfnamefont {J.~M.}\ \bibnamefont {Nagy}}, \bibinfo {author} {\bibfnamefont {C.~B.}\ \bibnamefont {Netterfield}}, \bibinfo {author} {\bibfnamefont {M.}~\bibnamefont {Nolta}}, \bibinfo {author} {\bibfnamefont {R.}~\bibnamefont
  {O’Brient}}, \bibinfo {author} {\bibfnamefont {B.}~\bibnamefont {Osherson}}, \bibinfo {author} {\bibfnamefont {I.~L.}\ \bibnamefont {Padilla}}, \bibinfo {author} {\bibfnamefont {B.}~\bibnamefont {Racine}}, \bibinfo {author} {\bibfnamefont {A.~S.}\ \bibnamefont {Rahlin}}, \bibinfo {author} {\bibfnamefont {C.}~\bibnamefont {Reintsema}}, \bibinfo {author} {\bibfnamefont {J.~E.}\ \bibnamefont {Ruhl}}, \bibinfo {author} {\bibfnamefont {M.~C.}\ \bibnamefont {Runyan}}, \bibinfo {author} {\bibfnamefont {T.~M.}\ \bibnamefont {Ruud}}, \bibinfo {author} {\bibfnamefont {J.~A.}\ \bibnamefont {Shariff}}, \bibinfo {author} {\bibfnamefont {E.~C.}\ \bibnamefont {Shaw}}, \bibinfo {author} {\bibfnamefont {C.}~\bibnamefont {Shiu}}, \bibinfo {author} {\bibfnamefont {J.~D.}\ \bibnamefont {Soler}}, \bibinfo {author} {\bibfnamefont {X.}~\bibnamefont {Song}}, \bibinfo {author} {\bibfnamefont {A.}~\bibnamefont {Trangsrud}}, \bibinfo {author} {\bibfnamefont {C.}~\bibnamefont {Tucker}}, \bibinfo {author} {\bibfnamefont {R.~S.}\
  \bibnamefont {Tucker}}, \bibinfo {author} {\bibfnamefont {A.~D.}\ \bibnamefont {Turner}}, \bibinfo {author} {\bibfnamefont {J.~F.}\ \bibnamefont {van~der List}}, \bibinfo {author} {\bibfnamefont {A.~C.}\ \bibnamefont {Weber}}, \bibinfo {author} {\bibfnamefont {I.~K.}\ \bibnamefont {Wehus}}, \bibinfo {author} {\bibfnamefont {S.}~\bibnamefont {Wen}}, \bibinfo {author} {\bibfnamefont {D.~V.}\ \bibnamefont {Wiebe}}, \bibinfo {author} {\bibfnamefont {E.~Y.}\ \bibnamefont {Young}},\ and\ \bibinfo {author} {\bibfnamefont {S.}~\bibnamefont {Collaboration}},\ }\bibfield  {title} {\bibinfo {title} {A constraint on primordial b-modes from the first flight of the spider balloon-borne telescope},\ }\href {https://doi.org/10.3847/1538-4357/ac20df} {\bibfield  {journal} {\bibinfo  {journal} {The Astrophysical Journal}\ }\textbf {\bibinfo {volume} {927}},\ \bibinfo {pages} {174} (\bibinfo {year} {2022})}\BibitemShut {NoStop}%
\bibitem [{\citenamefont {Arkani–Hamed}\ \emph {et~al.}(1998)\citenamefont {Arkani–Hamed}, \citenamefont {Dimopoulos},\ and\ \citenamefont {Dvali}}]{add}%
  \BibitemOpen
  \bibfield  {author} {\bibinfo {author} {\bibfnamefont {N.}~\bibnamefont {Arkani–Hamed}}, \bibinfo {author} {\bibfnamefont {S.}~\bibnamefont {Dimopoulos}},\ and\ \bibinfo {author} {\bibfnamefont {G.}~\bibnamefont {Dvali}},\ }\bibfield  {title} {\bibinfo {title} {The hierarchy problem and new dimensions at a millimeter},\ }\href {https://doi.org/https://doi.org/10.1016/S0370-2693(98)00466-3} {\bibfield  {journal} {\bibinfo  {journal} {Physics Letters B}\ }\textbf {\bibinfo {volume} {429}},\ \bibinfo {pages} {263} (\bibinfo {year} {1998})}\BibitemShut {NoStop}%
\bibitem [{\citenamefont {Randall}\ and\ \citenamefont {Sundrum}(1999{\natexlab{a}})}]{RS}%
  \BibitemOpen
  \bibfield  {author} {\bibinfo {author} {\bibfnamefont {L.}~\bibnamefont {Randall}}\ and\ \bibinfo {author} {\bibfnamefont {R.}~\bibnamefont {Sundrum}},\ }\bibfield  {title} {\bibinfo {title} {Large mass hierarchy from a small extra dimension},\ }\href {https://doi.org/10.1103/PhysRevLett.83.3370} {\bibfield  {journal} {\bibinfo  {journal} {Phys. Rev. Lett.}\ }\textbf {\bibinfo {volume} {83}},\ \bibinfo {pages} {3370} (\bibinfo {year} {1999}{\natexlab{a}})}\BibitemShut {NoStop}%
\bibitem [{\citenamefont {Randall}\ and\ \citenamefont {Sundrum}(1999{\natexlab{b}})}]{RS1}%
  \BibitemOpen
  \bibfield  {author} {\bibinfo {author} {\bibfnamefont {L.}~\bibnamefont {Randall}}\ and\ \bibinfo {author} {\bibfnamefont {R.}~\bibnamefont {Sundrum}},\ }\bibfield  {title} {\bibinfo {title} {An alternative to compactification},\ }\href {https://doi.org/10.1103/PhysRevLett.83.4690} {\bibfield  {journal} {\bibinfo  {journal} {Phys. Rev. Lett.}\ }\textbf {\bibinfo {volume} {83}},\ \bibinfo {pages} {4690} (\bibinfo {year} {1999}{\natexlab{b}})}\BibitemShut {NoStop}%
\bibitem [{\citenamefont {Dvali}\ \emph {et~al.}(2000)\citenamefont {Dvali}, \citenamefont {Gabadadze},\ and\ \citenamefont {Porrati}}]{dgp}%
  \BibitemOpen
  \bibfield  {author} {\bibinfo {author} {\bibfnamefont {G.}~\bibnamefont {Dvali}}, \bibinfo {author} {\bibfnamefont {G.}~\bibnamefont {Gabadadze}},\ and\ \bibinfo {author} {\bibfnamefont {M.}~\bibnamefont {Porrati}},\ }\bibfield  {title} {\bibinfo {title} {4d gravity on a brane in 5d minkowski space},\ }\href {https://doi.org/https://doi.org/10.1016/S0370-2693(00)00669-9} {\bibfield  {journal} {\bibinfo  {journal} {Physics Letters B}\ }\textbf {\bibinfo {volume} {485}},\ \bibinfo {pages} {208} (\bibinfo {year} {2000})}\BibitemShut {NoStop}%
\bibitem [{\citenamefont {Deruelle}\ \emph {et~al.}(2001)\citenamefont {Deruelle}, \citenamefont {Dole\ifmmode~\check{z}\else \v{z}\fi{}el},\ and\ \citenamefont {Katz}}]{Deruelle}%
  \BibitemOpen
  \bibfield  {author} {\bibinfo {author} {\bibfnamefont {N.}~\bibnamefont {Deruelle}}, \bibinfo {author} {\bibfnamefont {T.~c.~v.}\ \bibnamefont {Dole\ifmmode~\check{z}\else \v{z}\fi{}el}},\ and\ \bibinfo {author} {\bibfnamefont {J.}~\bibnamefont {Katz}},\ }\bibfield  {title} {\bibinfo {title} {Perturbations of brane worlds},\ }\href {https://doi.org/10.1103/PhysRevD.63.083513} {\bibfield  {journal} {\bibinfo  {journal} {Phys. Rev. D}\ }\textbf {\bibinfo {volume} {63}},\ \bibinfo {pages} {083513} (\bibinfo {year} {2001})}\BibitemShut {NoStop}%
\bibitem [{\citenamefont {Battye}\ \emph {et~al.}(2001)\citenamefont {Battye}, \citenamefont {Carter}, \citenamefont {Mennim},\ and\ \citenamefont {Uzan}}]{battyecarterPRD}%
  \BibitemOpen
  \bibfield  {author} {\bibinfo {author} {\bibfnamefont {R.~A.}\ \bibnamefont {Battye}}, \bibinfo {author} {\bibfnamefont {B.}~\bibnamefont {Carter}}, \bibinfo {author} {\bibfnamefont {A.}~\bibnamefont {Mennim}},\ and\ \bibinfo {author} {\bibfnamefont {J.-P.}\ \bibnamefont {Uzan}},\ }\bibfield  {title} {\bibinfo {title} {Einstein equations for an asymmetric brane-world},\ }\href {https://doi.org/10.1103/PhysRevD.64.124007} {\bibfield  {journal} {\bibinfo  {journal} {Phys. Rev. D}\ }\textbf {\bibinfo {volume} {64}},\ \bibinfo {pages} {124007} (\bibinfo {year} {2001})}\BibitemShut {NoStop}%
\bibitem [{\citenamefont {Battye}\ and\ \citenamefont {Carter}(2001{\natexlab{a}})}]{bayttecarterPLB}%
  \BibitemOpen
  \bibfield  {author} {\bibinfo {author} {\bibfnamefont {R.~A.}\ \bibnamefont {Battye}}\ and\ \bibinfo {author} {\bibfnamefont {B.}~\bibnamefont {Carter}},\ }\bibfield  {title} {\bibinfo {title} {Generic junction conditions in brane-world scenarios},\ }\href {https://doi.org/https://doi.org/10.1016/S0370-2693(01)00495-6} {\bibfield  {journal} {\bibinfo  {journal} {Physics Letters B}\ }\textbf {\bibinfo {volume} {509}},\ \bibinfo {pages} {331} (\bibinfo {year} {2001}{\natexlab{a}})}\BibitemShut {NoStop}%
\bibitem [{\citenamefont {Nash}(1956)}]{nash56}%
  \BibitemOpen
  \bibfield  {author} {\bibinfo {author} {\bibfnamefont {J.}~\bibnamefont {Nash}},\ }\bibfield  {title} {\bibinfo {title} {The imbedding problem for riemannian manifolds},\ }\href {http://www.jstor.org/stable/1969989} {\bibfield  {journal} {\bibinfo  {journal} {Annals of Mathematics}\ }\textbf {\bibinfo {volume} {63}},\ \bibinfo {pages} {20} (\bibinfo {year} {1956})}\BibitemShut {NoStop}%
\bibitem [{\citenamefont {Greene}(1970)}]{greene70}%
  \BibitemOpen
  \bibfield  {author} {\bibinfo {author} {\bibfnamefont {R.~E.}\ \bibnamefont {Greene}},\ }\bibfield  {title} {\bibinfo {title} {Isometric embeddings of riemannian and pseudo-riemannian manifolds},\ }\bibfield  {journal} {\bibinfo  {journal} {Memoirs of the American Mathematical Society}\ }\textbf {\bibinfo {volume} {63}},\ \href {https://doi.org/10.1090/memo/0097} {10.1090/memo/0097} (\bibinfo {year} {1970})\BibitemShut {NoStop}%
\bibitem [{\citenamefont {Maia}\ and\ \citenamefont {Monte}(2002)}]{MAIA20029}%
  \BibitemOpen
  \bibfield  {author} {\bibinfo {author} {\bibfnamefont {M.}~\bibnamefont {Maia}}\ and\ \bibinfo {author} {\bibfnamefont {E.~M.}\ \bibnamefont {Monte}},\ }\bibfield  {title} {\bibinfo {title} {Geometry of brane-worlds},\ }\href {https://doi.org/https://doi.org/10.1016/S0375-9601(02)00182-2} {\bibfield  {journal} {\bibinfo  {journal} {Physics Letters A}\ }\textbf {\bibinfo {volume} {297}},\ \bibinfo {pages} {9} (\bibinfo {year} {2002})}\BibitemShut {NoStop}%
\bibitem [{\citenamefont {Maia}\ \emph {et~al.}(2005)\citenamefont {Maia}, \citenamefont {Monte}, \citenamefont {Maia},\ and\ \citenamefont {Alcaniz}}]{GDE}%
  \BibitemOpen
  \bibfield  {author} {\bibinfo {author} {\bibfnamefont {M.~D.}\ \bibnamefont {Maia}}, \bibinfo {author} {\bibfnamefont {E.~M.}\ \bibnamefont {Monte}}, \bibinfo {author} {\bibfnamefont {J.~M.~F.}\ \bibnamefont {Maia}},\ and\ \bibinfo {author} {\bibfnamefont {J.~S.}\ \bibnamefont {Alcaniz}},\ }\bibfield  {title} {\bibinfo {title} {On the geometry of dark energy},\ }\href {https://doi.org/10.1088/0264-9381/22/9/010} {\bibfield  {journal} {\bibinfo  {journal} {Classical and Quantum Gravity}\ }\textbf {\bibinfo {volume} {22}},\ \bibinfo {pages} {1623} (\bibinfo {year} {2005})}\BibitemShut {NoStop}%
\bibitem [{\citenamefont {Maia}\ \emph {et~al.}(2007)\citenamefont {Maia}, \citenamefont {Silva},\ and\ \citenamefont {Fernandes}}]{Maia_2007}%
  \BibitemOpen
  \bibfield  {author} {\bibinfo {author} {\bibfnamefont {M.~D.}\ \bibnamefont {Maia}}, \bibinfo {author} {\bibfnamefont {N.}~\bibnamefont {Silva}},\ and\ \bibinfo {author} {\bibfnamefont {M.~C.~B.}\ \bibnamefont {Fernandes}},\ }\bibfield  {title} {\bibinfo {title} {Brane-world quantum gravity},\ }\href {https://doi.org/10.1088/1126-6708/2007/04/047} {\bibfield  {journal} {\bibinfo  {journal} {Journal of High Energy Physics}\ }\textbf {\bibinfo {volume} {2007}},\ \bibinfo {pages} {047} (\bibinfo {year} {2007})}\BibitemShut {NoStop}%
\bibitem [{\citenamefont {Maia}\ \emph {et~al.}(2011)\citenamefont {Maia}, \citenamefont {Capistrano}, \citenamefont {Alcaniz},\ and\ \citenamefont {Monte}}]{gde2}%
  \BibitemOpen
  \bibfield  {author} {\bibinfo {author} {\bibfnamefont {M.~D.}\ \bibnamefont {Maia}}, \bibinfo {author} {\bibfnamefont {A.~J.~S.}\ \bibnamefont {Capistrano}}, \bibinfo {author} {\bibfnamefont {J.~S.}\ \bibnamefont {Alcaniz}},\ and\ \bibinfo {author} {\bibfnamefont {E.~M.}\ \bibnamefont {Monte}},\ }\bibfield  {title} {\bibinfo {title} {{The Deformable Universe}},\ }\href {https://doi.org/10.1007/s10714-011-1192-8} {\bibfield  {journal} {\bibinfo  {journal} {Gen. Rel. Grav.}\ }\textbf {\bibinfo {volume} {43}},\ \bibinfo {pages} {2685} (\bibinfo {year} {2011})},\ \Eprint {https://arxiv.org/abs/1101.3951} {arXiv:1101.3951 [gr-qc]} \BibitemShut {NoStop}%
\bibitem [{\citenamefont {Heydari-Fard}\ and\ \citenamefont {Sepangi}(2007)}]{sepangi}%
  \BibitemOpen
  \bibfield  {author} {\bibinfo {author} {\bibfnamefont {M.}~\bibnamefont {Heydari-Fard}}\ and\ \bibinfo {author} {\bibfnamefont {H.}~\bibnamefont {Sepangi}},\ }\bibfield  {title} {\bibinfo {title} {Anisotropic brane gravity with a confining potential},\ }\href {https://doi.org/https://doi.org/10.1016/j.physletb.2007.04.008} {\bibfield  {journal} {\bibinfo  {journal} {Physics Letters B}\ }\textbf {\bibinfo {volume} {649}},\ \bibinfo {pages} {1} (\bibinfo {year} {2007})}\BibitemShut {NoStop}%
\bibitem [{\citenamefont {Jalalzadeh}\ \emph {et~al.}(2009)\citenamefont {Jalalzadeh}, \citenamefont {Mehrnia},\ and\ \citenamefont {Sepangi}}]{sepangi1}%
  \BibitemOpen
  \bibfield  {author} {\bibinfo {author} {\bibfnamefont {S.}~\bibnamefont {Jalalzadeh}}, \bibinfo {author} {\bibfnamefont {M.}~\bibnamefont {Mehrnia}},\ and\ \bibinfo {author} {\bibfnamefont {H.~R.}\ \bibnamefont {Sepangi}},\ }\bibfield  {title} {\bibinfo {title} {Classical tests in brane gravity},\ }\href {https://doi.org/10.1088/0264-9381/26/15/155007} {\bibfield  {journal} {\bibinfo  {journal} {Classical and Quantum Gravity}\ }\textbf {\bibinfo {volume} {26}},\ \bibinfo {pages} {155007} (\bibinfo {year} {2009})}\BibitemShut {NoStop}%
\bibitem [{\citenamefont {Ranjbar}\ \emph {et~al.}(2012)\citenamefont {Ranjbar}, \citenamefont {Sepangi},\ and\ \citenamefont {Shahidi}}]{sepangi2}%
  \BibitemOpen
  \bibfield  {author} {\bibinfo {author} {\bibfnamefont {A.}~\bibnamefont {Ranjbar}}, \bibinfo {author} {\bibfnamefont {H.~R.}\ \bibnamefont {Sepangi}},\ and\ \bibinfo {author} {\bibfnamefont {S.}~\bibnamefont {Shahidi}},\ }\bibfield  {title} {\bibinfo {title} {{Asymptotically Lifshitz Brane-World Black Holes}},\ }\href {https://doi.org/10.1016/j.aop.2012.08.002} {\bibfield  {journal} {\bibinfo  {journal} {Annals Phys.}\ }\textbf {\bibinfo {volume} {327}},\ \bibinfo {pages} {3170} (\bibinfo {year} {2012})},\ \Eprint {https://arxiv.org/abs/1108.4562} {arXiv:1108.4562 [hep-th]} \BibitemShut {NoStop}%
\bibitem [{\citenamefont {Capistrano}\ \emph {et~al.}(2021)\citenamefont {Capistrano}, \citenamefont {Seidel},\ and\ \citenamefont {Duarte}}]{capistrano2021}%
  \BibitemOpen
  \bibfield  {author} {\bibinfo {author} {\bibfnamefont {A.~J.}\ \bibnamefont {Capistrano}}, \bibinfo {author} {\bibfnamefont {P.~T.}\ \bibnamefont {Seidel}},\ and\ \bibinfo {author} {\bibfnamefont {H.~R.}\ \bibnamefont {Duarte}},\ }\bibfield  {title} {\bibinfo {title} {Subhorizon linear nash–greene perturbations with constraints on h(z) and the deceleration parameter q(z)},\ }\href {https://doi.org/https://doi.org/10.1016/j.dark.2020.100760} {\bibfield  {journal} {\bibinfo  {journal} {Physics of the Dark Universe}\ }\textbf {\bibinfo {volume} {31}},\ \bibinfo {pages} {100760} (\bibinfo {year} {2021})}\BibitemShut {NoStop}%
\bibitem [{\citenamefont {Capistrano}\ \emph {et~al.}(2022)\citenamefont {Capistrano}, \citenamefont {Cabral}, \citenamefont {Marão},\ and\ \citenamefont {Coimbra-Araújo}}]{capistrano2022}%
  \BibitemOpen
  \bibfield  {author} {\bibinfo {author} {\bibfnamefont {A.~J.~S.}\ \bibnamefont {Capistrano}}, \bibinfo {author} {\bibfnamefont {L.~A.}\ \bibnamefont {Cabral}}, \bibinfo {author} {\bibfnamefont {J.~A. P.~F.}\ \bibnamefont {Marão}},\ and\ \bibinfo {author} {\bibfnamefont {C.~H.}\ \bibnamefont {Coimbra-Araújo}},\ }\bibfield  {title} {\bibinfo {title} {{Linear Nash-Greene fluctuations on the evolution of S8 and H0 tensions}},\ }\href {https://doi.org/10.1140/epjc/s10052-022-10431-9} {\bibfield  {journal} {\bibinfo  {journal} {The European Physical Journal C}\ }\textbf {\bibinfo {volume} {82}},\ \bibinfo {pages} {1434} (\bibinfo {year} {2022})}\BibitemShut {NoStop}%
\bibitem [{\citenamefont {Capistrano}\ and\ \citenamefont {Cabral}(2023)}]{capistrano_inflation2023}%
  \BibitemOpen
  \bibfield  {author} {\bibinfo {author} {\bibfnamefont {A.~J.~S.}\ \bibnamefont {Capistrano}}\ and\ \bibinfo {author} {\bibfnamefont {L.~A.}\ \bibnamefont {Cabral}},\ }\bibfield  {title} {\bibinfo {title} {{Effective Potential for Quintessential Inflation Driven by Extrinsic Gravity }},\ }\href {https://doi.org/10.3390/universe9120497} {\bibfield  {journal} {\bibinfo  {journal} {Universe}\ }\textbf {\bibinfo {volume} {9}},\ \bibinfo {pages} {1} (\bibinfo {year} {2023})}\BibitemShut {NoStop}%
\bibitem [{\citenamefont {Israel}(1966)}]{israel}%
  \BibitemOpen
  \bibfield  {author} {\bibinfo {author} {\bibfnamefont {W.}~\bibnamefont {Israel}},\ }\bibfield  {title} {\bibinfo {title} {Singular hypersurfaces and thin shells in general relativity},\ }\href {https://doi.org/10.1007/BF02710419} {\bibfield  {journal} {\bibinfo  {journal} {Il Nuovo Cimento B (1965-1970)}\ }\textbf {\bibinfo {volume} {44}},\ \bibinfo {pages} {1} (\bibinfo {year} {1966})}\BibitemShut {NoStop}%
\bibitem [{\citenamefont {Battye}\ and\ \citenamefont {Carter}(2001{\natexlab{b}})}]{BATTYE2001331}%
  \BibitemOpen
  \bibfield  {author} {\bibinfo {author} {\bibfnamefont {R.~A.}\ \bibnamefont {Battye}}\ and\ \bibinfo {author} {\bibfnamefont {B.}~\bibnamefont {Carter}},\ }\bibfield  {title} {\bibinfo {title} {Generic junction conditions in brane-world scenarios},\ }\href {https://doi.org/https://doi.org/10.1016/S0370-2693(01)00495-6} {\bibfield  {journal} {\bibinfo  {journal} {Physics Letters B}\ }\textbf {\bibinfo {volume} {509}},\ \bibinfo {pages} {331} (\bibinfo {year} {2001}{\natexlab{b}})}\BibitemShut {NoStop}%
\bibitem [{\citenamefont {Tsujikawa}\ \emph {et~al.}(2004)\citenamefont {Tsujikawa}, \citenamefont {Sami},\ and\ \citenamefont {Maartens}}]{tsujikawa}%
  \BibitemOpen
  \bibfield  {author} {\bibinfo {author} {\bibfnamefont {S.}~\bibnamefont {Tsujikawa}}, \bibinfo {author} {\bibfnamefont {M.}~\bibnamefont {Sami}},\ and\ \bibinfo {author} {\bibfnamefont {R.}~\bibnamefont {Maartens}},\ }\bibfield  {title} {\bibinfo {title} {Observational constraints on braneworld inflation: The effect of a gauss-bonnet term},\ }\href {https://doi.org/10.1103/PhysRevD.70.063525} {\bibfield  {journal} {\bibinfo  {journal} {Phys. Rev. D}\ }\textbf {\bibinfo {volume} {70}},\ \bibinfo {pages} {063525} (\bibinfo {year} {2004})}\BibitemShut {NoStop}%
\bibitem [{\citenamefont {Carron}\ \emph {et~al.}(2022)\citenamefont {Carron}, \citenamefont {Mirmelstein},\ and\ \citenamefont {Lewis}}]{Carron_2022}%
  \BibitemOpen
  \bibfield  {author} {\bibinfo {author} {\bibfnamefont {J.}~\bibnamefont {Carron}}, \bibinfo {author} {\bibfnamefont {M.}~\bibnamefont {Mirmelstein}},\ and\ \bibinfo {author} {\bibfnamefont {A.}~\bibnamefont {Lewis}},\ }\bibfield  {title} {\bibinfo {title} {Cmb lensing from planck pr4 maps},\ }\href {https://doi.org/10.1088/1475-7516/2022/09/039} {\bibfield  {journal} {\bibinfo  {journal} {Journal of Cosmology and Astroparticle Physics}\ }\textbf {\bibinfo {volume} {2022}}\bibinfo  {number} { (09)},\ \bibinfo {pages} {039}}\BibitemShut {NoStop}%
\bibitem [{\citenamefont {Rosenberg}\ \emph {et~al.}(2022)\citenamefont {Rosenberg}, \citenamefont {Gratton},\ and\ \citenamefont {Efstathiou}}]{Rosenberg:2022sdy}%
  \BibitemOpen
\bibfield  {number} {  }\bibfield  {author} {\bibinfo {author} {\bibfnamefont {E.}~\bibnamefont {Rosenberg}}, \bibinfo {author} {\bibfnamefont {S.}~\bibnamefont {Gratton}},\ and\ \bibinfo {author} {\bibfnamefont {G.}~\bibnamefont {Efstathiou}},\ }\bibfield  {title} {\bibinfo {title} {Cmb power spectra and cosmological parameters from planck pr4 with camspec},\ }\href {https://doi.org/10.1093/mnras/stac2744} {\bibfield  {journal} {\bibinfo  {journal} {Mon. Not. Roy. Astron. Soc.}\ }\textbf {\bibinfo {volume} {517}},\ \bibinfo {pages} {4620} (\bibinfo {year} {2022})},\ \Eprint {https://arxiv.org/abs/2205.10869} {arXiv:2205.10869 [astro-ph.CO]} \BibitemShut {NoStop}%
\bibitem [{\citenamefont {Ade}\ \emph {et~al.}(2021{\natexlab{b}})\citenamefont {Ade} \emph {et~al.}}]{BICEPKeck}%
  \BibitemOpen
  \bibfield  {author} {\bibinfo {author} {\bibfnamefont {P.~A.~R.}\ \bibnamefont {Ade}} \emph {et~al.} (\bibinfo {collaboration} {BICEP/Keck Collaboration}),\ }\bibfield  {title} {\bibinfo {title} {Improved constraints on primordial gravitational waves using planck, wmap, and bicep/keck observations through the 2018 observing season},\ }\href {https://doi.org/10.1103/PhysRevLett.127.151301} {\bibfield  {journal} {\bibinfo  {journal} {Phys. Rev. Lett.}\ }\textbf {\bibinfo {volume} {127}},\ \bibinfo {pages} {151301} (\bibinfo {year} {2021}{\natexlab{b}})}\BibitemShut {NoStop}%
\bibitem [{\citenamefont {Beutler}\ \emph {et~al.}(2011)\citenamefont {Beutler}, \citenamefont {Blake}, \citenamefont {Colless}, \citenamefont {Jones}, \citenamefont {Staveley-Smith}, \citenamefont {Campbell}, \citenamefont {Parker}, \citenamefont {Saunders},\ and\ \citenamefont {Watson}}]{6dFGalaxy}%
  \BibitemOpen
  \bibfield  {author} {\bibinfo {author} {\bibfnamefont {F.}~\bibnamefont {Beutler}}, \bibinfo {author} {\bibfnamefont {C.}~\bibnamefont {Blake}}, \bibinfo {author} {\bibfnamefont {M.}~\bibnamefont {Colless}}, \bibinfo {author} {\bibfnamefont {D.~H.}\ \bibnamefont {Jones}}, \bibinfo {author} {\bibfnamefont {L.}~\bibnamefont {Staveley-Smith}}, \bibinfo {author} {\bibfnamefont {L.}~\bibnamefont {Campbell}}, \bibinfo {author} {\bibfnamefont {Q.}~\bibnamefont {Parker}}, \bibinfo {author} {\bibfnamefont {W.}~\bibnamefont {Saunders}},\ and\ \bibinfo {author} {\bibfnamefont {F.}~\bibnamefont {Watson}},\ }\bibfield  {title} {\bibinfo {title} {{The 6dF Galaxy Survey: baryon acoustic oscillations and the local Hubble constant}},\ }\href {https://doi.org/10.1111/j.1365-2966.2011.19250.x} {\bibfield  {journal} {\bibinfo  {journal} {Monthly Notices of the Royal Astronomical Society}\ }\textbf {\bibinfo {volume} {416}},\ \bibinfo {pages} {3017} (\bibinfo {year} {2011})},\ \Eprint
  {https://arxiv.org/abs/https://academic.oup.com/mnras/article-pdf/416/4/3017/2985042/mnras0416-3017.pdf} {https://academic.oup.com/mnras/article-pdf/416/4/3017/2985042/mnras0416-3017.pdf} \BibitemShut {NoStop}%
\bibitem [{\citenamefont {Ross}\ \emph {et~al.}(2015)\citenamefont {Ross}, \citenamefont {Samushia}, \citenamefont {Howlett}, \citenamefont {Percival}, \citenamefont {Burden},\ and\ \citenamefont {Manera}}]{Ross:2014qpa}%
  \BibitemOpen
  \bibfield  {author} {\bibinfo {author} {\bibfnamefont {A.~J.}\ \bibnamefont {Ross}}, \bibinfo {author} {\bibfnamefont {L.}~\bibnamefont {Samushia}}, \bibinfo {author} {\bibfnamefont {C.}~\bibnamefont {Howlett}}, \bibinfo {author} {\bibfnamefont {W.~J.}\ \bibnamefont {Percival}}, \bibinfo {author} {\bibfnamefont {A.}~\bibnamefont {Burden}},\ and\ \bibinfo {author} {\bibfnamefont {M.}~\bibnamefont {Manera}},\ }\bibfield  {title} {\bibinfo {title} {{The clustering of the SDSS DR7 main Galaxy sample \textendash{} I. A 4 per cent distance measure at $z = 0.15$}},\ }\href {https://doi.org/10.1093/mnras/stv154} {\bibfield  {journal} {\bibinfo  {journal} {Mon. Not. Roy. Astron. Soc.}\ }\textbf {\bibinfo {volume} {449}},\ \bibinfo {pages} {835} (\bibinfo {year} {2015})},\ \Eprint {https://arxiv.org/abs/1409.3242} {arXiv:1409.3242 [astro-ph.CO]} \BibitemShut {NoStop}%
\bibitem [{\citenamefont {Alam}\ \emph {et~al.}(2017)\citenamefont {Alam} \emph {et~al.}}]{Alam:2016hwk}%
  \BibitemOpen
  \bibfield  {author} {\bibinfo {author} {\bibfnamefont {S.}~\bibnamefont {Alam}} \emph {et~al.} (\bibinfo {collaboration} {BOSS}),\ }\bibfield  {title} {\bibinfo {title} {{The clustering of galaxies in the completed SDSS-III Baryon Oscillation Spectroscopic Survey: cosmological analysis of the DR12 galaxy sample}},\ }\href {https://doi.org/10.1093/mnras/stx721} {\bibfield  {journal} {\bibinfo  {journal} {Mon. Not. Roy. Astron. Soc.}\ }\textbf {\bibinfo {volume} {470}},\ \bibinfo {pages} {2617} (\bibinfo {year} {2017})},\ \Eprint {https://arxiv.org/abs/1607.03155} {arXiv:1607.03155 [astro-ph.CO]} \BibitemShut {NoStop}%
\bibitem [{\citenamefont {Wang}\ \emph {et~al.}(2023)\citenamefont {Wang}, \citenamefont {Mirpoorian}, \citenamefont {Pogosian}, \citenamefont {Silvestri},\ and\ \citenamefont {Zhao}}]{mgcamb2023}%
  \BibitemOpen
  \bibfield  {author} {\bibinfo {author} {\bibfnamefont {Z.}~\bibnamefont {Wang}}, \bibinfo {author} {\bibfnamefont {S.~H.}\ \bibnamefont {Mirpoorian}}, \bibinfo {author} {\bibfnamefont {L.}~\bibnamefont {Pogosian}}, \bibinfo {author} {\bibfnamefont {A.}~\bibnamefont {Silvestri}},\ and\ \bibinfo {author} {\bibfnamefont {G.-B.}\ \bibnamefont {Zhao}},\ }\bibfield  {title} {\bibinfo {title} {New mgcamb tests of gravity with cosmomc and cobaya},\ }\href {https://doi.org/10.1088/1475-7516/2023/08/038} {\bibfield  {journal} {\bibinfo  {journal} {Journal of Cosmology and Astroparticle Physics}\ }\textbf {\bibinfo {volume} {2023}}\bibinfo  {number} { (08)},\ \bibinfo {pages} {038}}\BibitemShut {NoStop}%
\bibitem [{\citenamefont {Torrado}\ and\ \citenamefont {Lewis}(2021)}]{cobaya}%
  \BibitemOpen
\bibfield  {number} {  }\bibfield  {author} {\bibinfo {author} {\bibfnamefont {J.}~\bibnamefont {Torrado}}\ and\ \bibinfo {author} {\bibfnamefont {A.}~\bibnamefont {Lewis}},\ }\bibfield  {title} {\bibinfo {title} {Cobaya: code for bayesian analysis of hierarchical physical models},\ }\href {https://doi.org/10.1088/1475-7516/2021/05/057} {\bibfield  {journal} {\bibinfo  {journal} {Journal of Cosmology and Astroparticle Physics}\ }\textbf {\bibinfo {volume} {2021}}\bibinfo  {number} { (05)},\ \bibinfo {pages} {057}}\BibitemShut {NoStop}%
\bibitem [{\citenamefont {Eisenhart}(2005)}]{eisen}%
  \BibitemOpen
\bibfield  {number} {  }\bibfield  {author} {\bibinfo {author} {\bibfnamefont {L.~P.}\ \bibnamefont {Eisenhart}},\ }\href@noop {} {\emph {\bibinfo {title} {On Riemannian Geometry}}}\ (\bibinfo  {publisher} {Dover Publications},\ \bibinfo {address} {New {Y}ork},\ \bibinfo {year} {2005})\BibitemShut {NoStop}%
\bibitem [{\citenamefont {Lim}(2014)}]{Lim2014}%
  \BibitemOpen
  \bibfield  {author} {\bibinfo {author} {\bibfnamefont {C.~S.}\ \bibnamefont {Lim}},\ }\bibfield  {title} {\bibinfo {title} {{The Higgs particle and higher-dimensional theories}},\ }\href {https://doi.org/10.1093/ptep/ptt083} {\bibfield  {journal} {\bibinfo  {journal} {Progress of Theoretical and Experimental Physics}\ }\textbf {\bibinfo {volume} {2014}},\ \bibinfo {pages} {02A101} (\bibinfo {year} {2014})},\ \Eprint {https://arxiv.org/abs/https://academic.oup.com/ptep/article-pdf/2014/2/02A101/9719228/ptt083.pdf} {https://academic.oup.com/ptep/article-pdf/2014/2/02A101/9719228/ptt083.pdf} \BibitemShut {NoStop}%
\bibitem [{\citenamefont {Mieling}\ \emph {et~al.}(2022)\citenamefont {Mieling}, \citenamefont {Hilweg},\ and\ \citenamefont {Walther}}]{Mieling2022}%
  \BibitemOpen
  \bibfield  {author} {\bibinfo {author} {\bibfnamefont {T.~B.}\ \bibnamefont {Mieling}}, \bibinfo {author} {\bibfnamefont {C.}~\bibnamefont {Hilweg}},\ and\ \bibinfo {author} {\bibfnamefont {P.}~\bibnamefont {Walther}},\ }\bibfield  {title} {\bibinfo {title} {Measuring space-time curvature using maximally path-entangled quantum states},\ }\href {https://doi.org/10.1103/PhysRevA.106.L031701} {\bibfield  {journal} {\bibinfo  {journal} {Phys. Rev. A}\ }\textbf {\bibinfo {volume} {106}},\ \bibinfo {pages} {L031701} (\bibinfo {year} {2022})}\BibitemShut {NoStop}%
\bibitem [{\citenamefont {Feng}\ \emph {et~al.}(2024)\citenamefont {Feng}, \citenamefont {Gu},\ and\ \citenamefont {Shu}}]{Feng2024}%
  \BibitemOpen
  \bibfield  {author} {\bibinfo {author} {\bibfnamefont {S.}~\bibnamefont {Feng}}, \bibinfo {author} {\bibfnamefont {B.-M.}\ \bibnamefont {Gu}},\ and\ \bibinfo {author} {\bibfnamefont {F.-W.}\ \bibnamefont {Shu}},\ }\bibfield  {title} {\bibinfo {title} {Quantum gravity induced entanglement of masses with extra dimensions},\ }\href {https://doi.org/10.1140/epjc/s10052-024-12413-5} {\bibfield  {journal} {\bibinfo  {journal} {The European Physical Journal C}\ }\textbf {\bibinfo {volume} {84}},\ \bibinfo {pages} {1} (\bibinfo {year} {2024})}\BibitemShut {NoStop}%
\bibitem [{\citenamefont {Gupta}(1954)}]{Gupta}%
  \BibitemOpen
  \bibfield  {author} {\bibinfo {author} {\bibfnamefont {S.~N.}\ \bibnamefont {Gupta}},\ }\bibfield  {title} {\bibinfo {title} {Gravitation and electromagnetism},\ }\href {https://doi.org/10.1103/PhysRev.96.1683} {\bibfield  {journal} {\bibinfo  {journal} {Phys. Rev.}\ }\textbf {\bibinfo {volume} {96}},\ \bibinfo {pages} {1683} (\bibinfo {year} {1954})}\BibitemShut {NoStop}%
\bibitem [{\citenamefont {Isham}\ \emph {et~al.}(1971)\citenamefont {Isham}, \citenamefont {Salam},\ and\ \citenamefont {Strathdee}}]{salam}%
  \BibitemOpen
  \bibfield  {author} {\bibinfo {author} {\bibfnamefont {C.~J.}\ \bibnamefont {Isham}}, \bibinfo {author} {\bibfnamefont {A.}~\bibnamefont {Salam}},\ and\ \bibinfo {author} {\bibfnamefont {J.}~\bibnamefont {Strathdee}},\ }\bibfield  {title} {\bibinfo {title} {$f$-dominance of gravity},\ }\href {https://doi.org/10.1103/PhysRevD.3.867} {\bibfield  {journal} {\bibinfo  {journal} {Phys. Rev. D}\ }\textbf {\bibinfo {volume} {3}},\ \bibinfo {pages} {867} (\bibinfo {year} {1971})}\BibitemShut {NoStop}%
\bibitem [{\citenamefont {Sapone}\ and\ \citenamefont {Kunz}(2009)}]{sapone2009}%
  \BibitemOpen
  \bibfield  {author} {\bibinfo {author} {\bibfnamefont {D.}~\bibnamefont {Sapone}}\ and\ \bibinfo {author} {\bibfnamefont {M.}~\bibnamefont {Kunz}},\ }\bibfield  {title} {\bibinfo {title} {Fingerprinting dark energy},\ }\href {https://doi.org/10.1103/PhysRevD.80.083519} {\bibfield  {journal} {\bibinfo  {journal} {Phys. Rev. D}\ }\textbf {\bibinfo {volume} {80}},\ \bibinfo {pages} {083519} (\bibinfo {year} {2009})}\BibitemShut {NoStop}%
\bibitem [{\citenamefont {Arjona}\ \emph {et~al.}(2019)\citenamefont {Arjona}, \citenamefont {Cardona},\ and\ \citenamefont {Nesseris}}]{nesseris2019}%
  \BibitemOpen
  \bibfield  {author} {\bibinfo {author} {\bibfnamefont {R.}~\bibnamefont {Arjona}}, \bibinfo {author} {\bibfnamefont {W.}~\bibnamefont {Cardona}},\ and\ \bibinfo {author} {\bibfnamefont {S.}~\bibnamefont {Nesseris}},\ }\bibfield  {title} {\bibinfo {title} {Unraveling the effective fluid approach for $f(r)$ models in the subhorizon approximation},\ }\href {https://doi.org/10.1103/PhysRevD.99.043516} {\bibfield  {journal} {\bibinfo  {journal} {Phys. Rev. D}\ }\textbf {\bibinfo {volume} {99}},\ \bibinfo {pages} {043516} (\bibinfo {year} {2019})}\BibitemShut {NoStop}%
\bibitem [{\citenamefont {Abbott}\ and\ \citenamefont {Wise}(1984)}]{Abbott1984}%
  \BibitemOpen
  \bibfield  {author} {\bibinfo {author} {\bibfnamefont {L.~F.}\ \bibnamefont {Abbott}}\ and\ \bibinfo {author} {\bibfnamefont {M.~B.}\ \bibnamefont {Wise}},\ }\bibfield  {title} {\bibinfo {title} {{Constraints on Generalized Inflationary Cosmologies}},\ }\href {https://doi.org/10.1016/0550-3213(84)90329-8} {\bibfield  {journal} {\bibinfo  {journal} {Nucl. Phys. B}\ }\textbf {\bibinfo {volume} {244}},\ \bibinfo {pages} {541} (\bibinfo {year} {1984})}\BibitemShut {NoStop}%
\bibitem [{\citenamefont {Lucchin}\ and\ \citenamefont {Matarrese}(1985)}]{Lucchin1984}%
  \BibitemOpen
  \bibfield  {author} {\bibinfo {author} {\bibfnamefont {F.}~\bibnamefont {Lucchin}}\ and\ \bibinfo {author} {\bibfnamefont {S.}~\bibnamefont {Matarrese}},\ }\bibfield  {title} {\bibinfo {title} {{Power Law Inflation}},\ }\href {https://doi.org/10.1103/PhysRevD.32.1316} {\bibfield  {journal} {\bibinfo  {journal} {Phys. Rev. D}\ }\textbf {\bibinfo {volume} {32}},\ \bibinfo {pages} {1316} (\bibinfo {year} {1985})}\BibitemShut {NoStop}%
\bibitem [{\citenamefont {Davies}\ and\ \citenamefont {Sahni}(1988)}]{sahni1988}%
  \BibitemOpen
  \bibfield  {author} {\bibinfo {author} {\bibfnamefont {P.~C.~W.}\ \bibnamefont {Davies}}\ and\ \bibinfo {author} {\bibfnamefont {V.}~\bibnamefont {Sahni}},\ }\bibfield  {title} {\bibinfo {title} {Quantum gravitational effects near cosmic strings},\ }\href {https://doi.org/10.1088/0264-9381/5/1/009} {\bibfield  {journal} {\bibinfo  {journal} {Classical and Quantum Gravity}\ }\textbf {\bibinfo {volume} {5}},\ \bibinfo {pages} {1} (\bibinfo {year} {1988})}\BibitemShut {NoStop}%
\bibitem [{\citenamefont {Martin}\ \emph {et~al.}(2014)\citenamefont {Martin}, \citenamefont {Ringeval},\ and\ \citenamefont {Vennin}}]{MARTIN201475}%
  \BibitemOpen
  \bibfield  {author} {\bibinfo {author} {\bibfnamefont {J.}~\bibnamefont {Martin}}, \bibinfo {author} {\bibfnamefont {C.}~\bibnamefont {Ringeval}},\ and\ \bibinfo {author} {\bibfnamefont {V.}~\bibnamefont {Vennin}},\ }\bibfield  {title} {\bibinfo {title} {Encyclopaedia inflationaris},\ }\href {https://doi.org/https://doi.org/10.1016/j.dark.2014.01.003} {\bibfield  {journal} {\bibinfo  {journal} {Physics of the Dark Universe}\ }\textbf {\bibinfo {volume} {5-6}},\ \bibinfo {pages} {75} (\bibinfo {year} {2014})},\ \bibinfo {note} {hunt for Dark Matter}\BibitemShut {NoStop}%
\bibitem [{\citenamefont {Becker}\ \emph {et~al.}(2005)\citenamefont {Becker}, \citenamefont {Becker},\ and\ \citenamefont {Krause}}]{Becker2005}%
  \BibitemOpen
  \bibfield  {author} {\bibinfo {author} {\bibfnamefont {K.}~\bibnamefont {Becker}}, \bibinfo {author} {\bibfnamefont {M.}~\bibnamefont {Becker}},\ and\ \bibinfo {author} {\bibfnamefont {A.}~\bibnamefont {Krause}},\ }\bibfield  {title} {\bibinfo {title} {{M-theory inflation from multi M5-brane dynamics}},\ }\href {https://doi.org/10.1016/j.nuclphysb.2005.03.011} {\bibfield  {journal} {\bibinfo  {journal} {Nucl. Phys. B}\ }\textbf {\bibinfo {volume} {715}},\ \bibinfo {pages} {349} (\bibinfo {year} {2005})},\ \Eprint {https://arxiv.org/abs/hep-th/0501130} {arXiv:hep-th/0501130} \BibitemShut {NoStop}%
\bibitem [{\citenamefont {Bennai}\ \emph {et~al.}(2006)\citenamefont {Bennai}, \citenamefont {Chakir},\ and\ \citenamefont {Sakhi}}]{Bennai2006}%
  \BibitemOpen
  \bibfield  {author} {\bibinfo {author} {\bibfnamefont {M.}~\bibnamefont {Bennai}}, \bibinfo {author} {\bibfnamefont {H.}~\bibnamefont {Chakir}},\ and\ \bibinfo {author} {\bibfnamefont {Z.}~\bibnamefont {Sakhi}},\ }\bibfield  {title} {\bibinfo {title} {{On Inflation Potentials in Randall-Sundrum Braneworld Model}},\ }\href@noop {} {\bibfield  {journal} {\bibinfo  {journal} {Eur. J. Phys.}\ }\textbf {\bibinfo {volume} {9}},\ \bibinfo {pages} {84} (\bibinfo {year} {2006})},\ \Eprint {https://arxiv.org/abs/0806.1137} {arXiv:0806.1137 [gr-qc]} \BibitemShut {NoStop}%
\bibitem [{\citenamefont {{Tristram, M.}}\ \emph {et~al.}(2021)\citenamefont {{Tristram, M.}} \emph {et~al.}}]{planckinflation}%
  \BibitemOpen
  \bibfield  {author} {\bibinfo {author} {\bibnamefont {{Tristram, M.}}} \emph {et~al.},\ }\bibfield  {title} {\bibinfo {title} {Planck constraints on the tensor-to-scalar ratio},\ }\href {https://doi.org/10.1051/0004-6361/202039585} {\bibfield  {journal} {\bibinfo  {journal} {A\&A}\ }\textbf {\bibinfo {volume} {647}},\ \bibinfo {pages} {A128} (\bibinfo {year} {2021})}\BibitemShut {NoStop}%
\bibitem [{\citenamefont {Alcaniz}\ and\ \citenamefont {Carvalho}(2007)}]{Alcaniz_2007}%
  \BibitemOpen
  \bibfield  {author} {\bibinfo {author} {\bibfnamefont {J.~S.}\ \bibnamefont {Alcaniz}}\ and\ \bibinfo {author} {\bibfnamefont {F.~C.}\ \bibnamefont {Carvalho}},\ }\bibfield  {title} {\bibinfo {title} {$\beta$-exponential inflation},\ }\href {https://doi.org/10.1209/0295-5075/79/39001} {\bibfield  {journal} {\bibinfo  {journal} {Europhysics Letters}\ }\textbf {\bibinfo {volume} {79}},\ \bibinfo {pages} {39001} (\bibinfo {year} {2007})}\BibitemShut {NoStop}%
\bibitem [{\citenamefont {Santos}\ \emph {et~al.}(2018)\citenamefont {Santos}, \citenamefont {Benetti}, \citenamefont {Alcaniz}, \citenamefont {Brito},\ and\ \citenamefont {Silva}}]{Santos_2018}%
  \BibitemOpen
  \bibfield  {author} {\bibinfo {author} {\bibfnamefont {M.}~\bibnamefont {Santos}}, \bibinfo {author} {\bibfnamefont {M.}~\bibnamefont {Benetti}}, \bibinfo {author} {\bibfnamefont {J.}~\bibnamefont {Alcaniz}}, \bibinfo {author} {\bibfnamefont {F.}~\bibnamefont {Brito}},\ and\ \bibinfo {author} {\bibfnamefont {R.}~\bibnamefont {Silva}},\ }\bibfield  {title} {\bibinfo {title} {Cmb constraints on $\beta$-exponential inflationary models},\ }\href {https://doi.org/10.1088/1475-7516/2018/03/023} {\bibfield  {journal} {\bibinfo  {journal} {Journal of Cosmology and Astroparticle Physics}\ }\textbf {\bibinfo {volume} {2018}}\bibinfo  {number} { (03)},\ \bibinfo {pages} {023}}\BibitemShut {NoStop}%
\bibitem [{\citenamefont {Alam}\ \emph {et~al.}(2021)\citenamefont {Alam}, \citenamefont {Aubert}, \citenamefont {Avila}, \citenamefont {Balland}, \citenamefont {Bautista}, \citenamefont {Bershady}, \citenamefont {Bizyaev}, \citenamefont {Blanton}, \citenamefont {Bolton}, \citenamefont {Bovy}, \citenamefont {Brinkmann}, \citenamefont {Brownstein}, \citenamefont {Burtin}, \citenamefont {Chabanier}, \citenamefont {Chapman}, \citenamefont {Choi}, \citenamefont {Chuang}, \citenamefont {Comparat}, \citenamefont {Cousinou}, \citenamefont {Cuceu}, \citenamefont {Dawson}, \citenamefont {de~la Torre}, \citenamefont {de~Mattia}, \citenamefont {Agathe}, \citenamefont {des Bourboux}, \citenamefont {Escoffier}, \citenamefont {Etourneau}, \citenamefont {Farr}, \citenamefont {Font-Ribera}, \citenamefont {Frinchaboy}, \citenamefont {Fromenteau}, \citenamefont {Gil-Mar\'{\i}n}, \citenamefont {Le~Goff}, \citenamefont {Gonzalez-Morales}, \citenamefont {Gonzalez-Perez}, \citenamefont {Grabowski}, \citenamefont {Guy},
  \citenamefont {Hawken}, \citenamefont {Hou}, \citenamefont {Kong}, \citenamefont {Parker}, \citenamefont {Klaene}, \citenamefont {Kneib}, \citenamefont {Lin}, \citenamefont {Long}, \citenamefont {Lyke}, \citenamefont {de~la Macorra}, \citenamefont {Martini}, \citenamefont {Masters}, \citenamefont {Mohammad}, \citenamefont {Moon}, \citenamefont {Mueller}, \citenamefont {Mu\~noz Guti\'errez}, \citenamefont {Myers}, \citenamefont {Nadathur}, \citenamefont {Neveux}, \citenamefont {Newman}, \citenamefont {Noterdaeme}, \citenamefont {Oravetz}, \citenamefont {Oravetz}, \citenamefont {Palanque-Delabrouille}, \citenamefont {Pan}, \citenamefont {Paviot}, \citenamefont {Percival}, \citenamefont {P\'erez-R\`afols}, \citenamefont {Petitjean}, \citenamefont {Pieri}, \citenamefont {Prakash}, \citenamefont {Raichoor}, \citenamefont {Ravoux}, \citenamefont {Rezaie}, \citenamefont {Rich}, \citenamefont {Ross}, \citenamefont {Rossi}, \citenamefont {Ruggeri}, \citenamefont {Ruhlmann-Kleider}, \citenamefont {S\'anchez},
  \citenamefont {S\'anchez}, \citenamefont {S\'anchez-Gallego}, \citenamefont {Sayres}, \citenamefont {Schneider}, \citenamefont {Seo}, \citenamefont {Shafieloo}, \citenamefont {Slosar}, \citenamefont {Smith}, \citenamefont {Stermer}, \citenamefont {Tamone}, \citenamefont {Tinker}, \citenamefont {Tojeiro}, \citenamefont {Vargas-Maga\~na}, \citenamefont {Variu}, \citenamefont {Wang}, \citenamefont {Weaver}, \citenamefont {Weijmans}, \citenamefont {Y\`eche}, \citenamefont {Zarrouk}, \citenamefont {Zhao}, \citenamefont {Zhao},\ and\ \citenamefont {Zheng}}]{Alam:2020sor}%
  \BibitemOpen
\bibfield  {number} {  }\bibfield  {author} {\bibinfo {author} {\bibfnamefont {S.}~\bibnamefont {Alam}}, \bibinfo {author} {\bibfnamefont {M.}~\bibnamefont {Aubert}}, \bibinfo {author} {\bibfnamefont {S.}~\bibnamefont {Avila}}, \bibinfo {author} {\bibfnamefont {C.}~\bibnamefont {Balland}}, \bibinfo {author} {\bibfnamefont {J.~E.}\ \bibnamefont {Bautista}}, \bibinfo {author} {\bibfnamefont {M.~A.}\ \bibnamefont {Bershady}}, \bibinfo {author} {\bibfnamefont {D.}~\bibnamefont {Bizyaev}}, \bibinfo {author} {\bibfnamefont {M.~R.}\ \bibnamefont {Blanton}}, \bibinfo {author} {\bibfnamefont {A.~S.}\ \bibnamefont {Bolton}}, \bibinfo {author} {\bibfnamefont {J.}~\bibnamefont {Bovy}}, \bibinfo {author} {\bibfnamefont {J.}~\bibnamefont {Brinkmann}}, \bibinfo {author} {\bibfnamefont {J.~R.}\ \bibnamefont {Brownstein}}, \bibinfo {author} {\bibfnamefont {E.}~\bibnamefont {Burtin}}, \bibinfo {author} {\bibfnamefont {S.}~\bibnamefont {Chabanier}}, \bibinfo {author} {\bibfnamefont {M.~J.}\ \bibnamefont {Chapman}}, \bibinfo
  {author} {\bibfnamefont {P.~D.}\ \bibnamefont {Choi}}, \bibinfo {author} {\bibfnamefont {C.-H.}\ \bibnamefont {Chuang}}, \bibinfo {author} {\bibfnamefont {J.}~\bibnamefont {Comparat}}, \bibinfo {author} {\bibfnamefont {M.-C.}\ \bibnamefont {Cousinou}}, \bibinfo {author} {\bibfnamefont {A.}~\bibnamefont {Cuceu}}, \bibinfo {author} {\bibfnamefont {K.~S.}\ \bibnamefont {Dawson}}, \bibinfo {author} {\bibfnamefont {S.}~\bibnamefont {de~la Torre}}, \bibinfo {author} {\bibfnamefont {A.}~\bibnamefont {de~Mattia}}, \bibinfo {author} {\bibfnamefont {V.~d.~S.}\ \bibnamefont {Agathe}}, \bibinfo {author} {\bibfnamefont {H.~d.~M.}\ \bibnamefont {des Bourboux}}, \bibinfo {author} {\bibfnamefont {S.}~\bibnamefont {Escoffier}}, \bibinfo {author} {\bibfnamefont {T.}~\bibnamefont {Etourneau}}, \bibinfo {author} {\bibfnamefont {J.}~\bibnamefont {Farr}}, \bibinfo {author} {\bibfnamefont {A.}~\bibnamefont {Font-Ribera}}, \bibinfo {author} {\bibfnamefont {P.~M.}\ \bibnamefont {Frinchaboy}}, \bibinfo {author} {\bibfnamefont
  {S.}~\bibnamefont {Fromenteau}}, \bibinfo {author} {\bibfnamefont {H.}~\bibnamefont {Gil-Mar\'{\i}n}}, \bibinfo {author} {\bibfnamefont {J.-M.}\ \bibnamefont {Le~Goff}}, \bibinfo {author} {\bibfnamefont {A.~X.}\ \bibnamefont {Gonzalez-Morales}}, \bibinfo {author} {\bibfnamefont {V.}~\bibnamefont {Gonzalez-Perez}}, \bibinfo {author} {\bibfnamefont {K.}~\bibnamefont {Grabowski}}, \bibinfo {author} {\bibfnamefont {J.}~\bibnamefont {Guy}}, \bibinfo {author} {\bibfnamefont {A.~J.}\ \bibnamefont {Hawken}}, \bibinfo {author} {\bibfnamefont {J.}~\bibnamefont {Hou}}, \bibinfo {author} {\bibfnamefont {H.}~\bibnamefont {Kong}}, \bibinfo {author} {\bibfnamefont {J.}~\bibnamefont {Parker}}, \bibinfo {author} {\bibfnamefont {M.}~\bibnamefont {Klaene}}, \bibinfo {author} {\bibfnamefont {J.-P.}\ \bibnamefont {Kneib}}, \bibinfo {author} {\bibfnamefont {S.}~\bibnamefont {Lin}}, \bibinfo {author} {\bibfnamefont {D.}~\bibnamefont {Long}}, \bibinfo {author} {\bibfnamefont {B.~W.}\ \bibnamefont {Lyke}}, \bibinfo {author}
  {\bibfnamefont {A.}~\bibnamefont {de~la Macorra}}, \bibinfo {author} {\bibfnamefont {P.}~\bibnamefont {Martini}}, \bibinfo {author} {\bibfnamefont {K.}~\bibnamefont {Masters}}, \bibinfo {author} {\bibfnamefont {F.~G.}\ \bibnamefont {Mohammad}}, \bibinfo {author} {\bibfnamefont {J.}~\bibnamefont {Moon}}, \bibinfo {author} {\bibfnamefont {E.-M.}\ \bibnamefont {Mueller}}, \bibinfo {author} {\bibfnamefont {A.}~\bibnamefont {Mu\~noz Guti\'errez}}, \bibinfo {author} {\bibfnamefont {A.~D.}\ \bibnamefont {Myers}}, \bibinfo {author} {\bibfnamefont {S.}~\bibnamefont {Nadathur}}, \bibinfo {author} {\bibfnamefont {R.}~\bibnamefont {Neveux}}, \bibinfo {author} {\bibfnamefont {J.~A.}\ \bibnamefont {Newman}}, \bibinfo {author} {\bibfnamefont {P.}~\bibnamefont {Noterdaeme}}, \bibinfo {author} {\bibfnamefont {A.}~\bibnamefont {Oravetz}}, \bibinfo {author} {\bibfnamefont {D.}~\bibnamefont {Oravetz}}, \bibinfo {author} {\bibfnamefont {N.}~\bibnamefont {Palanque-Delabrouille}}, \bibinfo {author} {\bibfnamefont
  {K.}~\bibnamefont {Pan}}, \bibinfo {author} {\bibfnamefont {R.}~\bibnamefont {Paviot}}, \bibinfo {author} {\bibfnamefont {W.~J.}\ \bibnamefont {Percival}}, \bibinfo {author} {\bibfnamefont {I.}~\bibnamefont {P\'erez-R\`afols}}, \bibinfo {author} {\bibfnamefont {P.}~\bibnamefont {Petitjean}}, \bibinfo {author} {\bibfnamefont {M.~M.}\ \bibnamefont {Pieri}}, \bibinfo {author} {\bibfnamefont {A.}~\bibnamefont {Prakash}}, \bibinfo {author} {\bibfnamefont {A.}~\bibnamefont {Raichoor}}, \bibinfo {author} {\bibfnamefont {C.}~\bibnamefont {Ravoux}}, \bibinfo {author} {\bibfnamefont {M.}~\bibnamefont {Rezaie}}, \bibinfo {author} {\bibfnamefont {J.}~\bibnamefont {Rich}}, \bibinfo {author} {\bibfnamefont {A.~J.}\ \bibnamefont {Ross}}, \bibinfo {author} {\bibfnamefont {G.}~\bibnamefont {Rossi}}, \bibinfo {author} {\bibfnamefont {R.}~\bibnamefont {Ruggeri}}, \bibinfo {author} {\bibfnamefont {V.}~\bibnamefont {Ruhlmann-Kleider}}, \bibinfo {author} {\bibfnamefont {A.~G.}\ \bibnamefont {S\'anchez}}, \bibinfo {author}
  {\bibfnamefont {F.~J.}\ \bibnamefont {S\'anchez}}, \bibinfo {author} {\bibfnamefont {J.~R.}\ \bibnamefont {S\'anchez-Gallego}}, \bibinfo {author} {\bibfnamefont {C.}~\bibnamefont {Sayres}}, \bibinfo {author} {\bibfnamefont {D.~P.}\ \bibnamefont {Schneider}}, \bibinfo {author} {\bibfnamefont {H.-J.}\ \bibnamefont {Seo}}, \bibinfo {author} {\bibfnamefont {A.}~\bibnamefont {Shafieloo}}, \bibinfo {author} {\bibfnamefont {A.~c.~v.}\ \bibnamefont {Slosar}}, \bibinfo {author} {\bibfnamefont {A.}~\bibnamefont {Smith}}, \bibinfo {author} {\bibfnamefont {J.}~\bibnamefont {Stermer}}, \bibinfo {author} {\bibfnamefont {A.}~\bibnamefont {Tamone}}, \bibinfo {author} {\bibfnamefont {J.~L.}\ \bibnamefont {Tinker}}, \bibinfo {author} {\bibfnamefont {R.}~\bibnamefont {Tojeiro}}, \bibinfo {author} {\bibfnamefont {M.}~\bibnamefont {Vargas-Maga\~na}}, \bibinfo {author} {\bibfnamefont {A.}~\bibnamefont {Variu}}, \bibinfo {author} {\bibfnamefont {Y.}~\bibnamefont {Wang}}, \bibinfo {author} {\bibfnamefont {B.~A.}\ \bibnamefont
  {Weaver}}, \bibinfo {author} {\bibfnamefont {A.-M.}\ \bibnamefont {Weijmans}}, \bibinfo {author} {\bibfnamefont {C.}~\bibnamefont {Y\`eche}}, \bibinfo {author} {\bibfnamefont {P.}~\bibnamefont {Zarrouk}}, \bibinfo {author} {\bibfnamefont {C.}~\bibnamefont {Zhao}}, \bibinfo {author} {\bibfnamefont {G.-B.}\ \bibnamefont {Zhao}},\ and\ \bibinfo {author} {\bibfnamefont {Z.}~\bibnamefont {Zheng}},\ }\bibfield  {title} {\bibinfo {title} {Completed sdss-iv extended baryon oscillation spectroscopic survey: Cosmological implications from two decades of spectroscopic surveys at the apache point observatory},\ }\href {https://doi.org/10.1103/PhysRevD.103.083533} {\bibfield  {journal} {\bibinfo  {journal} {Phys. Rev. D}\ }\textbf {\bibinfo {volume} {103}},\ \bibinfo {pages} {083533} (\bibinfo {year} {2021})}\BibitemShut {NoStop}%
\bibitem [{\citenamefont {Ade}\ \emph {et~al.}(2018)\citenamefont {Ade}, \citenamefont {Ahmed}, \citenamefont {Aikin}, \citenamefont {Alexander}, \citenamefont {Barkats}, \citenamefont {Benton}, \citenamefont {Bischoff}, \citenamefont {Bock}, \citenamefont {Bowens-Rubin}, \citenamefont {Brevik}, \citenamefont {Buder}, \citenamefont {Bullock}, \citenamefont {Buza}, \citenamefont {Connors}, \citenamefont {Cornelison}, \citenamefont {Crill}, \citenamefont {Crumrine}, \citenamefont {Dierickx}, \citenamefont {Duband}, \citenamefont {Dvorkin}, \citenamefont {Filippini}, \citenamefont {Fliescher}, \citenamefont {Grayson}, \citenamefont {Hall}, \citenamefont {Halpern}, \citenamefont {Harrison}, \citenamefont {Hildebrandt}, \citenamefont {Hilton}, \citenamefont {Hui}, \citenamefont {Irwin}, \citenamefont {Kang}, \citenamefont {Karkare}, \citenamefont {Karpel}, \citenamefont {Kaufman}, \citenamefont {Keating}, \citenamefont {Kefeli}, \citenamefont {Kernasovskiy}, \citenamefont {Kovac}, \citenamefont {Kuo},
  \citenamefont {Larsen}, \citenamefont {Lau}, \citenamefont {Leitch}, \citenamefont {Lueker}, \citenamefont {Megerian}, \citenamefont {Moncelsi}, \citenamefont {Namikawa}, \citenamefont {Netterfield}, \citenamefont {Nguyen}, \citenamefont {O'Brient}, \citenamefont {Ogburn}, \citenamefont {Palladino}, \citenamefont {Pryke}, \citenamefont {Racine}, \citenamefont {Richter}, \citenamefont {Schillaci}, \citenamefont {Schwarz}, \citenamefont {Sheehy}, \citenamefont {Soliman}, \citenamefont {St.~Germaine}, \citenamefont {Staniszewski}, \citenamefont {Steinbach}, \citenamefont {Sudiwala}, \citenamefont {Teply}, \citenamefont {Thompson}, \citenamefont {Tolan}, \citenamefont {Tucker}, \citenamefont {Turner}, \citenamefont {Umilt\'a}, \citenamefont {Vieregg}, \citenamefont {Wandui}, \citenamefont {Weber}, \citenamefont {Wiebe}, \citenamefont {Willmert}, \citenamefont {Wong}, \citenamefont {Wu}, \citenamefont {Yang}, \citenamefont {Yoon},\ and\ \citenamefont {Zhang}}]{BK15}%
  \BibitemOpen
  \bibfield  {author} {\bibinfo {author} {\bibfnamefont {P.~A.~R.}\ \bibnamefont {Ade}}, \bibinfo {author} {\bibfnamefont {Z.}~\bibnamefont {Ahmed}}, \bibinfo {author} {\bibfnamefont {R.~W.}\ \bibnamefont {Aikin}}, \bibinfo {author} {\bibfnamefont {K.~D.}\ \bibnamefont {Alexander}}, \bibinfo {author} {\bibfnamefont {D.}~\bibnamefont {Barkats}}, \bibinfo {author} {\bibfnamefont {S.~J.}\ \bibnamefont {Benton}}, \bibinfo {author} {\bibfnamefont {C.~A.}\ \bibnamefont {Bischoff}}, \bibinfo {author} {\bibfnamefont {J.~J.}\ \bibnamefont {Bock}}, \bibinfo {author} {\bibfnamefont {R.}~\bibnamefont {Bowens-Rubin}}, \bibinfo {author} {\bibfnamefont {J.~A.}\ \bibnamefont {Brevik}}, \bibinfo {author} {\bibfnamefont {I.}~\bibnamefont {Buder}}, \bibinfo {author} {\bibfnamefont {E.}~\bibnamefont {Bullock}}, \bibinfo {author} {\bibfnamefont {V.}~\bibnamefont {Buza}}, \bibinfo {author} {\bibfnamefont {J.}~\bibnamefont {Connors}}, \bibinfo {author} {\bibfnamefont {J.}~\bibnamefont {Cornelison}}, \bibinfo {author} {\bibfnamefont
  {B.~P.}\ \bibnamefont {Crill}}, \bibinfo {author} {\bibfnamefont {M.}~\bibnamefont {Crumrine}}, \bibinfo {author} {\bibfnamefont {M.}~\bibnamefont {Dierickx}}, \bibinfo {author} {\bibfnamefont {L.}~\bibnamefont {Duband}}, \bibinfo {author} {\bibfnamefont {C.}~\bibnamefont {Dvorkin}}, \bibinfo {author} {\bibfnamefont {J.~P.}\ \bibnamefont {Filippini}}, \bibinfo {author} {\bibfnamefont {S.}~\bibnamefont {Fliescher}}, \bibinfo {author} {\bibfnamefont {J.}~\bibnamefont {Grayson}}, \bibinfo {author} {\bibfnamefont {G.}~\bibnamefont {Hall}}, \bibinfo {author} {\bibfnamefont {M.}~\bibnamefont {Halpern}}, \bibinfo {author} {\bibfnamefont {S.}~\bibnamefont {Harrison}}, \bibinfo {author} {\bibfnamefont {S.~R.}\ \bibnamefont {Hildebrandt}}, \bibinfo {author} {\bibfnamefont {G.~C.}\ \bibnamefont {Hilton}}, \bibinfo {author} {\bibfnamefont {H.}~\bibnamefont {Hui}}, \bibinfo {author} {\bibfnamefont {K.~D.}\ \bibnamefont {Irwin}}, \bibinfo {author} {\bibfnamefont {J.}~\bibnamefont {Kang}}, \bibinfo {author} {\bibfnamefont
  {K.~S.}\ \bibnamefont {Karkare}}, \bibinfo {author} {\bibfnamefont {E.}~\bibnamefont {Karpel}}, \bibinfo {author} {\bibfnamefont {J.~P.}\ \bibnamefont {Kaufman}}, \bibinfo {author} {\bibfnamefont {B.~G.}\ \bibnamefont {Keating}}, \bibinfo {author} {\bibfnamefont {S.}~\bibnamefont {Kefeli}}, \bibinfo {author} {\bibfnamefont {S.~A.}\ \bibnamefont {Kernasovskiy}}, \bibinfo {author} {\bibfnamefont {J.~M.}\ \bibnamefont {Kovac}}, \bibinfo {author} {\bibfnamefont {C.~L.}\ \bibnamefont {Kuo}}, \bibinfo {author} {\bibfnamefont {N.~A.}\ \bibnamefont {Larsen}}, \bibinfo {author} {\bibfnamefont {K.}~\bibnamefont {Lau}}, \bibinfo {author} {\bibfnamefont {E.~M.}\ \bibnamefont {Leitch}}, \bibinfo {author} {\bibfnamefont {M.}~\bibnamefont {Lueker}}, \bibinfo {author} {\bibfnamefont {K.~G.}\ \bibnamefont {Megerian}}, \bibinfo {author} {\bibfnamefont {L.}~\bibnamefont {Moncelsi}}, \bibinfo {author} {\bibfnamefont {T.}~\bibnamefont {Namikawa}}, \bibinfo {author} {\bibfnamefont {C.~B.}\ \bibnamefont {Netterfield}}, \bibinfo
  {author} {\bibfnamefont {H.~T.}\ \bibnamefont {Nguyen}}, \bibinfo {author} {\bibfnamefont {R.}~\bibnamefont {O'Brient}}, \bibinfo {author} {\bibfnamefont {R.~W.}\ \bibnamefont {Ogburn}}, \bibinfo {author} {\bibfnamefont {S.}~\bibnamefont {Palladino}}, \bibinfo {author} {\bibfnamefont {C.}~\bibnamefont {Pryke}}, \bibinfo {author} {\bibfnamefont {B.}~\bibnamefont {Racine}}, \bibinfo {author} {\bibfnamefont {S.}~\bibnamefont {Richter}}, \bibinfo {author} {\bibfnamefont {A.}~\bibnamefont {Schillaci}}, \bibinfo {author} {\bibfnamefont {R.}~\bibnamefont {Schwarz}}, \bibinfo {author} {\bibfnamefont {C.~D.}\ \bibnamefont {Sheehy}}, \bibinfo {author} {\bibfnamefont {A.}~\bibnamefont {Soliman}}, \bibinfo {author} {\bibfnamefont {T.}~\bibnamefont {St.~Germaine}}, \bibinfo {author} {\bibfnamefont {Z.~K.}\ \bibnamefont {Staniszewski}}, \bibinfo {author} {\bibfnamefont {B.}~\bibnamefont {Steinbach}}, \bibinfo {author} {\bibfnamefont {R.~V.}\ \bibnamefont {Sudiwala}}, \bibinfo {author} {\bibfnamefont {G.~P.}\ \bibnamefont
  {Teply}}, \bibinfo {author} {\bibfnamefont {K.~L.}\ \bibnamefont {Thompson}}, \bibinfo {author} {\bibfnamefont {J.~E.}\ \bibnamefont {Tolan}}, \bibinfo {author} {\bibfnamefont {C.}~\bibnamefont {Tucker}}, \bibinfo {author} {\bibfnamefont {A.~D.}\ \bibnamefont {Turner}}, \bibinfo {author} {\bibfnamefont {C.}~\bibnamefont {Umilt\'a}}, \bibinfo {author} {\bibfnamefont {A.~G.}\ \bibnamefont {Vieregg}}, \bibinfo {author} {\bibfnamefont {A.}~\bibnamefont {Wandui}}, \bibinfo {author} {\bibfnamefont {A.~C.}\ \bibnamefont {Weber}}, \bibinfo {author} {\bibfnamefont {D.~V.}\ \bibnamefont {Wiebe}}, \bibinfo {author} {\bibfnamefont {J.}~\bibnamefont {Willmert}}, \bibinfo {author} {\bibfnamefont {C.~L.}\ \bibnamefont {Wong}}, \bibinfo {author} {\bibfnamefont {W.~L.~K.}\ \bibnamefont {Wu}}, \bibinfo {author} {\bibfnamefont {H.}~\bibnamefont {Yang}}, \bibinfo {author} {\bibfnamefont {K.~W.}\ \bibnamefont {Yoon}},\ and\ \bibinfo {author} {\bibfnamefont {C.}~\bibnamefont {Zhang}} (\bibinfo {collaboration} {Keck Array and
  bicep2 Collaborations}),\ }\bibfield  {title} {\bibinfo {title} {Constraints on primordial gravitational waves using planck, wmap, and new bicep2/keck observations through the 2015 season},\ }\href {https://doi.org/10.1103/PhysRevLett.121.221301} {\bibfield  {journal} {\bibinfo  {journal} {Phys. Rev. Lett.}\ }\textbf {\bibinfo {volume} {121}},\ \bibinfo {pages} {221301} (\bibinfo {year} {2018})}\BibitemShut {NoStop}%
\bibitem [{\citenamefont {Sailer}\ \emph {et~al.}(2022)\citenamefont {Sailer}, \citenamefont {Chen},\ and\ \citenamefont {White}}]{Sailer_2022}%
  \BibitemOpen
  \bibfield  {author} {\bibinfo {author} {\bibfnamefont {N.}~\bibnamefont {Sailer}}, \bibinfo {author} {\bibfnamefont {S.-F.}\ \bibnamefont {Chen}},\ and\ \bibinfo {author} {\bibfnamefont {M.}~\bibnamefont {White}},\ }\bibfield  {title} {\bibinfo {title} {Optical depth to reionization from perturbative 21 cm clustering},\ }\href {https://doi.org/10.1088/1475-7516/2022/10/007} {\bibfield  {journal} {\bibinfo  {journal} {Journal of Cosmology and Astroparticle Physics}\ }\textbf {\bibinfo {volume} {2022}}\bibinfo  {number} { (10)},\ \bibinfo {pages} {007}}\BibitemShut {NoStop}%
\end{thebibliography}%
\end{document}